\documentclass[12pt,a4paper]{article}
\usepackage{amssymb,amsmath}

\usepackage{graphicx}

\newcounter{mathe}[section]

\newtheorem{thm}{Theorem}[section]
\newtheorem{cor}[thm]{Corollary}
\newtheorem{lem}[thm]{Lemma}

\newtheorem{defn}[thm]{Definition}
\newtheorem{rem}[thm]{Remark}

\newtheorem{nota}[thm]{Notation}

\begin{document}
\begin{titlepage}
\begin{center}

\begin{Huge}
Existence of Spontaneous Pair Creation \end{Huge}

\vspace{4cm}

\begin{Large}
Dissertation von Peter Pickl \vspace{0.5cm}
\end{Large}

Vorgelegt am 8.7.2005 \vspace{0.5cm}

an der Fakult\"at f\"ur Mathematik, Informatik und Statistik der Ludwig-Maximilians-Universit\"at M\"unchen

\vspace{2cm}

Berichterstatter: Prof. Dr. D. D\"urr, Prof. Dr. S. Teufel, Prof. Dr. G. Nenciu

\vspace{0.5cm} Tag des Rigorosums: 15.12.2005

\end{center}

\cleardoublepage
\end{titlepage}

\section*{}
\thispagestyle{empty}

\cleardoublepage
 \tableofcontents

\cleardoublepage

\section*{}
\thispagestyle{empty}

\cleardoublepage

\section{Introduction}

\vspace{0.5cm}

Pair creation - particularly electron-positron-pair creation - is
a often discussed phenomenon in modern physics. In this work we
want to restrict ourselves to the creation of electron positron
pairs from the vacuum in strong external fields which vary slowly
in time.

It is a well known fact (see \cite{thaller} for example) that pair
creation is not possible in static fields, so at first sight it
would be natural to assume, that the probability of creating a
pair from the vacuum goes to zero as the time derivative of the
external field goes to zero, too. As we will see later, this is
not true in general. If the external field strength exceeds a
critical threshold, there will be a sudden jump in the probability
of adiabatic pair creation. This sudden jump led to the notion of
"spontaneous pair creation".

The problem of spontaneous pair creation has been discussed in the physical literature (see for example
\cite{greiner}). Several attempts have been made to verify or falsify the existence of spontaneous pair creation
by experiments. So far there exist no clear experimental data neither in favor nor against the existence of
spontaneous par creation. The reason for the lack of such experiments is that the critical strength of the field
which is needed to create pairs spontaneously is very big. One possible way to produce such strong potentials is
by colliding heavy ions. The total charge of the collided ions has to be greater than 180e - so one has to
collide two atoms not much smaller than Uranium. Such a collision leads two many other reactions in the nuclei,
so it is not easy to distinguish where the detected electrons and positrons come from.

In the physical literature the Dirac equation with the so called ''Dirac sea'' interpretation is used to
describe pair creation. We will do the same since this interpretation gives a picturesque description of the
situation which leads us also through the mathematical argument. Therefore we will formulate the theorem of
spontaneous pair creation as a one particle Dirac problem. Non rigourously the problem can be described as
follows: Is it possible to get transitions between the negative and the positive continuous spectrum by
adiabatically turning on an external potential beyond the critical value and adiabatically turning it off again?
After that we introduce the Dirac equation in second quantization and give a heuristical connection between the
Dirac equation with ''Dirac sea'' interpretation and the Dirac equation in second quantization, which gives us
also the right heuristics how to formulate and prove a Corollary which translates the Theorem of spontaneous
pair creation into the formalism of second quantized Dirac equation. Then we will prove the Theorem of
spontaneous pair creation itself.

For this proof we will need some spectral properties of the Dirac operators and properties of its bound states
and generalized eigenfunctions. Increasing the coupling constant of the potential one will see, that any bound
state vanishes in the positive continuous spectrum. For the proof we will need that the time derivative of the
energy of the bound state is non zero as it reaches $m$ - the threshold of the positive continuous spectrum. In
this case there always exists a bound state with energy equal to $m$. In our first attempt to prove the
existence of spontaneous pair creation we made the mistake in assuming, that there exists no bound state with
energy $m$, hence our old proof was worthless.

That mistake was revealed by Gheorge Nenciu, who also suggested some literature that was very helpful in
understanding the true situation.

\vspace{0.5cm}

\subsection{The free Dirac equation}\label{diracsection}

\vspace{0.5cm}

 The Dirac equation was one of the first equations to describe particle-antiparticle-creation and
-annihilation effects.

Originally the goal of the Dirac equation was to have a covariant
relativistic wave equation - generalizing the non-relativistic
Schr\"odinger equation.

\begin{equation}\label{dirac0}
  i\frac{\partial\psi_{t}}{\partial t}=-i\sum_{l=1}^{3}\alpha_{l}\partial_{l}\psi_{t}+\beta m\psi_{t}\equiv D^0\psi_{t}
\end{equation}

With complex valued $\alpha_l$ and $\beta$ this is obviously not
possible as, to get the right ''dispersion relation'' (i.e. the
relativistic energy-momentum relation $E^2=k^2+m^2$ under Fourier
transforming $\psi_t(x)$ in $\mathbf{t}$ and $\mathbf{x}$), the
$\alpha_l$ and $\beta$ may not commute, but with matrix valued
$\alpha_l$ and $\beta$ it is. One possible choice is

\begin{eqnarray}\label{alphas}
\alpha_{l}=
\begin{pmatrix}
  _{0} & _{\sigma_{l}}\\
  _{\sigma_{l}} & _{0}
\end{pmatrix}; \beta=
\begin{pmatrix}
  _{\mathbf{1}} & _{0}\\
  _{0} & _{-\mathbf{1}}
\end{pmatrix}; l=1,2,3
\end{eqnarray} with $\sigma_{l}$ being the Pauli matrices:
\begin{eqnarray*}
 \sigma_{1}=\begin{pmatrix}
  _{1} & _{0} \\
  _{0} & _{-1}
\end{pmatrix}; \sigma_{2}=\begin{pmatrix}
  _{0} & _{1} \\
  _{1} & _{0}
\end{pmatrix}; \sigma_{3}=\begin{pmatrix}
  _{0} & _{-i} \\
  _{i} & _{0}\;
\end{pmatrix}\;.
\end{eqnarray*}

This choice is called the "standard representation" and was
introduced by Dirac.

So $\psi$ is not a complex valued function, but a 4-vector valued
function and the underlying Hilbert space is
$\mathcal{H}=L^2(\mathbb{R}^3)^4=L^2(\mathbb{R}^3)\otimes\mathbb{C}^4$.

The ''generalized'' Eigenfunctions $\phi$ of $D^0$

$$D^0\phi=E_k\phi=\pm\sqrt{m^2+k^2}\phi$$

are of the form $e^{i\mathbf{k}\cdot \mathbf{x}}\gamma$, where gamma is a (complex valued) 4-vector. For any
$\mathbf{k}$ four different choices of $\gamma$ are possible, so for each $\mathbf{k}$ we get four different
Eigenfunctions. Two of them have positive energy, two of them negative energy. So we denote the Eigenfunctions
of $D^0$ by $\phi^{\pm,j,\mathbf{k}}$, where the sign stands for the sign of the energy, $j\in\{1,2\}$ for the
two different spins.

As the sign of the energy will play an important role in the
following sections, we define the subspaces $\mathcal{H}_+$ which
is the span of the eigenfunctions with positive energy,
$\mathcal{H}_-$ which is the span of the eigenfunctions with
negative energy and the projectors $P^+$ and $P^-$ into these
spaces. As these subspaces are orthogonal we can write

$$\mathcal{H}=\mathcal{H}_+\oplus\mathcal{H}_-$$

So in contrast to the free Schr\"odinger Hamiltonian $H_0$, the essential spectrum of $D^0$ is not bounded from
below. It consists of two absolutely continuous parts, a positive part reaching from $m$ to $\infty$ and a
negative part reaching from $-m$ to $-\infty$. In the case of static external fields this does not lead to any
problems, though the physical interpretation of electrons with negative energy may be difficult. But introducing
coupling of the electron to a radiation field would lead to a radiation catastrophe: Any electron would fall
down the negative energy continuum emitting radiation.

This radiation catastrophe can be heuristically overcome by using antisymmetrized wave functions. The wave
function $\Omega$ describing the ''vacuum'' is the antisymmetrized product of all the eigenfunctions of $D^0$
with negative energy (the so called Dirac sea)

$$\Omega=\prod^{antisym}_{k\in\mathbb{R};\, j=1;2}\phi^{\pm,j,\mathbf{k}}\;.$$

The antisymmetrized product of $\Omega$ and an additional wave
function $\psi\in\mathcal{H}_+$ describes an electron with
position probability density $\mid\psi\mid^2$. Introducing an
external interaction which - to keep things simple for the moment
- is supposed to have no influence on $\Omega$, any one electron
wave function $\psi_i\times\Omega$ with $\psi_i\in\mathcal{H}_+$
will after the interaction stay a wave function of the form
$\psi_f\times\Omega$ with $\psi_f\in\mathcal{H}_+$ since all
possible one particle states in $\Omega$ are occupied and adding
one more state would by antisymmetrization yield zero. This
phenomenon is called Pauli's exclusion principle. Although this
heuristic picture has not been made rigorous we use it anyhow, as
it is very descriptive.

Starting with the vacuum $\Omega$, it is now possible that a ''photon'' with energy bigger than $2m$ is absorbed
by one of the electrons with negative energy, which is thus lifted into the positive energy spectrum leaving a
"hole" in the Dirac sea behind. We end up with a wave function, which is the antisymmetrized product of all
eigenfunctions with negative energy except one and some $\psi\in\mathcal{H}_+$. The hole propagates - the
propagation is given indirectly by the propagation of all the other electrons in the Dirac sea - like a particle
with the same mass but opposite charge as the electron and is physically interpreted as a positron - the
antiparticle of the electron. Therefore we interpret the factors of our multi particle wave functions with
positive energy as electrons, missing factors with negative energy (''holes in the Dirac sea'') as positrons.
This is the so called ''sea interpretation'' of the Dirac equation, originally found by Dirac.

\vspace{0.5cm}

\subsection{The Dirac sea under influence of an external potential}

\vspace{0.5cm}

We shall now give a more quantitative description.  To keep the description of the interaction as simple as
possible we neglect interaction between the different particles - so the differential equations describing our
multi particle wave function decouple and we can use the Dirac equation for each factor $\psi^r_s$ of our multi
particle wave function $\Psi_t=\prod^{antisym}_{r\in \mathbb{R}}\psi^r_t$ separately.

The one particle Dirac equation with external potential reads:

\begin{equation}\label{dirac1}
  i\frac{\partial\psi_{t}}{\partial t}=-i\sum_{l=1}^{3}\alpha_{l}\partial_{l}\psi_{t}+A\hspace{-0.2cm}/\psi_{t}+\beta m\psi_{t}\equiv(D^0+A\hspace{-0.2cm}/)\psi_{t}
\end{equation}

where

\begin{equation}\label{potentialA}
A\hspace{-0.2cm}/=\mathbf{1}A_0 +\sum_{l=1}^{3}\alpha_{l}A_{l}\;.
\end{equation}

To get a physical interpretation of $\Psi_t$ (for example to
calculate the amplitude of pair creation) one has to write it down
in the ''free eigenbasis'' or more to say as linear combination of
products of eigenfunctions of the free Dirac equation, though it
may make sense to use another basis calculating the $\psi_t$ than
the set of eigenfunctions of the free Dirac equation.

\cleardoublepage

\section{Spontaneous pair creation in Dirac theory}

\vspace{0.5cm}

\subsection{Formulation of the problem}

\vspace{0.5cm}

One way of phrasing our result is that we show the existence of
slowly varying potentials which create pairs. A very simple way to
describe such a potential is the so called adiabatic switching
formalism: We consider a potential which can be factorized into a
purely space-dependent part and a purely time dependent part - the
so called switching factor.

\begin{eqnarray}\label{potential}
A\hspace{-0.2cm}/\hspace{0.03cm}_{s}^\lambda(\mathbf{x})&=&\lambda
\varphi(s)A\hspace{-0.2cm}/\hspace{0.03cm}(\mathbf{x}) \;.
\end{eqnarray}

Here $A_l\in C^{\infty}$ (see connection between $A\hspace{-0.2cm}/\hspace{0.03cm}(\mathbf{x})$ and $A_l$ in
(\ref{potentialA})) for all $l\in\{0,1,2,3\}$, $A\hspace{-0.2cm}/\hspace{0.03cm}(\mathbf{x})$ has compact
support $\mathcal{C}$ and can (due to CPT-symmetry without loss of generality) be defined to be repulsive for
electrons. For $\varphi$ we assume: $\varphi(s)\in C^1$ with

\begin{eqnarray}\label{switch}
\lim_{s\rightarrow\pm\infty}\varphi(s)&=&0\\\nonumber
\partial_{s}\varphi(s)&<&0\hspace{0.5cm} \text{for}\hspace{0.2cm}
s<0\\\nonumber
\partial_{s}\varphi(s)&>&0\hspace{0.5cm} \text{for}\hspace{0.2cm}
s>0\\\nonumber
\partial_{s}\varphi(s)&=&0\hspace{0.5cm} \text{for}\hspace{0.2cm}
s=0\\\nonumber \mid\partial_{s}\varphi(s)\mid&<&C \hspace{0.4cm}
\text{for some}\hspace{0.2cm} C\in\mathbb{R}^{+}\\
\nonumber\varphi_0&=&1
 \;.
\end{eqnarray}

We consider now a slowly varying function $A_{t\varepsilon}$ in (\ref{dirac1}). Then going to the macroscopic
time scale $s=t\varepsilon$ and introducing $\psi^\varepsilon_s=\psi_{\frac{s}{\varepsilon}}$ we obtain

\begin{equation}\label{dirac}
  i\varepsilon\frac{\partial\psi^\varepsilon_{s}}{\partial
  s}=(D^0+A\hspace{-0.2cm}/\hspace{0.03cm}^\lambda_{s}(\mathbf{x}))\psi_{s}^\varepsilon=:D_s\psi_{s}^\varepsilon
\end{equation}

By our description of the Dirac sea we expect (see \cite{greiner})
that for sufficiently strong potentials the probability of
creating (at least) one pair from the vacuum is one in the
adiabatic limit ($\lim_{\varepsilon\rightarrow0}$).

Heuristically to get a multi particle wave function
$\Psi_s^\varepsilon$ which describes adiabatic pair creation from
the vacuum with probability one (which means
$\lim_{\varepsilon\rightarrow0}\lim_{s\rightarrow-\infty}\Psi_s^\varepsilon=\Omega$
and
$\lim_{\varepsilon\rightarrow0}\lim_{s\rightarrow+\infty}\langle\Psi_s^\varepsilon\mid\Omega\rangle=0$)
at least one of the factors $\psi_s^{\varepsilon,r}$ of
$\Psi_s^\varepsilon$ has to lie in the positive energy subspace
$\mathcal{H}_+$ in the limit
$\lim_{\varepsilon\rightarrow0}\lim_{s\rightarrow+\infty}$:

\begin{equation}\label{eins}
\lim_{\varepsilon\rightarrow0}\lim_{s\rightarrow+\infty}\|
P^+\psi_s^{\varepsilon,r}\|=1\;.
\end{equation}

Since
$\lim_{\varepsilon\rightarrow0}\lim_{s\rightarrow-\infty}\Psi_s^\varepsilon=\Omega$
all the $\psi_s^{\varepsilon,r}$ lie in the negative energy
subspace $\mathcal{H}_-$ in the limit
$\lim_{\varepsilon\rightarrow0}\lim_{s\rightarrow-\infty}$:

\begin{equation}\label{zwei}
\lim_{\varepsilon\rightarrow0}\lim_{s\rightarrow-\infty}\|
P^+\psi_s^{\varepsilon,r}\|=0\;.
\end{equation}

So heuristically spontaneous pair creation can be treated as a
problem of the one particle Dirac equation: Find solutions of the
one particle Dirac equation with external potential (\ref{dirac})
which satisfy (\ref{eins}) and (\ref{zwei}).

We shall prove that such solutions exist. We shall also show that this result holds correspondingly in the so
called second quantized Dirac field with external field setting, whenever the latter makes sense, i.e. that pair
creation holds with exactly the same parameters in Fock space.

\vspace{0.5cm}

\subsection{SPC as one particle Dirac problem}

\vspace{0.5cm}

Dealing with the Dirac equation with static external potentials it is helpful to choose the eigenbasis of the
Dirac operator $D^0+A\hspace{-0.2cm}/$.

Since the time variation of the potentials is very slow, presenting the wave function in the eigenbasis of the
Dirac operator will still be helpful. But we have a family of Dirac operators indexed by $s$ and thus a family
of eigenenbasis'. One expects that under fairly general conditions on the potential, the eigenfunctions of the
Dirac operator $D^0+\mu A\hspace{-0.2cm}/$ and their eigenvalues vary smoothly with $\mu$, thus in the adiabatic
case the wave function representation in the ''time dependent basis'' will remain invariant (up to a certain
error). In particular no jumping over spectral gaps is possible.

The only way to get transition from the negative to the positive continuous spectrum is by ''lifting'' bound
states. Take a potential without bound states which is repulsive for electrons. Increasing $\mu$ at least one
bound state emerges from the negative continuous spectrum, is transported through the spectral gap between the
positive and negative continuous spectrum and vanishes - let us say at $\mu=\mu_1$ - in the positive continuous
spectrum. An example for such a potential is a scalar potential

$$
 A(\mathbf{x})=\bigg{\{}
\begin{tabular}{llll}
 1 \hspace{1cm}&for&$x<R$
\\
  0 \hspace{1cm}&for&$x>R$\\
\end{tabular}
$$

for some $R\in\mathbb{R}$ (see \cite{nenciu2}).

With this picture in mind, we are able to understand the cause of
spontaneous pair creation: As long as $\lambda$ is smaller than
$\mu_1$ (undercritical regime), the positive continuous energy
spectrum is isolated from the rest of the spectrum for all times
and no transition from negative to positive energies is possible.


\begin{figure}
\begin{center}
\includegraphics*[width=300pt,height=150pt]{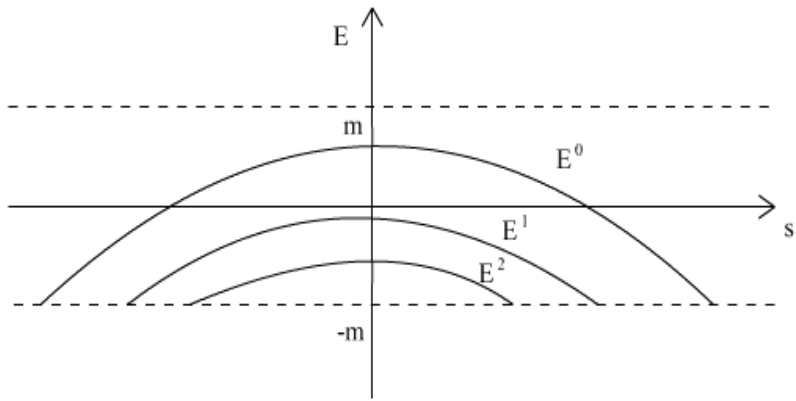} \caption[under]{Bound
spectrum of the Dirac operator in the undercritical case}
\end{center}
\end{figure}

Choosing $\lambda>\mu_1$ (overcritical regime) transitions might be possible.

\begin{figure}
\begin{center}
\includegraphics*[width=300pt,height=100pt]{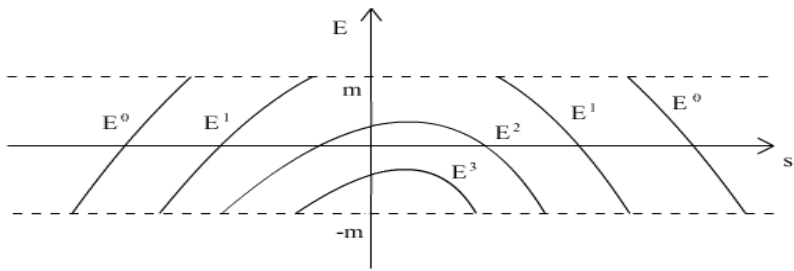}

\caption[over]{Bound spectrum of the Dirac operator in the overcritical case}
\end{center}
\end{figure}

So under fairly general conditions on the potential
$A\hspace{-0.2cm}/\hspace{0.03cm}(\mathbf{x})$ we expect a sudden
change in the probability of creating a pair in the adiabatic
limit: For any switching factor satisfying (\ref{switch}) one
expects, that the probability of creating a pair from the vacuum
is zero as long as $\lambda$ is smaller than $\mu_1$ and one for
$\lambda>\mu_1$. The critical value $\mu_1$ for $\lambda$ is
referred to as $\lambda_c$ in the literature.

\vspace{0.5cm}

\section{Existence of spontaneous pair creation}

\vspace{0.5cm}

Let $\lambda>\lambda_c$. As $\lambda$ will be fixed in the following, we drop the index ''$\lambda$'' for the
potential. We assume that only one eigenvalue $E_s$ - which may be degenerated - disappears in the upper
continuous spectrum. Let $\mathcal{N}_{s}$ denote the respective set of bound states, i.e. for all $\phi_s\in
\mathcal{N}_s$

$$D_s\phi_{s}:=(D^0+A\hspace{-0.2cm}/\hspace{0.03cm}_{s}(\mathbf{x}))\phi_{s}=E_s\phi_{s}\;.$$

Let $s_{m1}<0$ the time the bound states disappear in the continuous positive spectrum and $s_{m2}>0$ the time,
the bound states evolve again

\begin{eqnarray}\label{qualitativ}
\lim_{s\nearrow s_{m1}}E_s&=&m\\\nonumber \lim_{s\searrow s_{m2}}E_s&=&m\;.
\end{eqnarray}

Any normalized bound state $\phi_s\in\mathcal{N}_s$ could in principle lead to a pair creation. In the following
these bound states will be called ''overcritical''.

\begin{defn}\label{Properdef}

We use the notation ''properly dives into the positive continuous spectrum'' for overcritical bound states if
there exists a $s_0<s_{m1}$ such that

\begin{equation}\label{diveproperlys}
0<\partial_s E_s<C
\end{equation}

for all $s_0\leq s\leq s_{m1}$.

\end{defn}

For the heuristic treatment we assume that $E_s$ is not
degenerated, i.e. that (up to a phase factor) only one bound state
is overcritical.  A generalization to overcritical function which
do not dive into the positive continuous spectrum properly is
possible, too, but laborious and - as the class of overcritical
functions of bound states which do not dive into the positive
continuous spectrum probably is very special - is left out in this
work.

The theorem later will be formulated for an arbitrary number of
 overcritical
functions.

\begin{figure}\label{Bild3}
\begin{center}
\includegraphics*[width=300pt,height=150pt]{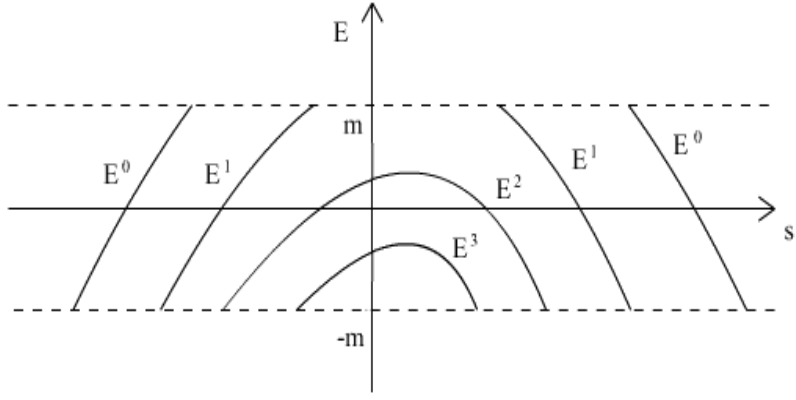}
\caption[zeichnung]{Energy eigenvalue of $\phi_{\varphi(s)}$}
\end{center}
\end{figure}

Let $s_{m1}$ and $s_{m2}$ denote the values of $s$ where the bound
state disappears in the positive continuous spectrum and where it
evolves again

\begin{equation}\label{Esm1}
\lim_{s\nearrow s_{m1}}E_s=m \hspace{2cm}\lim_{s\searrow
s_{m2}}E_s=m\;.
\end{equation}

$s_i<s_{m1}$ and $s_f>s_{m2}$ be values of $s$ where the overcritical bound state already / still exist.
Furthermore we choose $s_i$ and $s_f$ so that there is no crossing of the energy of the eigenstates with the
energy of other eigenstates for $s_i<s<s_f$.

Let us consider a wave function $\psi^\varepsilon_s$ which is
solution of the Dirac equation (\ref{dirac}) with potential
$A\hspace{-0.2cm}/\hspace{0.03cm}_{s}(\mathbf{x})$ and which is at
time $t_i=\frac{s_i}{\varepsilon}$ equal (up to a phase factor) to
the overcritical bound state of the Dirac operator with potential
$A\hspace{-0.2cm}/\hspace{0.03cm}_{s_i}(\mathbf{x})$.

\begin{equation}\label{psi}
\psi^\varepsilon_{s_i}=\phi_{s_i}
\end{equation}

Due to the adiabatic switching formalism, $\psi^\varepsilon_s$
will lie in the negative continuous energy spectrum for all times
$s<s_0$, so $\psi^\varepsilon_{s_i}$ will satisfy (\ref{eins}).
Assume, that $\psi^\varepsilon_{s_f}$ is orthogonal to
$\phi_{s_f}$ (so $\psi^\varepsilon_s$ ''missed'' the only way,
that would lead it back into the negative continuous spectrum).
Such a $\psi^\varepsilon_s$ will satisfy (\ref{zwei}), too. As we
heuristically assumed that wave functions which satisfy
(\ref{eins}) and (\ref{zwei}) describe the creation of a pair, we
define the probability of adiabatic pair creation as

\begin{eqnarray}\label{haupt}
p(A\hspace{-0.2cm}/\hspace{0.03cm})=1-\lim_{\varepsilon\rightarrow0}\langle\psi_{s_f}^\varepsilon\mid\phi_{s_f}\rangle\;.
\end{eqnarray}

($\langle f\mid g\rangle$ denotes the scalar product $\int f^t g d^3x$. In our case $f$ and $g$ are vector
valued.)

We shall show later on, that this is in fact equal to the
probability of adiabatic pair creation in a Fock space setting.

\vspace{0.5cm}

\subsection{The Theorem}

\vspace{0.5cm}

 For technical reasons we will restrict ourselves to pure electric potentials (in this case
$A\hspace{-0.2cm}/\hspace{0.03cm}$ is a multiple of the unit matrix, so we can treat it as a scalar and write
$A$ instead of $A\hspace{-0.2cm}/\hspace{0.03cm}$) of the form defined in (\ref{potential}) where
$A(\mathbf{x})$ is such that

\begin{equation}\label{restricta1}
\int A(\mathbf{x})\phi_{\lambda_c}(\mathbf{x})d^3x\neq0
\end{equation}

or

\begin{equation}\label{restricta2}
(1-i\beta)\int A(\mathbf{x})\phi_{\lambda_c}(\mathbf{x})\mathbf{x}d^3x\neq0
\end{equation}

\begin{rem}

The set of potentials $A$ which satisfy (\ref{restricta1}) or (\ref{restricta2}) is not empty. One can easily
prove that the s-wave of a spherical step potential satisfies (\ref{restricta1}): In this case
$\phi_{\lambda_c}$ can be split into two two-spinors which are both eigenstates of the Schr\"odinger equation
with certain potential and energy. The lowest energy state of the Schr\"odinger equation is always positive, so
is $A$ and (\ref{restricta1}) follows.

\end{rem}\vspace{1cm}

\begin{thm}\label{onecrit}
Let $\lambda A_{s}(\mathbf{x})$ be of the form defined in (\ref{potential}) with $A_l=0$ for $l\neq 0$ and
$A>0$, where

\begin{tabular}{rl}
  - & $A(\mathbf{x})\in C^\infty$ is compactly supported and satisfies (\ref{restricta1}) or (\ref{restricta2}), \\
  - & $\lambda$ is such that only one eigenvalue $E_s$ disappears in the upper continuous\\
  &  spectrum and \\
  - & $\varphi(s)$ is such that $E_{s}$ dives properly into the positive continuous spectrum. \\
\end{tabular}

Let  $\phi_{s}$ and $\widetilde{\phi}_{s}$ be overcritical bound states of $A_s(\mathbf{x})$, let $s_i<s_{m1}$
and $s_f>s_{m2}$ be such that $\phi_{s_i}$ and $\widetilde{\phi}_{s_f}$ already / still exist. Let
$\psi_{s}^\varepsilon$ be solution of the Dirac equation (\ref{dirac}) with $\psi_{s_i}^\varepsilon=\phi_{s_i}$,
then

\begin{equation}\label{onecriteq}
\lim_{\varepsilon\rightarrow0}\langle\psi_{s_f}^\varepsilon\mid\widetilde{\phi}_{s_f}\rangle=0\;.
\end{equation}

\end{thm}\vspace{1cm}

One direct result of Theorem \ref{onecrit} and the time adiabatic theorem is

\begin{cor}\label{psiprop}
Let $\widetilde{\psi}_s^\varepsilon$ be any solution of the dirac equation
 where
$\widetilde{\psi}_{s_i}^\varepsilon$ is orthogonal to all
overcritical bound states $\phi_{s_i}$ and
$\lim_{s\rightarrow-\infty}\|
P^+\widetilde{\psi}_s^\varepsilon\|=0$. Then under the conditions
of Theorem \ref{onecrit}

\begin{eqnarray*}
\lim_{\varepsilon\rightarrow0}\lim_{s\rightarrow\pm\infty}\|
P^\pm\psi_s^\varepsilon\|&=&1\\
\lim_{\varepsilon\rightarrow0}\lim_{s\rightarrow+\infty}\|
P^+\widetilde{\psi}_s^\varepsilon\|&=&0
\end{eqnarray*}

\end{cor}\vspace{1cm}\vspace{1cm}

To prove the existence of spontaneous pair creation we follow the propagation of $\psi^\varepsilon_s$. For
sufficiently small $\varepsilon$, $\psi^\varepsilon_s$ follows more or less the bound states $\phi_{s}$.
Reaching the time $s_{m1}$ the bound state ''vanishes'' in the positive continuous spectrum of the Hamiltonian.
It is left to assure, that $\psi^\varepsilon_s$ will stay in the positive continuous spectrum after removing the
potential again and not fall back via the overcritical bound state into the negative energy spectrum.

\vspace{0.5cm}

\subsection{Nenciu's contribution}

\vspace{0.5cm}

The picture we gave so far has been developed by Nenciu \cite{nenciu1}. He did not prove all of Theorem
\ref{onecrit} but conjectured it. Instead he used a switching factor with a jump of height $\delta$ at
$s=s_{m1}-\mathcal{O}(\delta)$ chosen in a way that for $s<s_{m1}-\mathcal{O}(\delta)$ the bound state is
isolated from the positive energy spectrum and for $s_{m1}-\mathcal{O}(\delta)\leq s<s_{m2}$ the bound state
disappears. This jump of course is a violation of the adiabatic idea. But it gives a regime with a sudden change
in the pair creation amplitude when the coupling constant reaches the critical value.

For $\delta\ll1$ the part of the wave function which does not lie in the upper continuous spectrum is
negligible. So for any fixed $0<\delta\ll1$ the wave function will show a typical scattering state behavior and
thus propagate away from the range of the potential. Hence it is orthogonal to any bound state which may
reappear at times $s>s_{m2}$. Hence we have pair creation with probability one in the limit
$\lim_{\delta\rightarrow0}\lim_{\varepsilon\rightarrow0}$.

\cleardoublepage

\vspace{0.5cm}

\section{Fock space formulation of the theorem}

\vspace{0.5cm}

The usually accepted way of describing pair creation and annihilation effects is by virtue of the so called
Second Quantized Dirac Equation with external potential.

We shall show that our main Theorem \ref{onecrit} yields as corollary the corresponding statement in the setting
of second quantized formulation.

\vspace{0.5cm}
\subsection{Heuristic connection between Fock space and Dirac sea}

\vspace{0.5cm}

The Dirac sea uses the ideas of a wave function describing an infinite number of particles. This idea has not
been made mathematically rigorous. We want to give a phenomenological description: Any heuristically
antisymmetrized multi particle wave function $\Psi_t=\prod^{antisym}_{r\in \mathbb{R}}\psi^r_t$ where all the
$\psi^r_t$ are eigenfunctions of the free Dirac operator can be clearly characterized by giving the number of
states with positive energy and the number of holes in the Dirac sea and their spins and momenta. Generalization
to wave functions $\Psi_t=\prod^{antisym}_{r\in \mathbb{R}}\psi^r_t$ which are not products of eigenfunctions is
possible via linear combination of the $\Psi_t$.

Let us introduce the Fock space $\mathcal{F}$. This space
essentially focuses on the arbitrary number of electron positron
$\mathbf{pairs}$ which may be present. One takes the direct sum of
all spaces $\mathcal{F}^{(n)}$ describing $n$ electron-positron
pairs

$$\mathcal{F}=\oplus_{n=0}^{\infty}\mathcal{F}^{(n)}$$

where the wave functions in $\mathcal{F}^{(n)}$ describe $n$
particles with positive energy and $n$ holes in the Dirac sea (see
the interpretation of the multi particle wave functions given in
section \ref{diracsection}):

$$\mathcal{F}^{(n)}=\mathcal{F}^{(n)}_{+}\otimes\mathcal{F}^{(n)}_{-}\;.$$

$\mathcal{F}^{(n)}_{\pm}$ are the antisymmetrized tensor products
of $n$ copies of $\mathcal{H}_{\pm}$.

Here we have defined the Fock space for zero total charge. It is
possible to allow more electrons than positrons or vice versa. But
as we are dealing with pair creation from the vacuum (the state
where no electrons and positrons are present), the total charge of
our system is always zero.

\vspace{0.5cm}

\subsection{Spontaneous pair creation in Second Quantized Dirac Theory}\label{quedsection}

\vspace{0.5cm}

What happens in the presence of a compactly supported time dependent potential
$A\hspace{-0.2cm}/\hspace{0.03cm}_{t}$? In contrast to the free case the vacuum is in general not stable anymore
(see discussion in section \ref{diracsection}). Electrons with negative energy may propagate into the positive
energy spectrum, leaving holes in the Dirac sea, so pair creation may occur.

We want to estimate the amplitude of pair creations and the wave
functions of the created electrons and positrons during a small
time interval $\Delta t$.

Let us start at time $t$ with an arbitrary number $n$ of particles with positive energy. As we only deal with an
uncharged system, there are also $n$ holes in the Dirac sea - or in other words all but $n$ negative energy
states are occupied. Assume that at time $t=0$ all the wave functions $\psi_t^r$  are eigenfunctions of $D^0$.
In general $\mid\Psi_t\rangle$ will of course be a superposition of such wave functions, also with different
numbers of pairs, but due to linearity the generalization is easy. Since there is no interaction between the
particles, the differential equations for the different particles decouple and the propagation of the system is
given by the propagation of each single particle. In general there might be transition from $\mathcal{H}^+$ to
$\mathcal{H}^-$, so pair creation may occur.

To calculate the amplitude of the pair creations and the wave functions of the electrons and positrons at time
$t+\Delta t$, one has to observe the propagation of all particles which are - as mentioned above - of infinite
number. For each factor $\psi^r_t$ of the multi particle wave function $\Psi_t=\prod^{antisym}_{r\in
\mathbb{R}}\psi^r_t$ the propagation is given by the Dirac equation (\ref{dirac1}). So for small $\Delta t$ we
get (using the notation $D_t$ for the Dirac operator $D^0+A\hspace{-0.2cm}/\hspace{0.03cm}_{t}$)

$$\psi^r_{t+\Delta t}=(1-i\Delta tD_t)\psi^r_{t}$$

as approximation.

Look at the situation in Fock space, identifying particles with
positive energy as electrons and holes in the Dirac sea as
positrons.

Let us start with a vector in Fock space describing $n$ electrons and $n$ positrons whose wave functions are
eigenfunctions of $D^0$.

From the propagation of the particles which are present in the
Dirac-sea picture we construct the new Fock space vector. We
separate each Dirac particle, calculate its propagation and put it
back again. Taking away a particle with negative energy leaves a
hole in the Dirac sea. In Fock-space language, taking away a
particle with negative energy means to create a positron, taking
away a particle with positive energy means to annihilate an
electron.

Therefore we define the operators $a_{\pm,j,\mathbf{k}}$ as the creators of a electron/positron with spin $j$
and momentum $\mathbf{k}$, $a^\dag_{\pm,j,\mathbf{k}}$ as the annihilators. As the multi particle wave function
was antisymmetrized, these operators satisfy the anti-commutation relations

\begin{eqnarray}\label{anticom}
a_{\pm,j,\mathbf{k}}a_{\pm,j^{\prime},\mathbf{k}^{\prime}}^{\dag}+a_{\pm,j^{\prime},\mathbf{k}^{\prime}}^{\dag}a_{\pm,j,\mathbf{k}}=
\delta(\mathbf{k},\mathbf{k}^\prime)\delta(s,j^\prime)\nonumber\\
a_{\mp,j,\mathbf{k}}a_{\pm,j^{\prime},\mathbf{k}^{\prime}}^{\dag}+a_{\pm,j^{\prime},\mathbf{k}^{\prime}}^{\dag}a_{\mp,j,\mathbf{k}}=0
\end{eqnarray}

Otherwise it would be possible to create two electrons or two
positrons with same spin and same momentum, which is a violation
of Pauli's exclusion principle.

We start with the propagation of the particles with negative
energy. As described above, we take away a particle with negative
energy and ''quantum numbers'' $j$ and $\mathbf{k}$ for spin and
momentum. In Fock space language, taking away a electron with
negative energy is $\mathbf{creation}$ of a positron
\begin{equation}\label{focknu}
\mid\nu_t\rangle=a_{-,j,\mathbf{k}}\mid\Psi_t\rangle
\end{equation}

Then we calculate the propagation of the particle with the given
spin and momentum

$$\varphi_{t+\Delta t}=(1-i\Delta tD_t)\phi^{-,j,\mathbf{k}}$$

We calculate for each ''quantum number'' $\pm$, $j^\prime$ and
$\mathbf{k}^\prime$ the scalar product of
$\phi^{\pm,j^\prime,\mathbf{k}^\prime}$ and $\varphi_{t+\Delta t}$

$$\widehat{\varphi}^{\pm,j^{\prime},\mathbf{k}^\prime}_{t+\Delta t}=
\langle\phi^{\pm,j^\prime,\mathbf{k}^\prime}\mid(1-i\Delta
tD_t)\phi^{-,j,\mathbf{k}}\mid\rangle$$

 and create/annihilate all the electrons/positrons with
these quantum numbers in the given amplitude to $\nu$ (see
\ref{focknu}):

\begin{eqnarray*}
&\big(&\sum_{j^\prime=1,2}\int_{\mathbf{k}^\prime}a^\dag_{-,j,\mathbf{k}}\langle\phi^{-,j^\prime,\mathbf{k}^\prime}\mid1-i\Delta
tD_t\phi^{-,j,\mathbf{k}}\mid\rangle
d^3k^\prime\\&&+\sum_{j^\prime=1,2}\int_{\mathbf{k}^\prime}a_{+,j,\mathbf{k}}\langle\phi^{+,j^\prime,\mathbf{k}^\prime}\mid1-i\Delta
tD_t\phi^{-,j,\mathbf{k}}\mid\rangle d^3k^\prime\;\;\big)\;\;a_{-,j,\mathbf{k}}\mid\Psi_t\rangle\;.
\end{eqnarray*}

We do this procedure for any occupied state of the Dirac sea.
Since \nolinebreak{$a_{-,j,\mathbf{k}}\mid\Psi\rangle=0$} if there
is a positron with spin $j$ and $\mathbf{k}$ present in
$\mid\Psi\rangle$ we can generalize this formula to all states -
not caring whether they are occupied or not. This leads us to:

\begin{eqnarray*}
\sum_{j=1,2}\int_{\mathbf{k}}&\big(&\sum_{j^\prime=1,2}\int_{\mathbf{k}^\prime}a^\dag_{-,j,\mathbf{k}}\langle\phi^{-,j^\prime,\mathbf{k}^\prime}\mid1-i\Delta
tD_t\phi^{-,j,\mathbf{k}}\mid\rangle
d^3k^\prime\\&&+\sum_{j^\prime=1,2}\int_{\mathbf{k}^\prime}a_{+,j,\mathbf{k}}\langle\phi^{+,j^\prime,\mathbf{k}^\prime}\mid1-i\Delta
tD_t\phi^{-,j,\mathbf{k}}\mid\rangle d^3k^\prime\;\;\big)\;\;a_{-,j,\mathbf{k}}d^3k\mid\Psi_t\rangle\;.
\end{eqnarray*}

Observing the propagation of the particles with positive energy we
have to use - as described above - the annihilator of electrons.
Now we use the fact, that
$a^\dag_{+,j,\mathbf{k}}\mid\Psi\rangle=0$ if the electron with
the given quantum numbers is not present in $\mid\Psi\rangle$ to
generalize the formula to all $j$ and $\mathbf{k}$ without caring,
whether the particles with the given quantum numbers are present
in our system.

So observing the propagation of all particles we get for small
$\Delta t$:

\begin{eqnarray*}
\mid\Psi_{t+\Delta t}
\rangle&=&\big(\sum_{j=1,2}\int_{\mathbf{k}}\big(\sum_{j^\prime=1,2}\int_{\mathbf{k}^\prime}a^\dag_{-,j,\mathbf{k}}\langle\phi^{-,j^\prime,\mathbf{k}^\prime}\mid
1-i\Delta tD_t\phi^{-,j,\mathbf{k}}\mid\rangle
d^3k^\prime\\&&+\sum_{j^\prime=1,2}\int_{\mathbf{k}^\prime}a_{+,j,\mathbf{k}}\langle\phi^{+,j^\prime,\mathbf{k}^\prime}\mid1-i\Delta
tD_t\phi^{+,j,\mathbf{k}}\mid\rangle
d^3k^\prime\;\;\big)\;\;a^\dag_{+,j,\mathbf{k}}d^3k\\&&+\sum_{j=1,2}\int_{\mathbf{k}}\big(\sum_{j^\prime=1,2}\int_{\mathbf{k}^\prime}a^\dag_{-,j,\mathbf{k}}\langle\phi^{-,j^\prime,\mathbf{k}^\prime}\mid1-i\Delta
tD_t\phi^{-,j,\mathbf{k}}\mid\rangle
d^3k^\prime\\&&+\sum_{j^\prime=1,2}\int_{\mathbf{k}^\prime}a_{+,j,\mathbf{k}}\langle\phi^{+,j^\prime,\mathbf{k}^\prime}\mid1-i\Delta
tD_t\phi^{-,j,\mathbf{k}}\mid\rangle
d^3k^\prime\;\big)\;a_{-,j,\mathbf{k}}d^3k\;\big)\mid\Psi_t\rangle\;.
\end{eqnarray*}

and

\begin{eqnarray*}
\mid\Psi_{t+\Delta t} \rangle-\mid\Psi_{t} \rangle&=&-i\Delta t
\;\;\big(\;\;\sum_{j=1,2}\int_{\mathbf{k}}\big(\sum_{j^\prime=1,2}\int_{\mathbf{k}^\prime}a^\dag_{-,j,\mathbf{k}}\langle\phi^{-,j^\prime,\mathbf{k}^\prime}\mid
D_t\phi^{-,j,\mathbf{k}}\mid\rangle
d^3k^\prime\\&&+\sum_{j^\prime=1,2}\int_{\mathbf{k}^\prime}a_{+,j,\mathbf{k}}\langle\phi^{+,j^\prime,\mathbf{k}^\prime}\mid
D_t\phi^{+,j,\mathbf{k}}\mid\rangle
d^3k^\prime\;\;\big)\;\;a^\dag_{+,j,\mathbf{k}}d^3k\\&&+\sum_{j=1,2}\int_{\mathbf{k}}\big(\sum_{j^\prime=1,2}\int_{\mathbf{k}^\prime}a^\dag_{-,j,\mathbf{k}}\langle\phi^{-,j^\prime,\mathbf{k}^\prime}\mid
D_t\phi^{-,j,\mathbf{k}}\mid\rangle
d^3k^\prime\\&&+\sum_{j^\prime=1,2}\int_{\mathbf{k}^\prime}a_{+,j,\mathbf{k}}\langle\phi^{+,j^\prime,\mathbf{k}^\prime}\mid
D_t\phi^{-,j,\mathbf{k}}\mid\rangle
d^3k^\prime\;\big)\;\;a_{-,j,\mathbf{k}}d^3k\;\big)\mid\Psi_t\rangle\;.
\end{eqnarray*}

Thus dividing by $\Delta t$ we get with $\Delta t\rightarrow0$

\begin{eqnarray*}
i\partial_t\mid\Psi_t
\rangle&=&\big(\;\sum_{j=1,2}\int_{\mathbf{k}}\;\big(\;\sum_{j^\prime=1,2}\int_{\mathbf{k}^\prime}a^\dag_{-,j,\mathbf{k}}\langle\phi^{-,j^\prime,\mathbf{k}^\prime}\mid
D_t\phi^{-,j,\mathbf{k}}\mid\rangle
d^3k^\prime\\&&+\sum_{j^\prime=1,2}\int_{\mathbf{k}^\prime}a_{+,j,\mathbf{k}}\langle\phi^{+,j^\prime,\mathbf{k}^\prime}\mid
D_t\phi^{+,j,\mathbf{k}}\mid\rangle
d^3k^\prime\;\;\big)\;\;a^\dag_{+,j,\mathbf{k}}d^3k\\&&+\sum_{j=1,2}\int_{\mathbf{k}}\;\big(\;\sum_{j^\prime=1,2}\int_{\mathbf{k}^\prime}a^\dag_{-,j,\mathbf{k}}\langle\phi^{-,j^\prime,\mathbf{k}^\prime}\mid
D_t\phi^{-,j,\mathbf{k}}\mid\rangle
d^3k^\prime\\&&+\sum_{j^\prime=1,2}\int_{\mathbf{k}^\prime}a_{+,j,\mathbf{k}}\langle\phi^{+,j^\prime,\mathbf{k}^\prime}\mid
D_t\phi^{-,j,\mathbf{k}}\mid\rangle
d^3k^\prime\;\;\big)\;\;a_{-,j,\mathbf{k}}d^3k\;\;\big)\;\;\mid\Psi_t\rangle\;.
\end{eqnarray*}

Defining the field operators as

\begin{equation}\label{fieldop}
\widehat{\chi}:=\sum_{j=1}^2\int
\phi^{+,j}_{k}a_{+,j,k}+\phi^{-,j}_{k}a^\dag_{-,j,k}d^3k
\end{equation}

we get

\begin{equation}\label{sqde}
i\partial_t \mid\Psi_t\rangle=\int d^3 x\widehat{\chi}^\dag D_t
\widehat{\chi}\hspace{0.3cm}\mid\Psi_t\rangle
\end{equation}

as second quantized Dirac equation with $\int d^3
x\widehat{\chi}^\dag D \widehat{\chi}\hspace{0.3cm}$ as Dirac
field Hamiltonian.





We want to draw from Corollary \ref{psiprop} another Corollary asserting the pair creation in the Second
Quantized Dirac Equation. That equation is not always well depending on the choice of $A$. We assume that $A$ be
such that the second quantized Dirac equation makes sense (see \cite{thaller} for references). We call such
$A$'s good. Using the notation $\mid\Omega\rangle$ for the vacuum and having the ideas of the previous sections
in mind, we have

\begin{cor}\label{qedmain}
Let $A\hspace{0.03cm}_{s}(\mathbf{x})$ be a good potential of the
form defined (\ref{potential}) with (at least) one overcritical
bound state. Let this overcritical bound state dive properly into
the positive continuous spectrum. Let
$\mid\Psi^\varepsilon_t\rangle$ be a solution of the Second
Quantized Dirac Equation with potential
$A\hspace{0.03cm}_{\frac{t}{\varepsilon}}(\mathbf{x})$ with
$\lim_{t\rightarrow-\infty}\mid\Psi^\varepsilon_t\rangle=\mid\Omega\rangle$.

Then

$$\lim_{\varepsilon\rightarrow0}\lim_{t\rightarrow\infty}\langle\Omega\mid\Psi^\varepsilon_t\rangle=0\;.$$

\end{cor}\vspace{1cm}

\vspace{0.5cm}
\noindent\textbf{Proof of Corollary \ref{qedmain}}\\

We shall prove this Corollary rigorously, following the intuition given by the sea picture. ''Adding'' an
electron with positive energy to the multi particle wave function $\Psi_t=\prod^{antisym}_{r\in
\mathbb{R}}\psi^r_t$ corresponds to the creation of an electron in Fock space, ''adding'' an electron with
negative energy corresponds to the annihilation of a positron.

''Subtracting'' an electron with positive energy corresponds to the annihilation of an electron, ''subtracting''
an electron with negative energy corresponds to the creation of a positron.

Hence we define for any $\omega\in L^{2}(\mathbb{R}^{3})\bigotimes\mathbb{C}^{4}$ the operators in Fock space

\begin{eqnarray}\label{add}
\widehat{\omega}^\dag&:=&\sum_{j=1}^2\int\langle\omega ,
\phi^{+,j}_{k}\rangle
a^\dag_{+,j,k}d^3k+\sum_{j=1}^2\int\langle\omega ,
\phi^{-,j}_{k}\rangle a_{-,j,k}d^3k\\\nonumber
\widehat{\omega}&:=&\sum_{j=1}^2\int\langle\omega ,
\phi^{+,j}_{k}\rangle a_{+,j,k}d^3k+\sum_{j=1}^2\int\langle\omega
, \phi^{-,j}_{k}\rangle a_{-,j,k}^\dag d^3k\;.
\end{eqnarray}

whereas $\phi^{+,j}_{k}$ and $\phi^{-,j}_{k}$ are the solutions of
the free Dirac equation with momentum $\mathbf{k}$ spin
$j\in\{1;2\}$ and positive and negative energy respectively.

Following the ideas above we get

\begin{lem}\label{tausend}
Let $\mid\Psi_t\rangle$ be solution of the Second Quantized Dirac
Equation (\ref{sqde}), $\xi_t\in
 L^{2}(\mathbb{R}^{3})\bigotimes\mathbb{C}^{4}$ be solution of the
 Dirac equation (\ref{dirac}).

 Then

\begin{eqnarray*}
\mid\widetilde{\Psi}_t\rangle&:=&\widehat{\xi_t}^\dag\mid\Psi_t\rangle\\
\mid\widetilde{\widetilde{\Psi}}_t\rangle&:=&\widehat{\xi_t}\mid\Psi_t\rangle
\end{eqnarray*}

are solutions of the Second Quantized Dirac Equation (\ref{sqde}).
\end{lem}\vspace{1cm}

The proof of this Lemma is given below.

We use this Lemma on $\mid\Psi^\varepsilon_t\rangle$, the special
$\lim_{t\rightarrow-\infty}\mid\Psi^\varepsilon_t\rangle=\mid\Omega\rangle$ solution of the second quantized
Dirac equation with potential $A\hspace{0.03cm}_{\frac{t}{\varepsilon}}(\mathbf{x})$ and on
$\psi_{\frac{t}{\varepsilon}}^\varepsilon$. It follows that

\begin{equation}\label{newwave}
\mid\overline{\Psi}_t^\varepsilon\rangle:=\widehat{\psi_{\frac{t}{\varepsilon}}^\varepsilon}^\dag\widehat{\psi_{\frac{t}{\varepsilon}}^\varepsilon}\mid\Psi_t^\varepsilon\rangle
\end{equation}
is solution of the Second Quantized Dirac Equation. Furthermore
one can show by direct calculation (using the commutator relations
of the $a_{\pm,j,\mathbf{k}}$ and $ a^{\dag}_{\pm,j,\mathbf{k}}$,
the fact, that $a^\dag_{\pm,j,\mathbf{k}}\mid\Omega\rangle=0$ and
(\ref{zwei})) that

\begin{eqnarray}\label{andersrum}
&&\lim_{t\rightarrow-\infty}\mid\overline{\Psi_t^\varepsilon}\rangle\;\;
\\&=&
\lim_{t\rightarrow-\infty}\widehat{\psi_{\frac{t}{\varepsilon}}^\varepsilon}^\dag\widehat{\psi_{\frac{t}{\varepsilon}}^\varepsilon}\mid\Psi_t^\varepsilon\rangle\;\;
\nonumber\nonumber\\&=&\lim_{t\rightarrow-\infty}\widehat{\psi}^{\varepsilon\dag}_t\big(\sum_{l=1}^2\int\langle\psi_{\frac{t}{\varepsilon}}^\varepsilon
\mid \phi^{+,j}_{k}\rangle\;\;
a_{+,j,k}d^3k\mid\Psi_t^\varepsilon\rangle\;\;\nonumber\\&&+\sum_{j=1}^2\int\langle\psi_{\frac{t}{\varepsilon}}^\varepsilon\mid
\phi^{-,j}_{k}\rangle\;\; a_{-,j,k}^\dag
d^3k\mid\Psi_t^\varepsilon\rangle\;\;\big)
\nonumber\nonumber\\&=&\lim_{t\rightarrow-\infty}\widehat{\psi}^{\varepsilon\dag}_t\big(\sum_{j=1}^2\int\langle\psi_{\frac{t}{\varepsilon}}^\varepsilon
\mid \phi^{-,j}_{k}\rangle\;\; a_{-,j,k}^\dag
d^3k\mid\Psi_t^\varepsilon\rangle\;\;\big)
\nonumber\nonumber\\&=&\lim_{t\rightarrow-\infty}\big(\sum_{l=1}^2\int\langle\psi_{\frac{t}{\varepsilon}}^\varepsilon
\mid \phi^{+,j^\prime}_{k}\rangle\;\;
a^\dag_{+,l,k}d^3k\nonumber\\&&+\sum_{l=1}^2\int\langle\psi_{\frac{t}{\varepsilon}}^\varepsilon
\mid \phi^{-,j^\prime}_{k}\rangle\;\;
a_{-,l,k}d^3k\big)\sum_{j=1}^2\int\langle\psi_{\frac{t}{\varepsilon}}^\varepsilon\mid
\phi^{-,j}_{k}\rangle\;\; a_{-,j,k}^\dag
d^3k\mid\Psi_t^\varepsilon\rangle\;\;
\nonumber\nonumber\\&=&\lim_{t\rightarrow-\infty}\sum_{l,j=1}^2\int\int\langle\psi_{\frac{t}{\varepsilon}}^\varepsilon
\mid \phi^{-,j^\prime}_{k^\prime}\rangle\;\;
\langle\psi_{\frac{t}{\varepsilon}}^\varepsilon\mid
\phi^{-,j}_{k}\rangle\;\; a_{-,l,k^\prime} a_{-,j,k}^\dag
\mid\Psi_t^\varepsilon\rangle\;\; d^3k^\prime d^3k
\nonumber\nonumber\\&=&\lim_{t\rightarrow-\infty}\big(\sum_{l,j=1}^2\int\int\langle\psi_{\frac{t}{\varepsilon}}^\varepsilon
\mid \phi^{-,j^\prime}_{k^\prime}\rangle\;\;
\langle\psi_{\frac{t}{\varepsilon}}^\varepsilon\mid
\phi^{-,j}_{k}\rangle\;\;
\delta(j,l)\delta(\mathbf{k},\mathbf{k}^\prime)\mid\Psi_t^\varepsilon\rangle\;\;
d^3k^\prime d^3k
\nonumber\\&&-\sum_{l,j=1}^2\int\int\langle\psi_{\frac{t}{\varepsilon}}^\varepsilon
\mid \phi^{-,j^\prime}_{k^\prime}\rangle\;\;
\langle\psi_{\frac{t}{\varepsilon}}^\varepsilon\mid
\phi^{-,j}_{k}\rangle\;\; a_{-,j,k}^\dag
a_{-,l,k^\prime}\mid\Psi_t^\varepsilon\rangle\;\; d^3k^\prime
d^3k\big)
\nonumber\nonumber\\&=&\lim_{t\rightarrow-\infty}\sum_{j=1}^2\int\langle\psi_{\frac{t}{\varepsilon}}^\varepsilon
\mid \phi^{-,j}_{\mathbf{k}}\rangle\;\;
\langle\psi_{\frac{t}{\varepsilon}}^\varepsilon\mid
\phi^{-,j}_{k}\rangle\;\; \mid\Psi_t^\varepsilon\rangle\;\; d^3k
\nonumber\nonumber\\&=&\lim_{t\rightarrow-\infty}\int\mid
P^-\mathcal{F}_0(\psi_{\frac{t}{\varepsilon}}^\varepsilon)\mid^2
 d^3k \mid\Psi_t^\varepsilon\rangle\;\;
\nonumber\nonumber\\&=&\lim_{t\rightarrow-\infty}\mid\Psi_t^\varepsilon\rangle\;\;\;.
\end{eqnarray}



Using the uniqueness of the Dirac propagation it follows, that

$$\mid\Psi_t^\varepsilon\rangle=\mid\overline{\Psi_t^\varepsilon}\rangle\;.$$

This equation,
$\lim_{t\rightarrow\infty}\langle\psi_{\frac{t}{\varepsilon}}^\varepsilon
\mid \phi^{-,j}_{k}\rangle=0$ (which follows directly from
Corollary \ref{psiprop}) and
$a^\dag_{\pm,j,\mathbf{k}}\mid\Omega\rangle=0$ yield

\begin{eqnarray*}
\lim_{\varepsilon\rightarrow0}\lim_{t\rightarrow\infty}\langle\Omega\mid\Psi^\varepsilon_t\rangle
&=&\lim_{\varepsilon\rightarrow0}\lim_{t\rightarrow\infty}\langle\Omega\mid\widehat{\psi_{\frac{t}{\varepsilon}}^\varepsilon}^\dag\mid\overline{\Psi_t^\varepsilon}\rangle
\\&=&\lim_{\varepsilon\rightarrow0}\lim_{t\rightarrow\infty}\langle\widehat{\psi_{\frac{t}{\varepsilon}}^\varepsilon}\Omega\mid\overline{\Psi_t^\varepsilon}\rangle
\\&=&\lim_{\varepsilon\rightarrow0}\lim_{t\rightarrow\infty}\langle\big(\sum_{j=1}^2\int\langle\psi_{\frac{t}{\varepsilon}}^\varepsilon
\mid \phi^{+,j}_{k}\rangle
a^\dag_{+,j,k}d^3k\\&&+\sum_{j=1}^2\int\langle\psi_{\frac{t}{\varepsilon}}^\varepsilon
\mid \phi^{-,j}_{k}\rangle
a_{-,j,k}d^3k\big)\Omega\mid\overline{\Psi_t^\varepsilon}\rangle
\\&=&\lim_{\varepsilon\rightarrow0}\lim_{t\rightarrow\infty}\langle\big(\sum_{j=1}^2\int\langle\psi_{\frac{t}{\varepsilon}}^\varepsilon
\mid \phi^{+,j}_{k}\rangle
a^\dag_{+,j,k}d^3k\big)\Omega\mid\overline{\Psi_t^\varepsilon}\rangle
\\&=&0
\end{eqnarray*}

which proves Corollary \ref{qedmain}.

\begin{flushright}$\Box$\end{flushright}

We shall now give the proof of Lemma \ref{tausend}. For ease of
reference we recall what the Lemma says.

\vspace{0.5cm}
\noindent\textbf{Lemma \ref{tausend}}\\

\it

Let $\mid\Psi_t\rangle$ be solution of the Second Quantized Dirac
Equation (\ref{sqde}), $\xi_t\in
 L^{2}(\mathbb{R}^{3})\bigotimes\mathbb{C}^{4}$ be solution of the
 Dirac equation (\ref{dirac}) with $\widehat{\xi_t}$ defined as in
 (\ref{add}).

 Then

\begin{eqnarray}
\label{firstpart}\mid\widetilde{\Psi}_t\rangle&:=&\widehat{\xi_t}^\dag\mid\Psi_t\rangle\\
\label{secondpart}\mid\widetilde{\widetilde{\Psi}}_t\rangle&:=&\widehat{\xi_t}\mid\Psi_t\rangle
\end{eqnarray}

are solutions of the Second Quantized Dirac Equation (\ref{sqde}).

\rm

\vspace{0.5cm}
\noindent\textbf{Proof}\\

We will only prove (\ref{firstpart}). (\ref{secondpart}) follows
analogously. We shall leave out the index $t$.

We need to show that

\begin{equation}\label{show}
\partial_t
\big(\widehat{\xi}^\dag\mid\Psi\rangle\big)=H\big(\widehat{\xi}^\dag\mid\Psi\rangle\big)
\end{equation}

We start with the right hand side:

\begin{eqnarray*}
H\big(\widehat{\xi}^\dag\mid\Psi_t\rangle\big)&=&\int \widehat{\psi}^\dag D \widehat{\psi}d^3
x\hspace{0.3cm}\widehat{\xi}^\dag\mid\Psi_t\rangle\;.
\end{eqnarray*}

Inserting the field operator(\ref{fieldop}) leads to
\begin{eqnarray*}
H\big(\widehat{\xi}^\dag\mid\Psi_t\rangle\big)
&=&\int \widehat{\psi}^\dag D \sum_{s=1}^2\int \phi^{+,s}_{k}a_{+,s,k}+\phi^{-,s}_{k}a^\dag_{-,s,k}d^3k d^3
x\hspace{0.3cm}\widehat{\xi}^\dag\mid\Psi_t\rangle\;.
\end{eqnarray*}
Using the definition (\ref{add}) of the operator
$\widehat{\xi}^\dag$ yields

\begin{eqnarray*}H\big(\widehat{\xi}^\dag\mid\Psi_t\rangle\big)&=&\int \widehat{\psi}^\dag D \sum_{s=1}^2\int
\phi^{+,s}_{k}a_{+,s,k}+\phi^{-,s}_{k}a^\dag_{-,s,k}d^3k
\sum_{s^{\prime}=1}^2\int\langle\xi,\phi^{+,s^{\prime}}_{k^{\prime}}\rangle
a^\dag_{+,s^{\prime},k^{\prime}}d^3k^{\prime}\\&&+\int\langle\xi ,
\phi^{-,s^{\prime}}_{k}\rangle
a_{-,s^{\prime},k^{\prime}}d^3k^{\prime}
d^3 x\hspace{0.3cm}\mid\Psi_t\rangle\;.
\end{eqnarray*}

Next we use the commutation relations for the
$a_{\pm,s,k}^{(\dag)}$ (\ref{anticom}). This leads to

\begin{eqnarray*}
H\big(\widehat{\xi}^\dag\mid\Psi_t\rangle\big)
&=&\widehat{\xi}^\dag\int \widehat{\psi}^\dag D \widehat{\psi}d^3 x\hspace{0.3cm}\mid\Psi_t\rangle\\&&+
\int \widehat{\psi}^\dag D \sum_{s,s^{\prime}=1}^2\int\int \langle\xi ,
\phi^{+,s^{\prime}}_{k^{\prime}}\rangle\phi^{+,s}_{k}\delta(k,k^\prime)\delta(s,s^\prime)\\&&\hspace{2cm}+\langle\xi
, \phi^{-,s^{\prime}}_{k^{\prime}}\rangle\phi^{-,s}_{k}\delta(k,k^\prime)\delta(s,s^\prime)d^3k d^3k^{\prime}
d^3 x\hspace{0.3cm}\mid\Psi_t\rangle\
\end{eqnarray*}

Executing the $d^3 k^{\prime}$-integration gives us

\begin{eqnarray*}
H\big(\widehat{\xi}^\dag\mid\Psi_t\rangle\big)
&=&\widehat{\xi}^\dag\int \widehat{\psi}^\dag D \widehat{\psi}d^3 x\hspace{0.3cm}\mid\Psi_t\rangle+
\int \widehat{\psi}^\dag D \sum_{s=1}^2\int \langle\xi ,
\phi^{+,s}_{k}\rangle\phi^{+,s}_{k}\\&&\hspace{5cm}+\langle\xi , \phi^{-,s}_{k}\rangle\phi^{-,s}_{k}d^3k d^3
x\hspace{0.3cm}\mid\Psi_t\rangle\\
&=&\widehat{\xi}^\dag\int \widehat{\psi}^\dag D \widehat{\psi}d^3 x\hspace{0.3cm}\mid\Psi_t\rangle+
\int \widehat{\psi}^\dag D \xi d^3 x\hspace{0.3cm}\mid\Psi_t\rangle\;.
\end{eqnarray*}

As $\xi$ is by definition solution of the Dirac equation
(\ref{dirac}) we can write

\begin{eqnarray*}
H\big(\widehat{\xi}^\dag\mid\Psi_t\rangle\big)
&=&\widehat{\xi}^\dag\int \widehat{\psi}^\dag D \widehat{\psi}d^3 x\hspace{0.3cm}\mid\Psi_t\rangle+
\int \widehat{\psi}^\dag \partial_t\xi d^3 x\hspace{0.3cm}\mid\Psi_t\rangle\;.
\end{eqnarray*}

We again use (\ref{fieldop}), now for $\widehat{\psi}^\dag $ and
get

\begin{eqnarray*}
H\big(\widehat{\xi}^\dag\mid\Psi_t\rangle\big)
&=&\widehat{\xi}^\dag\int \widehat{\psi}^\dag D \widehat{\psi}d^3 x\hspace{0.3cm}\mid\Psi_t\rangle\\&&+
\int \sum_{s=1}^2\int \phi^{+,s}_{k}a^\dag_{+,s,k}+\phi^{-,s}_{k}a_{-,s,k}d^3k
\partial_t\xi d^3
x\hspace{0.3cm}\mid\Psi_t\rangle\;.
\end{eqnarray*}

Executing the $d^3 x$-integration gives us

\begin{eqnarray*}
H\big(\widehat{\xi}^\dag\mid\Psi_t\rangle\big)
&=&\widehat{\xi}^\dag\int \widehat{\psi}^\dag D \widehat{\psi}d^3 x\hspace{0.3cm}\mid\Psi_t\rangle\\&&+
 \sum_{s=1}^2\int
\langle\phi^{+,s}_{k},\partial_t\xi\rangle a^\dag_{+,s,k}+\langle\phi^{-,s}_{k},\partial_t\xi\rangle
a_{-,s,k}d^3k \hspace{0.3cm}\mid\Psi_t\rangle\;.
\end{eqnarray*}

For the left hand side of (\ref{show}) we have

\begin{eqnarray*}
\partial_t
\big(\widehat{\xi}^\dag\mid\Psi_t\rangle\big)&=&\partial_t \widehat{\xi}^\dag\mid\Psi_t\rangle+
\widehat{\xi}^\dag\partial_t\mid\Psi_t\rangle\\
&=&\sum_{s=1}^2\int \langle\phi^{+,s}_{k},\partial_t\xi\rangle
a^\dag_{+,s,k}+\langle\phi^{-,s}_{k},\partial_t\xi\rangle a_{-,s,k}d^3k
\hspace{0.3cm}\mid\Psi_t\rangle\\&&+\widehat{\xi}^\dag\int \widehat{\psi}^\dag D \widehat{\psi}d^3
x\hspace{0.3cm}\mid\Psi_t\rangle
\end{eqnarray*}

and (\ref{show}) follows.

\begin{flushright}$\Box$\end{flushright}

\vspace{0.5cm}

\section{The ''Critical'' Bound State}\label{cbs}

\vspace{0.5cm}

We consider the Dirac operator

\begin{equation}\label{dsanders}
D_\mu:=D^0+\mu A(\mathbf{x})\;.
\end{equation}

We call a coupling constant $\lambda_c$ critical, if and only if there exists a solution $\phi_{\lambda_c}\in
L^2$ with

\begin{equation}\label{dgmp2}
\left(D_{\lambda_c}-m\right)\phi_{\lambda_c}\equiv0\;.
\end{equation}

Note that by our choice of the potential $A\in C^\infty$ the solution $\phi_{\lambda_c}$ is also $C^\infty$.



To see whether a critical state exists we invert (\ref{dgmp2}) and get formally

\begin{equation}\label{EH1}
 \phi_{\lambda_c}(\mathbf{x})=(m-D^0)^{-1}\lambda_c
A(\mathbf{x})\phi_{\lambda_c}(\mathbf{x})\;,
\end{equation}

and replacing the $(m-D^0)^{-1}=\lim_{\delta\rightarrow0}(m-D^0+i\delta)^{-1}$ by the integral kernel
$G^{+}_{k=0}=G^+$:

\begin{equation}\label{Green1}
 (m-D^0)G^{+}(\mathbf{x}-\mathbf{x'})=\delta(\mathbf{x}-\mathbf{x'})\;.
\end{equation}

where \cite{thaller}

\begin{equation}\label{kernel1}
G^{+}(\mathbf{x})=\frac{1}{4\pi}\left(-x^{-1}(m+\beta m)
-ix^{-2}\sum_{j=1}^{3}\alpha_{j}\frac{x_{j}}{x}\right)\;,
\end{equation}

we obtain the Lippmann Schwinger equation

\begin{eqnarray}\label{bseq}
 \phi_{\lambda_c}(\mathbf{x})&=&\int
 G^{+}(\mathbf{x'}) \lambda_cA(\mathbf{x}-\mathbf{x'})
 \phi_{\lambda_c}(\mathbf{x}-\mathbf{x'})d^{3}x'
 \\&=&-\int
\frac{1}{4\pi}x'^{-1}(m+\beta m) \lambda_cA(\mathbf{x}-\mathbf{x'})
 \phi_{\lambda_c}(\mathbf{x}-\mathbf{x'})d^{3}x'
\nonumber\\&&-\int \frac{1}{4\pi}x'^{-2}\sum_{j=1}^{3}\alpha_{j}\frac{x'_{j}}{x'}
\lambda_cA(\mathbf{x}-\mathbf{x'})
 \phi_{\lambda_c}(\mathbf{x}-\mathbf{x'})d^{3}x'
\nonumber\\&=:&\phi_{c,1}(\mathbf{x})+\phi_{c,2}(\mathbf{x})\;.
\end{eqnarray}

Since $A$ has compact support, $\phi_{c,2}(\mathbf{x})$ decays like $x^{-2}$ and thus is in $L^2$. For
$\phi_{c,1}(\mathbf{x})$ we can write

\begin{eqnarray}\label{vorwirdnull2}
\phi_{c,1}(\mathbf{x})&=&-x^{-1}\int \frac{1}{4\pi}(m+\beta m) \lambda_cA(\mathbf{x}-\mathbf{x'})
 \phi_{\lambda_c}(\mathbf{x}-\mathbf{x'})d^{3}x'
\nonumber\\&&-\int \frac{1}{4\pi}(x'^{-1}-x^{-1})(m+\beta m) \lambda_cA(\mathbf{x}-\mathbf{x'})
 \phi_{\lambda_c}(\mathbf{x}-\mathbf{x'})d^{3}x'
\nonumber\\&=:&\phi_{c,3}(\mathbf{x})+\phi_{c,4}(\mathbf{x})\;.
\end{eqnarray}

Using

\begin{eqnarray*}
x'^{-1}-x^{-1}&=&\frac{x-x'}{xx'}
\end{eqnarray*}

we see that for large $x$ and $\mathbf{x}-\mathbf{x}'\in\mathcal{S}_A$, the compact support of $A$,

$$\mid x'^{-1}-x^{-1}\mid\leq  \mbox{diam}(\mathcal{S}_A)\frac{1}{xx'}$$

is of order $x^{-2}$ and thus $\phi_{c,4}$ of order $x^{-2}$. Hence $\phi_{c,4}\in L^2$.

The decay of $\phi_{c,3}(\mathbf{x})$ depends on the spinor components of $\phi_{\lambda_c}(\mathbf{y})$. There
are two possibilities:

Either the spinor components of $\phi_{\lambda_c}(\mathbf{y})$ are such that

$$\int (1+\beta)
A(\mathbf{y})
 \phi_{\lambda_c}(\mathbf{y})d^{3}y\neq 0$$

and thus $\phi_{c,1}(\mathbf{x})$ is of order $x^{-1}$ and thus not in $L^2$ or such that the spinor

\begin{equation}\label{wirdnull}
\int(1+\beta) A(\mathbf{y})
 \phi_{\lambda_c}(\mathbf{y})d^{3}y= 0
\end{equation}

and thus $\phi_{c,1}(\mathbf{x})$ is of order $x^{-2}$ and thus in $L^2$. The identity (\ref{wirdnull}) will
play a crucial role later on.

It has been proven by Klaus \cite{klaus} that for ''critical'' bound states that dive into the positive
continuous spectrum properly $\phi_{\lambda_c}$ is in $L^2$. Thus (\ref{wirdnull}) holds in our case and we have
that $\phi_{\lambda_c}=\phi_{c,2}+\phi_{c,4}$, and thus

\begin{equation}\label{phicdecay}
\mid\phi_{\lambda_c}\mid\leq Cx^{-2}
\end{equation}
for some appropriate $C<\infty$.

\begin{nota}

In what follows the letters $C$ and $C_n$, $n\in\mathbb{N}_0$ will be used for various constants that need
 not be identical even within the same equation.

\end{nota}

In the following we denote the set of bound states present at $\mu=\lambda_c$ by $\mathcal{N}$:

\begin{equation}\label{defmengen}
\mathcal{N}:=\{\phi_{\lambda_c}\in L^2:(D_{\lambda_c}-m)\phi_{\lambda_c}=0\}\;.
\end{equation}

\vspace{0.5cm}

\section{Generalized Eigenfunctions}

\vspace{0.5cm}





In the following we will use with slight abuse of notation the spin index $j\in\{1,2,3,4\}$, where $j=1,3$
stands for the different spins of the eigenfunctions with negative energy, $j=2,4$ stands for the different
spins of the eigenfunctions with positive energy.

The generalized eigenfunctions $\phi^{j}(\mathbf{k},\mu,\mathbf{x})$ are solutions of

\begin{equation}\label{dgmp}
(-1)^j E_k\phi^{j}(\mathbf{k},\mu,\mathbf{x})=D_{\mu}\phi^{j}(\mathbf{k},\mu,\mathbf{x})
\end{equation}

with $E_k=\sqrt{m^2+k^2}$.

We change to the Lippmann Schwinger equation

\begin{equation}\label{weobtain}
\phi^{j}(\mathbf{k},\mu,\mathbf{x})=\phi^{j}(\mathbf{k},0,\mathbf{x})+\int
 G^{+}_{k}(\mathbf{x}-\mathbf{x'})\mu A(\mathbf{x'})
 \phi^{j}(\mathbf{k},\mu,\mathbf{x'})d^{3}x'
 \end{equation}

with $\phi^{j}(\mathbf{k},0,\mathbf{x})$ being the well known generalized eigenfunctions of the free Dirac
operator $D^0$ \cite{thaller} and $G^{+}_{k}$ being the kernel of
$(E_{k}-D^0)^{-1}=\lim_{\delta\rightarrow0}(E_{k}-D^0+i\delta)^{-1}$

\begin{equation}\label{kernel}
G^{+}_{k}(\mathbf{x})=\frac{1}{4\pi}e^{ikx}\left(-x^{-1}(E_{k}+\sum_{j=1}^{3}\alpha_{j}k\frac{x_{j}}{x}+\beta m)
-ix^{-2}\sum_{j=1}^{3}\alpha_{j}\frac{x_{j}}{x}\right)\;.
\end{equation}

\begin{lem} \label{properties}

Let $A\hspace{-0.2cm}/\in C^\infty$ be compactly supported, $A>0$, $\overline{\mu}>\lambda_c$ be such, that
$\lambda_c$ is the only critical coupling constant in $[\lambda_c,\overline{\mu}]$. Let
$\mathcal{P}:=\mathbb{R}^3\times[\lambda_c,\overline{\mu}]\backslash(0,\lambda_c)$, $j=2,4$.

 Then

\begin{description}
\item[(a)] there exist unique solutions $\phi^{j}(\mathbf{k},\mu,\cdot)$ of (\ref{weobtain}) in $L^\infty$ for
all $(\mathbf{k},\mu)\in\mathcal{P}$ such that

     \item[(b)] for any $(\mathbf{k},\mu)\in\mathcal{P}$,
     these
solutions $\phi^{j}(\mathbf{k},\mu,\cdot)$ are H\"older continuous of degree 1 in $\mathbf{x}$,

\item[(c)] any such solution $\phi^{j}(\mathbf{k},\mu,\cdot)$ satisfies (\ref{dgmp}),

\item[(d)] for any $\mu\in[\lambda_c,\overline{\mu}]$ the set of $\{\phi^{j}(\mathbf{k},\mu,\cdot)\}$
    defines a generalized Fourier transform in the space of scattering
    states in the positive continuous subspace by
\begin{equation}\label{her0}
\mathcal{F}_{\mu}(\psi)(\mathbf{k},j):=\int(2\pi)^{-\frac{3}{2}}\langle\phi^{j}(\mathbf{k},\mu,\mathbf{x}),\psi(\mathbf{x})\rangle
d^{3}x
\end{equation}

and

\begin{equation}\label{hin0}
    \psi(\mathbf{x})=\sum_{j=1}^{4}\int(2\pi)^{-\frac{3}{2}}
  \phi^{j}(\mathbf{k},\mu,\mathbf{x})\mathcal{F}_{\mu}(\psi)(\mathbf{k},j)d^{3}k\;.
 \end{equation}

The so defined $\mathcal{F}_{\mu}(\psi)$ is isometric to $\psi$, i.e.

\begin{eqnarray*}
\|\psi\|:=
\left(\int
\mid\psi(\mathbf{x})\mid^2d^3x\right)^{\frac{1}{2}}=\left(\sum_{j=1}^4\int
\mid\mathcal{F}_{\mu}(\psi)(\mathbf{k},j)\mid^2d^3k\right)^{\frac{1}{2}}
=:\|\mathcal{F}_{\mu}(\psi)\|
\end{eqnarray*}

\item[(e)]  the functions $\phi^{j}(\mathbf{k},\mu,\mathbf{x})$
are infinitely often continuously differentiable with respect to
$k$ for $(\mathbf{k},\mu)\in\mathcal{P}$. Furthermore there exists
$0<\alpha<\infty$ uniform in $(\mathbf{k},\mu)\in\mathcal{P}$ and
for all $n\in\mathbb{N}_0$ constants $C_n<\infty$ uniform in
$(\mathbf{k},\mu)\in\mathcal{P}$ and functions
$f^{n}(\mathbf{k},\mu,\mathbf{x})\in\mathcal{N}$ (see
(\ref{defmengen})) with ($\|
f(\cdot)\|_\infty:=\sup_{\mathbf{x}\in\mathbb{R}^3}\mid
f(\mathbf{x})\mid$)

\begin{eqnarray}\label{lemgeprobe1}
\| f^{n}(\mathbf{k},\mu,\cdot)\|_\infty
&<&C_n\left(1+k^n\left(\mid\mu-\lambda_c-\alpha k^{2}\mid+k^{3}\right)^{-n-1}\right)
\end{eqnarray}

such that

\begin{eqnarray}\label{lemgeprobe}
&&\|(\mid\cdot\mid+1)^{-n}\left(\partial_{k}^{n}\phi^{j}(\mathbf{k},\mu,\cdot)
-f^{n}(\mathbf{k},\mu,\cdot)\right)\|_{\infty}\nonumber\\&&\hspace{2cm}<C_n\left(1+(\mu-\lambda_c+
k^2)\| f^{n}(\mathbf{k},\mu,\cdot)\|_\infty\right)\;.
\end{eqnarray}

\end{description}
\end{lem}

\begin{rem}
For $j=1,3$ we can use the results in \cite{pickl}.

The divergent behavior of the generalized eigenfunctions expressed
by (\ref{lemgeprobe1}) is related to the fact, that there exist
solutions $\phi_{\lambda_c}$ of (\ref{dgmp2}). Non rigorously such
solutions can be seen as solutions of (\ref{weobtain}) with
$\|\phi_{\lambda_c}\|_\infty\gg1$ so that
$\phi^{j}(\mathbf{k},0,\mathbf{x})$ becomes negligible. Since the
generalized eigenfunctions are continuous in $\mu$ and
$\mathbf{k}$ it is reasonable to assume that the divergent part of
$\phi^{j}(\mathbf{k},\mu,\mathbf{x})$ as $\mu\rightarrow\lambda_c$
and $k\rightarrow0$ lies in $\mathcal{N}$.

Equation (\ref{lemgeprobe1}) and (\ref{lemgeprobe}) give us a quantitative estimate of the divergent behavior of
$\phi^{j}(\mathbf{k},\mu,\cdot)$ and its derivatives with respect to $k$ for small $k$ and $\mu$ close to
$\lambda_c$.

Observe for example the case $n=0$. Adding (\ref{lemgeprobe1}) and (\ref{lemgeprobe}) yields, that
$\phi^{j}(\mathbf{k},\mu,\cdot)$ diverges like the right hand side of (\ref{lemgeprobe1}) (note, that for small
$k$ and small $\mu-\lambda_c$ the right hand side of (\ref{lemgeprobe}) is much smaller than the right hand side
of (\ref{lemgeprobe1})).

Hence subtracting a sufficient multiple of the critical bound states $f^{0}(\mathbf{k},\mu,\mathbf{x})$ the
divergence of the generalized eigenfunctions becomes weaker, as can be seen on (\ref{lemgeprobe}).

\end{rem}\vspace{1cm}

\newpage

\vspace{0.5cm}
\noindent\textbf{Proof of Lemma \ref{properties}}\\

For (a) - (d) with $\mu\neq\lambda_c$ one can use Lemma 3.4 in \cite{pickl}. The $\mu=\lambda_c$ with
$\mathbf{k}\neq0$ can be proven equivalently (one needs the invertibility of $1-\mu T_k$ which is given in that
case).

We shall therefore only prove

\vspace{0.5cm}
\noindent\textbf{Part (e) of the Lemma}\\

We start with a short summary of proof of part (a) of the Lemma, following the proof in \cite{pickl}.

We first show that for any $(\mathbf{k},\mu)\in \mathcal{P}$ there exists a unique solution
$\phi^{j}(\mathbf{k},\mu,\cdot)$ of (\ref{LSE}).

Let $\mathcal{B}$ be the Banach space of all continuous functions tending uniformly to zero as
$x\rightarrow\infty$ (equipped with the supremum norm).

Defining the family of operators $T_k:L^\infty\rightarrow \mathcal{B}$ by

\begin{eqnarray}\label{deft}
T_k f(\mathbf{x})&:=&\int
 G^{+}_{k}(\mathbf{x'})A(\mathbf{x}-\mathbf{x'})f(\mathbf{x}-\mathbf{x}')d^3x'
 \\\label{deft2}
&=&\int
 G^{+}_{k}(\mathbf{x}-\mathbf{x}')A(\mathbf{x'})f(\mathbf{x'})d^3x'
\end{eqnarray}

(\ref{weobtain}) can be written as

\begin{equation}\label{LSE}
\phi^{j}(\mathbf{k},\mu,\mathbf{x})=\phi^{j}(\mathbf{k},0,\mathbf{x})+\mu
T_\mathbf{k}\phi^{j}(\mathbf{k},\mu,\mathbf{x})\;.
\end{equation}

The proof that $T_k$ maps $C^\infty$ into $\mathcal{B}$ can be
found in \cite{pickl}.  Note that the definition of $T_k$ yields
that $\| T_k \|_\infty^{op}$ exists. Using the continuity of $T_k$
it follows that

\begin{eqnarray}\label{tkuniform}
\sup_{k<k_0}\| T_k \|_\infty
 < \infty
\end{eqnarray}
for any $k_0<\infty$. In view of (\ref{kernel1}) and (\ref{kernel}) we have that

\begin{eqnarray*}
T_0 \phi_{\lambda_c}&=&\int
 G^{+}(\mathbf{x'})A(\mathbf{x}-\mathbf{x'})\phi_{\lambda_c}(\mathbf{x}-\mathbf{x}')d^3x'
 \end{eqnarray*}

It follows with (\ref{bseq}) that

\begin{equation}\label{taufn}
\lambda_c T_0\phi_{\lambda_c}=\phi_{\lambda_c}
\end{equation}

Furthermore defining

\begin{equation}\label{defzetatilde}
\zeta^{j}(\mathbf{k},\mu,\mathbf{x}):=\phi^{j}(\mathbf{k},\mu,\mathbf{x})-\phi^{j}(\mathbf{k},0,\mathbf{x})
\end{equation}

and

\begin{equation}\label{defg}
g^j(\mathbf{k},\mu,\mathbf{x}):=-\mu T_\mathbf{k}\phi^{j}(\mathbf{k},0,\mathbf{x})
\end{equation}

(\ref{LSE}) becomes

\begin{eqnarray}\label{zetatildelse}
\zeta^{j}(\mathbf{k},\mu,\mathbf{x})&=&
%
\mu T_\mathbf{k} \zeta^{j}(\mathbf{k},\mu,\mathbf{x})+g^j(\mathbf{k},\mu,\mathbf{x})\;.
\end{eqnarray}

We wish to show that (\ref{zetatildelse}) has a unique solution in $\mathcal{B}$ for any $(\mathbf{k},\mu)\in
\mathcal{P}$. For the Schr\"odinger Greens-function, this has been proven by Ikebe \cite{ikebe}. We want to
proceed in the same way.

Note that by \cite{pickl} $g(\mathbf{k},\mu,\cdot)\in\mathcal{B}$ for any $(\mathbf{k},\mu)\in \mathcal{P}$.
Ikebe uses the Riesz-Schauder theory of completely continuous operators in a Banach space \cite{riesz}:

If T is a completely continuous operator in $\mathcal{B}$, then for any given $g\in\mathcal{B}$ the equation

\begin{equation}\label{zetatildelse5}
f=g+Tf
\end{equation}
has a unique solution in $\mathcal{B}$ if $\widetilde{f}=T\widetilde{f}$ implies that $\widetilde{f}\equiv0$.

Since $T_k$ is a ''nice'' integral operator it is completely continuous. We wish to assert that

\begin{equation}\label{ftilde}
\widetilde{f}(\mathbf{x})=-\mu T_\mathbf{k}\widetilde{f}(\mathbf{x})
\end{equation}

has for $(\mathbf{k},\mu)\in\mathcal{P}$ only the trivial solution. Due to \cite{klaus} there are no zero energy
resonances for the class of Dirac operators we consider, so the only non trivial solutions are the bound states
with energy $m$. Since we assumed that there is no bound state with energy $m$ in
$\mu\in]\lambda_c,\overline{\mu}]$, (\ref{ftilde}) has for $(\mathbf{k},\mu)\in\mathcal{P}$ the unique solution
$\widetilde{f}\equiv0$.

Now we are in the position to prove (e). We formulate (e) for $\zeta^{j}(\mathbf{k},\mu,\cdot)$, which can
straight forwardly be done.

\begin{lem}\label{equive1}

Let $A\in C^\infty$ be compactly supported, $A>0$, $\overline{\mu}$ be such, that $\lambda_c$ is the only
critical coupling constant in $[\lambda_c,\overline{\mu}]$. Let $\mathcal{B}$ be the Banach space of all
continuous functions tending uniformly to zero as $x\rightarrow\infty$ (equipped with the supremum norm). Then
on $\mathcal{P}$ the functions $\zeta^{j}(\mathbf{k},\mu,\cdot)\in\mathcal{B}$ are infinitely often continuously
differentiable with respect to $k$, furthermore there exists a $0<\alpha<\infty$ and for every
$n\in\mathbb{N}_0$ a constant $C^n<\infty$ uniform in $(\mathbf{k},\mu)\in\mathcal{P}$ and a function
$f^{n}(\mathbf{k},\mu,\mathbf{x})\in\mathcal{N}$ (see (\ref{defmengen})) with

\begin{eqnarray}\label{1geprobe1}
\|
f^{n}(\mathbf{k},\mu,\cdot)\|_\infty<C^n\left(1+k^n\left(\mid\mu-\lambda_c-\alpha
k^{2}\mid+k^{3}\right)^{-n-1}\right)
\end{eqnarray}

such that

\begin{eqnarray}\label{1geprobe}
&&\|(x+1)^{-n}\left(\partial_{k}^{n}\zeta^{j}(\mathbf{k},\mu,\cdot)
-f^{n}(\mathbf{k},\mu,\cdot)\right)\|_{\infty}
\nonumber\\&&\hspace{2cm}<C^n\left(1+(\mu-\lambda_c+ k^2)\|
f^{n}(\mathbf{k},\mu,\cdot)\|_\infty\right)\;.
\end{eqnarray}

\end{lem}

From (\ref{1geprobe})

\begin{eqnarray*}
&&\|(\mid\cdot\mid+1)^{-n}\left(\partial_{k}^{n}\phi^{j}(\mathbf{k},\mu,\cdot)
-f^{n}(\mathbf{k},\mu,\cdot)\right)\|_{\infty}\nonumber
\\&&\hspace{2cm}\leq\|(\mid\cdot\mid+1)^{-n}\left(\partial_{k}^{n}\zeta^{j}(\mathbf{k},\mu,\cdot)
-f^{n}(\mathbf{k},\mu,\cdot)\right)\|_{\infty}\nonumber
\\&&\hspace{2.3cm}+\|(\mid\cdot\mid+1)^{-n}\left(\partial_{k}^{n}\phi^{j}(\mathbf{k},0,\cdot)
-f^{n}(\mathbf{k},\mu,\cdot)\right)\|_{\infty}\nonumber
\\&&\hspace{2cm}<C_n\left(1+(\mu-\lambda_c+
k^2)\| f^{n}(\mathbf{k},\mu,\cdot)\|_\infty\right)+C_n
\end{eqnarray*}

Lemma \ref{properties} (e) follows.

\begin{flushright}$\Box$\end{flushright}

The main difficulty in proving Lemma \ref{equive1} arises from small $k$. Therefore we show first

\begin{lem}\label{equive}

Under the conditions of Lemma \ref{equive1} there exists a $k_0>0$ such that on
$\mathcal{P}_{k_0}:=\{(\mathbf{k},\mu)\in \mathbb{R}^3\times[\lambda_c;\overline{\mu}]: k<
k_0\}\backslash(0,\lambda_c)$ the functions $\zeta^{j}(\mathbf{k},\mu,\mathbf{x})$ are infinitely often
continuously differentiable with respect to $k$, furthermore there exists a $0<\alpha<\infty$ and for every
$n\in\mathbb{N}_0$ a constant $C<\infty$ uniform in $(\mathbf{k},\mu)\in\mathcal{P}_{k_0}$  and a function
$f^{n}(\mathbf{k},\mu,\mathbf{x})\in\mathcal{N}$  with

\begin{eqnarray*}
\|
f^{n}(\mathbf{k},\mu,\cdot)\|_\infty&<&C^n\left(1+k^n\left(\mid\mu-\lambda_c-\alpha
k^{2}\mid+k^{3}\right)^{-n-1}\right)
\end{eqnarray*}

such that

\begin{eqnarray*}
\|(x+1)^{-n}\left(\partial_{k}^{n}\zeta^{j}(\mathbf{k},\mu,\cdot)
-f^{n}(\mathbf{k},\mu,\cdot)\right)\|_{\infty}<C^n\left(1+(\mu-\lambda_c+
k^2)\| f^{n}\|_\infty\right)\;.
\end{eqnarray*}

\end{lem}

\begin{rem}
From the proof we shall see that $k_0$ is small.
\end{rem}\vspace{1cm}

We first prove Lemma \ref{equive}, later we will show that  Lemma \ref{equive} implies Lemma \ref{equive1}.

\vspace{0.5cm}
\noindent\textbf{Proof of Lemma \ref{equive}}\\

For ease of notation we shall drop the spin index $j$.

We define a split of $\mathcal{B}$ into a direct sum of two ''orthogonal'' linear subspaces.

\begin{defn}\label{defmbot}
Set
\begin{equation}\label{def}
\mathcal{M}:=A\mathcal{N}=\{f\mid f=A\phi, \phi\in\mathcal{N}\}\;.
\end{equation}

and let $\mathcal{M}^\bot\subset\mathcal{B}$ be the set of functions in $\mathcal{B}$ which are ''orthogonal''
to $\mathcal{M}$ in the sense that

$$f\in\mathcal{M}^\bot\Leftrightarrow\langle
Af,\phi\rangle=0\hspace{1cm}\forall\space \phi\in\mathcal{N}\;.$$

\end{defn}

\begin{rem}
Since $f$ need not be in $L^2$ we nevertheless write with slight abuse of notation $\langle f,A\phi\rangle$ for
$\langle Af,\phi\rangle$.

\end{rem}\vspace{1cm}

\begin{lem}\label{directsum}

\begin{equation}\label{bissum}
\mathcal{B}=\mathcal{M}\oplus\mathcal{M}^\bot\;,
\end{equation}

i.e. every $f\in\mathcal{B}$ can be uniquely decomposed in
$f^\|\in\mathcal{M}$ and $f^\bot\in\mathcal{M}^\bot$ such that

\begin{equation}\label{zerleg}
f=f^\|+f^\bot\;.
\end{equation}

\end{lem}

\vspace{0.5cm}
\noindent\textbf{Proof}\\

Let $f\in\mathcal{B}$, $1_{\mathcal{S}_A}$ be the characteristic function of the support of $A$. Set

\begin{eqnarray*}
f_1&:=&1_{\mathcal{S}_A}f\\
f_2&:=&(1-1_{\mathcal{S}_A})f\;.
\end{eqnarray*}

$f_2$ is zero on the support of $A$ hence it follows trivially that $f_2\in \mathcal{M}^\bot$.
$f_1\in\mathcal{B}$, being compactly supported, is in $L^2\cap\mathcal{B}$.

Since $\mathcal{M}$ is a linear subspace of $L^2\cap\mathcal{B}$ and $\mathcal{M}^\bot\cap L^2$ its orthogonal
complement it follows that

$$L^2\cap\mathcal{B}=\mathcal{M}\oplus(\mathcal{M}^\bot\cap L^2)\;.$$

Hence there exists a $f^\|\in\mathcal{M}$ and a
$f_3\in\mathcal{M}^\bot\cap L^2$ with $f_1=f^\|+f_3$. Setting
$f^\bot:=f_2+f_3$ (\ref{zerleg}) follows.

\begin{flushright}$\Box$\end{flushright}

We introduce now for any $\mathcal{A}\subseteq\mathcal{B}$ and any $k_0>0$ the sets
$\widetilde{\mathcal{A}}_{k_0}$ of functions
$f(\mathbf{k},\mu,\mathbf{x}):\mathcal{P}\times\mathbb{R}^3\rightarrow \mathbb{C}^4$:

\begin{defn}\label{defnm}

Let $\mathcal{P}_{k_0}:=\{(\mathbf{k},\mu)\in \mathbb{R}^3\times[\lambda_c;\overline{\mu}]: k<
k_0\}\backslash(0,\lambda_c)$, then

\begin{eqnarray*}
f(\mathbf{k},\mu,\mathbf{x})\in\widetilde{\mathcal{A}}_{k_0}\Leftrightarrow&(a)&\text f(\mathbf{k},\mu,\cdot)\in
\mathcal{A}\;\;\;\;\text{for any }\space (\mathbf{k},\mu)\in\mathcal{P}_{k_0}
%
%
\\
&(b)&f(\mathbf{k},\mu,\mathbf{x})\space\text{is for any }\mathbf{k}\in\mathbb{R}^3\text{ with }k\leq k_0\text{ continuous}\\&&\text{in }\space\mu\text{ with respect to the supremum norm in }\mathcal{B}\;,\\
&(c)&\|
f\|_\infty:=\sup_{(\mathbf{k},\mu,\mathbf{x})\in\mathcal{P}_{k_0}\times\mathbb{R}^3}\{\mid
f(\mathbf{k},\mu,\mathbf{x})\mid\}<\infty\;.
\end{eqnarray*}

\end{defn}

We shall first prove Lemma \ref{equive} for $n=0$. Choose $(\mathbf{k},\mu)\in\mathcal{P}$ and recall equation
(\ref{zetatildelse})

\begin{equation}\label{zetatildelse2}
\zeta(\mathbf{k},\mu,\mathbf{x})=\left(1-\mu T_\mathbf{k}\right)^{-1} g(\mathbf{k},\mu,\mathbf{x})\;.
\end{equation}

We wish to estimate the ''close to $k=0$'' behavior of $\zeta^{j}(\mathbf{k},\mu,\mathbf{x})$. For that we will
split $g(\mathbf{k},\mu,\cdot)$ into two parts

\begin{equation}\label{gsplit}
g(\mathbf{k},\mu,\cdot)=:g(\mathbf{k},\mu,\cdot)^\|+g^\bot(\mathbf{k},\mu,\cdot)
\end{equation}

where $g(\mathbf{k},\mu,\cdot)^\|\in\mathcal{M}$ and
$g^\bot(\mathbf{k},\mu,\cdot)\in\mathcal{M}^\bot$, i.e.
$g(\mathbf{k},\mu,\cdot)^\|$ can be written as (note that $A$ is
positive and $\phi$ is nonzero in the range of the potential, thus
$\| A\phi_{\lambda_c}\|^2\neq0$)

\begin{eqnarray}\label{defgparllel}
g(\mathbf{k},\mu,\cdot)^\|&=&A\phi_{\lambda_c}\frac{\langle\phi_{\lambda_c}\mid
Ag(\mathbf{k},\mu,\cdot)\rangle}{\| A\phi_{\lambda_c}\|^2}
\end{eqnarray}

for some $\phi_{\lambda_c}\in\mathcal{N}$. We choose the
normalization $\|\phi_{\lambda_c}\|_\infty=1$.

Letting now $(\mathbf{k},\mu)$ vary in $\mathcal{P}_{k_0}$ we
shall show that
$g(\mathbf{k},\mu,\cdot)^\|(\mathbf{x})\in\widetilde{\mathcal{M}}_{k_0}$
and
$g^\bot(\mathbf{k},\mu,\cdot)(\mathbf{x})\in\widetilde{\mathcal{M}}^\bot_{k_0}$
for any $k_0>0$.


 Since $\| T_{k}\|_\infty^{op}$ is bounded
uniformly in $k\leq k_0$ for any $k_0<\infty$ (see
(\ref{tkuniform})) it follows that $\|
g(\mathbf{k},\mu,\cdot)\|_\infty$ (see (\ref{defg})) 
is bounded uniformly in any $(\mathbf{k},\mu)\in\mathcal{P}_{k_0}$. Hence

\begin{eqnarray*}
\langle \phi_{\lambda_c}\mid A g(\mathbf{k},\mu,\mathbf{x})\rangle&=&\int\phi_{\lambda_c}(\mathbf{x})
A(\mathbf{x}) g(\mathbf{k},\mu,\mathbf{x})d^3x
\\&\leq&\| g(\mathbf{k},\mu,\cdot)\|_\infty\int\phi_{\lambda_c}(\mathbf{x})
A(\mathbf{x})d^3x
\\&\leq&C\| g(\mathbf{k},\mu,\cdot)\|_\infty
\end{eqnarray*}

is bounded uniformly in $(\mathbf{k},\mu)\in\mathcal{P}_{k_0}$ for any $k_0<\infty$ with an appropriate
$C<\infty$ since $\phi_{\lambda_c}$ is bounded uniformly in $(\mathbf{k},\mu)\in\mathcal{P}$.

Hence we have for $g^\|(\mathbf{k},\mu,\cdot)$ and
$g^\bot(\mathbf{k},\mu,\cdot):=g(\mathbf{k},\mu,\cdot)-g^\|(\mathbf{k},\mu,\cdot)$
that on $\mathcal{P}_{k_0}$

\begin{eqnarray}\label{supposethat}
\| g^\|\|_\infty&<&\infty\\
\label{supposethat2} \| g^\bot\|_\infty&<&\infty\;.
\end{eqnarray}

The continuity of the scalar product and the continuity of
$g(\mathbf{k},\mu,\cdot)$ in $\mu$ yield that
$g^\|(\mathbf{k},\mu,\cdot)$ and $g^\bot(\mathbf{k},\mu,\cdot)$
are continuous in $\mu$ for any $\mathbf{k}\in\mathbb{R}^3$, hence
$g^\|\in\widetilde{\mathcal{M}}_{k_0}$ and
$g^\bot\in\widetilde{\mathcal{M}}^\bot_{k_0}$ for any $k_0>0$.

We now return to (\ref{zetatildelse2}). We get with (\ref{gsplit})

\begin{eqnarray}
\label{splitinvers}\zeta^{j}(\mathbf{k},\mu,\cdot)&=&(\mu
T_{k}-1)^{-1} g^\|(\mathbf{k},\mu,\cdot)+(\mu
T_{k}-1)^{-1}g^\bot(\mathbf{k},\mu,\cdot)\;.
\end{eqnarray}

We shall determine $\zeta^{j}(\mathbf{k},\mu,\cdot)$ now more precisely by the following iteration procedure

\begin{lem}\label{iteration}

Let $k_0>0$. Then there exits a constant $0<C<\infty$ such that
for any $h_{0}^\|\in\widetilde{\mathcal{M}}_{k_0}$ and for any
$h_{0}^\bot\in\widetilde{\mathcal{M}}^\bot_{k_0}$ there exist
$f_{1}\in\widetilde{\mathcal{N}}_{k_0}$,
$\omega_{1}\in\widetilde{\mathcal{B}}_{k_0}$,
$h_{1}^\|\in\widetilde{\mathcal{M}}_{k_0}$ and
$h_{1}^\bot\in\widetilde{\mathcal{M}}^\bot_{k_0}$ with

\begin{eqnarray}\label{desform}
&&(\mu T_{k}-1)^{-1} h_{0}^\|+(\mu
T_{k}-1)^{-1}h_{0}^\bot\nonumber\\&=&f_{1}+\omega_{1}+(\mu
T_{k}-1)^{-1} h_{1}^\|+(\mu T_{k}-1)^{-1}h_{1}^\bot
\end{eqnarray}

and

\begin{eqnarray*}
\| f_{1} (\mathbf{k},\mu,\cdot)\|_\infty&\leq& C\|
h_{0}^\|(\mathbf{k},\mu,\cdot)\|_\infty \left(\mid \mu-\lambda_c
-\alpha k^2\mid+
k^3\right)^{-1}\\
\| \omega_{1} (\mathbf{k},\mu,\cdot)\|_\infty&\leq& Ck^2\|
f_{1}(\mathbf{k},\mu,\cdot)\|_\infty+ C\|
h_{0}^\bot(\mathbf{k},\mu,\cdot)\|_\infty\\
\|  h_{1}^\|(\mathbf{k},\mu,\cdot)\|_\infty&\leq& Ck\|
h_{0}^\|(\mathbf{k},\mu,\cdot)\|_\infty+ Ck^2\|
h_{0}^\bot(\mathbf{k},\mu,\cdot)\|_\infty\\
\|  h_{1}^\bot(\mathbf{k},\mu,\cdot)\|_\infty&\leq& C\|
h_{1}^\|(\mathbf{k},\mu,\cdot)\|_\infty +Ck\|
h_{0}^\bot(\mathbf{k},\mu,\cdot)\|_\infty\;.
\end{eqnarray*}

\end{lem}

The proof of this Lemma is the heart of this section and will be done later.

Iterating this Lemma $p$-times yields that

\begin{eqnarray}\label{iterationsforml}
&&(\mu T_{k}-1)^{-1} h_{0}^\|+(\mu
T_{k}-1)^{-1}h_{0}^\bot\nonumber\\ &&\hspace{2cm}=\sum_{j=1}^p
f_{j}+\sum_{j=1}^p \widetilde{\omega}_{j}+(\mu
T_{k}-1)^{-1}h_{p}^\|+(\mu T_{k}-1)^{-1}h_{p}^\bot
\end{eqnarray}

It follows iteratively that

\begin{eqnarray}\label{abschsummanden1}
\|  h_j^\|(\mathbf{k},\mu,\cdot)\|_\infty&\leq& C
(3Ck)^j\\\label{abschsummanden2}
\|  h_j^\bot(\mathbf{k},\mu,\cdot)\|_\infty&\leq&
C(3Ck)^{j-1}\\\label{abschsummanden3}
\| f_j(\mathbf{k},\mu,\cdot) \|_\infty&\leq&\left(\mid
\mu-\lambda_c -\alpha k^2\mid+
k^3\right)^{-1}C^{j+1}k^{j-1}\\\nonumber
\| \omega_j(\mathbf{k},\mu,\cdot) \|_\infty&\leq&C k^2\|
f_j(\mathbf{k},\mu,\cdot) \|_\infty+ C^{j+1}
k^{j-1}\\\label{abschsummanden4}&\leq&\left(C^2 \left(\mid
\mu-\lambda_c -\alpha k^2\mid+ k^3\right)^{-1}C^jk^2+ C^{j+1}
\right)k^{j-1}\;.
\end{eqnarray}

Next we prove the convergence of the four summands on the right hand side of (\ref{iterationsforml}). Choose
$k_0$ such that $C k_0<1$. Note that in fact $C$ does depend on $k_0$. But since $\mathcal{P}_{k_0}\subset
\mathcal{P}_{k_1}$ for all $k_0<k_1$, it follows that the $C$ one gets for $k_0$ is smaller or equal to the $C$
one gets for $k_1$, hence $k_0$ can in fact be chosen such that $C k_0<1$.

Then it follows with (\ref{abschsummanden1}) and (\ref{abschsummanden2}) that

\begin{eqnarray*}
\lim_{j\rightarrow\infty}\|
h_j^\bot\|_\infty&=&0\\
\lim_{j\rightarrow\infty}\| h_j^\|\|_\infty&=&0\;.
\end{eqnarray*}

Since the operator $(\mu T_{k}-1)^{-1}$ is for fixed $(\mathbf{k},\mu)\neq(0,\lambda_c)$ a bounded operator in
$\mathcal{B}$ it follows that

\begin{equation}\label{hatgefehlt}
\lim_{p\rightarrow \infty}\|(\mu T_{k}-1)^{-1}h_{p}^\|+(\mu
T_{k}-1)^{-1}h_{p}^\bot\|=0\;.
\end{equation}

Since $\sum_{j=1}^N f_j$ is geometric it is a Cauchy series for all $(\mathbf{k},\mu)\in \mathcal{P}_{k_0}$.
$\mathcal{N}$ is a finitely dimensional vector space hence it is complete. It follows that for any
$(\mathbf{k},\mu)\in \mathcal{P}_{k_0}$ there exists a $f(\mathbf{k},\mu,\cdot)$ such that

$$\lim_{N\rightarrow\infty} \| \sum_{j=1}^N f_j(\mathbf{k},\mu,\cdot)-f(\mathbf{k},\mu,\cdot)\|_\infty=0\;.$$

Using the completeness of $\mathcal{B}$ we get similarly with (\ref{abschsummanden4}) the convergence of
$\sum_{j=1}^N \omega_j$, i.e. for any $(\mathbf{k},\mu)\in \mathcal{P}_{k_0}$ there exists a
$\omega(\mathbf{k},\mu,\cdot)\in\widetilde{\mathcal{B}}_{k_0}$ such that

$$\lim_{N\rightarrow\infty} \| \sum_{j=1}^N \omega_j(\mathbf{k},\mu,\cdot)-\omega(\mathbf{k},\mu,\cdot)\|_\infty=0\;.$$

Using (\ref{abschsummanden3}) we have that

\begin{eqnarray}\label{stern}
\| f(\mathbf{k},\mu,\cdot)\|_\infty&=&\| \sum_{j=1}^\infty
f_j(\mathbf{k},\mu,\cdot)\|_\infty\nonumber\\&\leq&\sum_{j=1}^\infty
\| f_j(\mathbf{k},\mu,\cdot)\|_\infty
\nonumber\\&\leq&\sum_{j=1}^\infty \left(\mid \mu-\lambda_c
-\alpha k^2\mid+ k^3\right)^{-1}C^{j+1}k^{j-1} \nonumber\\&=&
\left(\mid \mu-\lambda_c -\alpha k^2\mid+
k^3\right)^{-1}\sum_{j=0}^\infty C^{j+1}k^{j} \nonumber\\&=&C
\left(\mid \mu-\lambda_c -\alpha k^2\mid+
k^3\right)^{-1}\frac{1}{1-Ck}\;,
\end{eqnarray}

using (\ref{abschsummanden4})

\begin{eqnarray}\label{stern2}
\|\omega(\mathbf{k},\mu,\cdot)\|_\infty&=&\|\sum_{j=1}^\infty
\omega_j(\mathbf{k},\mu,\cdot)\|_\infty\\&\leq&C_1k^2\sum_{j=1}^\infty\|
f_j(\mathbf{k},\mu,\cdot)\|_\infty + \sum_{j=1}^\infty C^{j+1}
k^{j-1}
\\&=&Ck^2\sum_{j=1}^\infty\| f_j(\mathbf{k},\mu,\cdot)\|_\infty+  \sum_{j=0}^\infty
C^{j+1} k^{j}
\\&=&Ck^2\sum_{j=1}^\infty\|
f_j(\mathbf{k},\mu,\cdot)\|_\infty+\frac{C}{1-Ck}\;.
\end{eqnarray}

So taking the limit $p\rightarrow\infty$ on the right hand side of (\ref{iterationsforml}) and using
(\ref{hatgefehlt}) yields

$$(\mu T_{k}-1)^{-1} h_{0}^\|+(\mu T_{k}-1)^{-1}h_{0}^\bot=f+\omega\;.$$

We apply this to $h_0^\|=g^\|$ and $h_0^\bot=g^\bot$ observing
(\ref{supposethat}) and (\ref{supposethat2}). With (\ref{gsplit})
and (\ref{zetatildelse2})

\begin{equation}\label{ergebnis}
\zeta^{j}=(\mu T_{k}-1)^{-1} g=f+\omega\;.
\end{equation}

(\ref{stern2}), (\ref{ergebnis}) and (\ref{stern}) yield Lemma \ref{equive} for $n=0$.

\begin{flushright}$\Box$\end{flushright}

\vspace{0.5cm}
\noindent\textbf{Proof of Lemma \ref{iteration}}\\

Let $h_0^\|\in\widetilde{\mathcal{M}}_{k_0}$,
$h_0^\bot\in\widetilde{\mathcal{M}}^\bot_{k_0}$. We denote

\begin{eqnarray}\label{c3undc4}
\| h_0^\|\|_\infty&=:&C^\|
\\\label{c3undc4b}\| h_0^\bot\|_\infty&=:&C^\bot\;.
\end{eqnarray}

To prove the Lemma we first control the term $(\mu T_{k}-1)^{-1}
h_0^\|$. Since the control of this term is involved we give the
result in

\begin{lem}\label{taufphic}

There exist $\alpha,k_0,C_1,C_2\in\mathbb{R}$, $C_3>0$ such that
for any $h^\|\in\widetilde{\mathcal{M}}_{k_0}$ there exists
$f\in\widetilde{\mathcal{N}}_{k_0}$ and
$h^\bot\in\widetilde{\mathcal{M}}^\bot_{k_0}$ so that

\begin{equation*}
\left(\mu T_{k}-1\right)^{-1}h^\|= f+ \left(\mu
T_{k}-1\right)^{-1}h^\bot
\end{equation*}

and

\begin{equation*}
\| f(\mathbf{k},\mu,\cdot)\|_\infty\leq C_1 \|
h^\|(\mathbf{k},\mu,\cdot)\|_\infty\left(\mid \mu-\lambda_c
-\alpha k^2\mid+ C_3 k^3\right)^{-1}
\end{equation*}

and

\begin{eqnarray*}
\|
h^\bot(\mathbf{k},\mu,\cdot)\|_\infty&\leq&C_2(\mu-\lambda_c+k^2)\|
f(\mathbf{k},\mu,\cdot)\|_\infty
\end{eqnarray*}

\end{lem}

which we will prove later on.

\vspace{0.2cm}

Using Lemma \ref{taufphic} for $h^\|=h_0^\|$ we get that there
exist a $f\in\widetilde{\mathcal{N}}_{k_0}$ and
$h^\bot\in\widetilde{\mathcal{M}}^\bot_{k_0}$ with

\begin{equation*}
(\mu T_{k} -1)^{-1}h_0^\|=f+(\mu T_{k}-1)^{-1}h^\bot
\end{equation*}

and with (\ref{c3undc4})

\begin{eqnarray}
\label{fiteration}\| f(\mathbf{k},\mu,\cdot)\|_\infty&\leq&C_1
C^\|\left(\mid \mu-\lambda_c -\alpha k^2\mid+ C_3
k^3\right)^{-1}\\\label{weiterunten}
\| h^\bot(\mathbf{k},\mu,\cdot)\|_\infty&\leq& C_2\|
f(\mathbf{k},\mu,\cdot)\|_\infty(\mu-\lambda_c+k^2)\;.
\end{eqnarray}

Hence setting $$\widetilde{h}^\bot=h^\bot+h_0^\bot$$ we have by (\ref{c3undc4}) and (\ref{weiterunten})

\begin{equation}\label{gtildenorm}\| \widetilde{h}^\bot(\mathbf{k},\mu,\cdot)\|_\infty\leq C_2\|
f(\mathbf{k},\mu,\cdot)\|_\infty(\mu-\lambda_c+k^2)+C^\bot\end{equation}

and we obtain

\begin{equation}\label{restzeta}
(\mu T_{k} -1)^{-1}\left(h_0^\|+h_0^\bot\right)-f=(\mu
T_{k}-1)^{-1}\widetilde{h}^\bot\;,
\end{equation}

(\ref{restzeta}) is almost of the form of (\ref{desform}) in Lemma \ref{iteration}. To obtain the desired result
we need to estimate now $(\mu T_{k}-1)^{-1}\widetilde{h}^\bot$. For that we consider first the operator $(\mu
T_{k}-1)^{-1}$ for $k=0$.

Our proof is based on the insight that $\lambda_c T_0-1$ is invertible on $\mathcal{M}^\bot$ which is spelled
out in Lemma \ref{senkrechtphic} below. For this we note that $\lambda_c T_0$ is symmetric with respect to the
scalar product $\langle f,Ag\rangle$, $f$, $g\in\mathcal{B}$ (see (\ref{symmetric}) below). This explains a
posteriori that $\mathcal{M}$ and $\mathcal{M}^\bot$ are the relevant spaces and not $\mathcal{N}$ and
$\mathcal{N}^\bot$ as one might think at first sight.

Furthermore $\mathcal{M}^\bot\cap\mathcal{N}=\{0\}$, which is immediate from (remember that $A>0$)

\begin{equation}\label{possum}
\langle A\phi_{\lambda_c}\mid\phi_{\lambda_c}\rangle=\int A(\mathbf{x})\mid \phi_{\lambda_c}(\mathbf{x})\mid^2
d^3x>0\;.
\end{equation}

\begin{lem}\label{senkrechtphic}

\begin{itemize}

\item[(a)] For any $\mu\in[\lambda_c,\overline{\mu}]$ we have that

$$
h^\bot\in\mathcal{M}^\bot\Leftrightarrow (\mu T_0-1) h^\bot\in\mathcal{M}^\bot\;.
$$

\item[(b)] For any $\mu\in[\lambda_c,\overline{\mu}]$ the map $\mu T_0-1:
\mathcal{M}^\bot\rightarrow\mathcal{M}^\bot$ is invertible.

\item[(c)] There exists a $C<\infty$ such that for all $h^\bot\in\mathcal{M}^\bot$ and all
$\mu\in[\lambda_c,\overline{\mu}]$

\begin{equation}\label{invert}
\| \left(\mu T_0-1\right)^{-1}h^\bot\|_\infty\leq C \|
h^\bot\|_\infty\;.
\end{equation}

\end{itemize}
\end{lem}

which will be proven below.

Setting now

\begin{equation}\label{defomega2}
\omega(\mathbf{k},\mu,\mathbf{x}):=(\mu T_{0}-1)^{-1}\widetilde{h}^\bot\;.
\end{equation}

we get with (\ref{gtildenorm}) in (\ref{invert}) for $h^\bot=\widetilde{h}^\bot$

\begin{equation}\label{noromegatilde}
\|\omega(\mathbf{k},\mu,\cdot)\|_\infty\leq CC_2\|
f(\mathbf{k},\mu,\cdot)\|_\infty(\mu-\lambda_c+k^2)+CC^\bot\;.
\end{equation}

Thus $\omega(\mathbf{k},\mu,\cdot)\in\mathcal{B}$.

Now we can estimate $(\mu T_{k}-1)^{-1}\widetilde{h}^\bot$. Writing

\begin{eqnarray*}
(\mu T_{k}-1)\omega(\mathbf{k},\mu,\cdot)&=&\widetilde{h}^\bot+\mu(T_{k}-T_0)\omega(\mathbf{k},\mu,\cdot)\;,
\end{eqnarray*}

we obtain

\begin{eqnarray}\label{fehleromega}
(\mu T_{k} -1)^{-1}\widetilde{h}^\bot&=&\omega(\mathbf{k},\mu,\cdot)-(\mu T_{k}
-1)^{-1}\mu(T_{k}-T_0)\omega(\mathbf{k},\mu,\cdot)\nonumber\\&=:&\omega(\mathbf{k},\mu,\cdot)+(\mu T_{k}
-1)^{-1}h_1
\end{eqnarray}

with

\begin{equation}\label{gk1}
h_1(\mathbf{k},\mu,\cdot):=-\mu(T_{k}-T_0)\omega(\mathbf{k},\mu,\cdot)\;.
\end{equation}

We now consider the usual splitting (\ref{gsplit})

\begin{eqnarray*}
h_{1}&=:&h_{1}^\|+h_{1}^\bot
\end{eqnarray*}

and estimate the $\| h_{1}^\|(\mathbf{k},\mu,\cdot)\|_\infty$ and
$\| h_{1}^\bot(\mathbf{k},\mu,\cdot)\|_\infty$ separately. This
can be done using

\begin{lem}\label{ordnungtk}

 There exists a $C\in\mathbb{R}$ such that
for all $\omega\in\mathcal{B}$

\begin{equation*}
\|(T_k-T_0)\omega\|_\infty<Ck\| \omega\|_\infty\;,
\end{equation*}

and

\begin{equation*}
\mid \langle
A(T_k-T_0)\omega\mid\phi_{\lambda_c}\rangle\mid<Ck^2\|
\omega\|_\infty
\end{equation*}

for all $\phi_{\lambda_c}\in\mathcal{N}$ with
$\|\phi_{\lambda_c}\|_\infty=1$.

\end{lem}

The proof will be given later.

We use the Lemma in (\ref{gk1}) for $\omega=\omega(\mathbf{k},\mu,\cdot)$ and get with (\ref{noromegatilde})
that

\begin{eqnarray*}
\|
h_{1}(\mathbf{k},\mu,\cdot)\|_\infty&=&\|\mu(T_{k}-T_0)\omega(\mathbf{k},\mu,\cdot)\|_\infty
\\&<&\mu CC_2 k \left(\|
f(\mathbf{k},\mu,\mathbf{x})\|_\infty
(\mu-\lambda_c+k^2)+C^\bot\right) \\
\| h^\|_{1}(\mathbf{k},\mu,\cdot)\|_\infty&\leq&
\sup_{\phi_{\lambda_c}\in\mathcal{N}}\mid\langle A
h_{1}(\mathbf{k},\mu,\cdot)\mid\phi_{\lambda_c}\rangle\mid\\&<&\mu
CC_2 k^2\left(\| f(\mathbf{k},\mu,\mathbf{x})\|_\infty
(\mu-\lambda_c+k^2)+C^\bot\right)
\end{eqnarray*}

It follows with (\ref{fiteration}) that

\begin{eqnarray}\label{gk1bound1}
\| h_{1}(\mathbf{k},\mu,\cdot)\|_\infty&<&\mu CC_2 k
\left(\frac{C_1 C^\|(\mu-\lambda_c+k^2)}{\mid \mu-\lambda_c
-\alpha k^2\mid+ C_3 k^3} +C^\bot\right)\\\label{gk1bound2}
\| h^\|_{1}(\mathbf{k},\mu,\cdot)\|_\infty&<&\mu CC_2
k^2\left(\frac{C_1 C^\|(\mu-\lambda_c+k^2)}{\mid \mu-\lambda_c
-\alpha k^2\mid+ C_3 k^3}+C^\bot\right)
\end{eqnarray}

Since for $(\mathbf{k},\mu)\in\mathcal{P}_{k_0}$ ($\Rightarrow k\leq k_0<1$).

\begin{eqnarray*}
\left|\frac{k(\mu-\lambda_c+k^2)}{| \mu-\lambda_c -\alpha k^2|+
C_3 k^3}\right|&=&\left|\frac{k(\mu-\lambda_c-\alpha k^2)+\alpha
k^3+k^3}{| \mu-\lambda_c -\alpha k^2|+ C_3
k^3}\right|\\&=&\left|\frac{k(\mu-\lambda_c-\alpha k^2)}{|
\mu-\lambda_c -\alpha k^2|+ C_3 k^3}+\frac{k^3(1+\alpha)}{|
\mu-\lambda_c -\alpha k^2|+ C_3 k^3}\right|
\\&\leq&1+\frac{(1+\alpha)}{C_3}
\end{eqnarray*}

it follows with (\ref{gk1bound1}) and (\ref{gk1bound2}) that there exists a $C<\infty$ such that

\begin{eqnarray}\label{estg2}
\| h_{1}^\|(\mathbf{k},\mu,\cdot)\|_\infty&\leq& C \left(C^\| k^2
+C^\bot k\right)
\\\label{estg1}
\| h_{1}^\bot(\mathbf{k},\mu,\cdot)\|_\infty&:=&\|
h_{1}(\mathbf{k},\mu,\cdot)-h_{1}^\|(\mathbf{k},\mu,\cdot)\|_\infty\leq
C\left(C^\| k +C^\bot \right)
\;.
\end{eqnarray}

With (\ref{fiteration}) and (\ref{noromegatilde}) Lemma \ref{iteration} follows.

\begin{flushright}$\Box$\end{flushright}


We shall now give the proofs of Lemma \ref{taufphic}, \ref{senkrechtphic} and \ref{ordnungtk} . For ease of
reference we recall each time what the Lemmata say.

\vspace{0.5cm}
\noindent\textbf{Lemma \ref{taufphic}}\\

\it

There exist $\alpha,k_0,C_1,C_2\in\mathbb{R}$ ,$C_3>0$ such that
for any $h^\|\in\widetilde{\mathcal{M}}_{k_0}$ there exists
$f\in\widetilde{\mathcal{N}}_{k_0}$ and
$h^\bot\in\widetilde{\mathcal{M}}^\bot_{k_0}$ so that

\begin{equation}\label{lemtauf}
\left(\mu T_{k}-1\right)^{-1}h^\|= f+ \left(\mu
T_{k}-1\right)^{-1}h^\bot
\end{equation}

and

\begin{equation}\label{lemtauf1}
\| f(\mathbf{k},\mu,\cdot)\|_\infty\leq C_1 \|
h^\|(\mathbf{k},\mu,\cdot)\|_\infty\left(\mid \mu-\lambda_c
-\alpha k^2\mid+ C_3 k^3\right)^{-1}
\end{equation}

and

\begin{eqnarray}\label{lemtauf2}
\|
h^\bot(\mathbf{k},\mu,\cdot)\|_\infty&\leq&C_2(\mu-\lambda_c+k^2)\|
f(\mathbf{k},\mu,\cdot)\|_\infty
\end{eqnarray}

\rm

\vspace{0.5cm}
\noindent\textbf{Proof}\\

Let $h^\|\in\widetilde{\mathcal{M}}_{k_0}$. Let the degeneracy of
the critical bound state be $n$.

(\ref{lemtauf}) is equivalent to

\begin{equation*}
\left(\mu T_{k}-1\right)f(\mathbf{k},\mu,\cdot)=h^\|+
h^\bot(\mathbf{k},\mu,\cdot)\;.
\end{equation*}

While the logic here is that $h^\|$ is given and $f$, $h^\bot$ are
to be found, we turn the argument around. We start with
controlling $\left(\mu T_{k}-1\right)\widetilde{f}$ for arbitrary
$\widetilde{f}\in\widetilde{\mathcal{N}}_{k_0}$, and show that
there exists a $\widetilde{h}^\|\in\widetilde{\mathcal{M}}_{k_0}$
and a $\widetilde{h}^\bot\in\widetilde{\mathcal{M}}^\bot_{k_0}$
such that

$$\left(\mu
T_{k}-1\right)\widetilde{f}=\widetilde{h}^\|+\widetilde{h}^\bot\;.$$

Since $\mu T_{k}-1$ is a linear operator and since the projectors from $\mathcal{B}$ onto $\mathcal{M}$ and
$\mathcal{M}^\bot$ are linear, it follows that for any $(\mathbf{k},\mu)\in\mathcal{P}_{k_0}$ there exists
linear operators $B(\mathbf{k},\mu)$ from $\mathcal{N}\rightarrow\mathcal{M}$ and $B^\bot(\mathbf{k},\mu)$ from
$\mathcal{N}\rightarrow\mathcal{M}^\bot$ such that for any $\widetilde{f}\in\widetilde{\mathcal{N}}_{k_0}$

\begin{eqnarray}\label{linop1}
\widetilde{h}^\|(\mathbf{k},\mu,\cdot)=B(\mathbf{k},\mu)\widetilde{f}(\mathbf{k},\mu,\cdot)
\\\label{linop2}
\widetilde{h}^\bot(\mathbf{k},\mu,\cdot)=B^\bot(\mathbf{k},\mu)\widetilde{f}(\mathbf{k},\mu,\cdot)
\end{eqnarray}

Note that $\mathcal{N}$ and $\mathcal{M}$ have finite dimension, so $B$ is a mapping between finite dimensional
vector spaces.

We now give some properties of $B(\mathbf{k},\mu)$ and $B^\bot(\mathbf{k},\mu)$. We will first show how they
imply the Lemma and then prove them one after another.

\begin{itemize}
    \item[(a)] \begin{equation}\label{punkta}B(\mathbf{k},\mu)\text{ is invertible for any }(\mathbf{k},\mu)\in\mathcal{P}_{k_0}\end{equation}
    \item[(b)] \begin{equation}\label{punktb}\| B^{-1}(\mathbf{k},\mu)\|_2^{op}<C_1\left(\mid
\mu-\lambda_c -\alpha k^2\mid+ C_3 k^3\right)^{-1}\end{equation}
    \item[(c)]  \begin{equation}\label{punktc}\|
    B^\bot(\mathbf{k},\mu)\|_\infty^{op}<\widetilde{C}_2(\mu-\lambda_c+k^2)\end{equation}
\end{itemize}

for appropriate $C_1<\infty$, $C_2<\infty$, $0<C_3<\infty$ uniform in $(\mathbf{k},\mu)\in\mathcal{P}_{k_0}$.

Assume that (a) holds, i.e. we can find for any
$h^\|\in\widetilde{\mathcal{M}}_{k_0}$ a
$f\in\widetilde{\mathcal{N}}_{k_0}$ such that the projection of
$(\mu T_{k}-1)f$ onto $\widetilde{\mathcal{M}}_{k_0}$ is equal to
$h^\|$. Let $h^\bot$ be the projection of $(\mu T_{k}-1)f$ onto
$\widetilde{\mathcal{M}}^\bot_{k_0}$. It follows that
(\ref{lemtauf}) is satisfied.

Assume that furthermore (b) holds. Using the equivalence of all
norms in a finitely dimensional vector space (i.e. replacing $\|
\cdot\|_\infty$ by $\|\cdot\|$ in the n-dimensional spaces
$\mathcal{M}$ and $\mathcal{N}$) it follows that $\|
B^{-1}(\mathbf{k},\mu)\|_2^{op}\leq C\|
B^{-1}(\mathbf{k},\mu)\|_\infty^{op}$. Since

\begin{eqnarray*}
\|\widetilde{f}(\mathbf{k},\mu,\cdot)\|_\infty&=& \|
B(\mathbf{k},\mu)^{-1}\widetilde{h}^\|(\mathbf{k},\mu,\cdot)\|_\infty
\\&\leq&\| B(\mathbf{k},\mu)^{-1}\|^{op}_\infty\;\|\widetilde{h}^\|(\mathbf{k},\mu,\cdot)\|_\infty
\end{eqnarray*}

(\ref{lemtauf1}) follows.

Assume that furthermore (c) holds. Since
$\widetilde{h}^\bot(\mathbf{k},\mu,\cdot)=B^\bot(\mathbf{k},\mu)\widetilde{f}(\mathbf{k},\mu,\cdot)$
(\ref{lemtauf2}) follows.

It is left to prove that (a)-(c) hold to verify Lemma \ref{taufphic}.

(a) holds if $B(\mathbf{k},\mu)f=0$ has no non trivial solution.

Furthermore for (b) if $B$ were invertible

\begin{eqnarray*}
\| B^{-1}(\mathbf{k},\mu)\|_2^{op}
:=\sup_{\widetilde{h}^\|\in\mathcal{M}\backslash\{0\}}\frac{\|
B^{-1}\widetilde{h}^\|\|}{\| \widetilde{h}^\|\|}
=\sup_{\widetilde{f}\in\mathcal{N}\backslash\{0\}}\frac{\|
\widetilde{f}\|}{\| B\widetilde{f}\|}
=\left(\inf_{\widetilde{f}\in\mathcal{N},\| \widetilde{f}\|=1}\|
B\widetilde{f}\|\right)^{-1}
\end{eqnarray*}

Hence (a) and (b) follow from

\begin{equation}\label{infmain}
\inf_{\phi_{\lambda_c}\in\mathcal{N},\| \phi_{\lambda_c}\|=1}\|
B\phi_{\lambda_c}\|\geq C\left(\mid \mu-\lambda_c -\alpha k^2\mid+
C_3 k^3\right)
\end{equation}

with $C>0$ uniform in $(\mathbf{k},\mu)\in\mathcal{P}_{k_0}$.

Using Schwartz inequality we have that

\begin{eqnarray*}
\| B\phi_{\lambda_c}\|&\geq&\frac{1}{\|
A\phi_{\lambda_c}\|}\left|\langle B\phi_{\lambda_c},
A\phi_{\lambda_c}\rangle\right|
=\frac{1}{\| A\phi_{\lambda_c}\|}\left|\langle P_\mathcal{M}(\mu
T_{k}-1)\phi_{\lambda_c},A \phi_{\lambda_c}\rangle\right|
\\&=&\frac{1}{\| A\phi_{\lambda_c}\|}\left|A(\mu T_{k}-1)\phi_{\lambda_c},
\phi_{\lambda_c}\rangle\right|
\end{eqnarray*}

where $P_\mathcal{M}$ is the projector on $\mathcal{M}$. Since $A$ is bounded it follows, that

$$\sup_{\phi_{\lambda_c}\in\mathcal{N},\|\phi_{\lambda_c}\|=1}\|
A\phi_{\lambda_c}\|<\infty$$

hence (\ref{infmain}) holds if there exists a $C>0$ uniform in $(\mathbf{k},\mu)\in\mathcal{P}_{k_0}$ with

\begin{eqnarray}\label{opnorma}
\inf_{\phi_{\lambda_c}\in\mathcal{N},\|\phi_{\lambda_c}\|=1}\{\left|\langle
A(\mu
T_{k}-1)\phi_{\lambda_c},\phi_{\lambda_c}\rangle\right|\}\geq C
\mid \mu-\lambda_c -\alpha k^2\mid+ k^3\;.
\end{eqnarray}

We shall show this now.


For this let $\phi_{\lambda_c}\in\mathcal{N}\backslash\{0\}$. We shall use Taylors formula to estimate
$\left(\mu T_{k}-1\right)\phi_{\lambda_c}$. In view of (\ref{deft})

$$(\mu T_\mathbf{k}-1)\phi_{\lambda_c}=\int
\left(G^{+}_{k}(\mathbf{x'})-
G^{+}_{0}(\mathbf{x'})\right)A(\mathbf{x}-\mathbf{x'})f(\mathbf{x}-\mathbf{x}')d^3x'\;,$$

i.e. we develop $G^{+}_{k}$ around $k=0$, so we need the following derivatives

\begin{eqnarray}\label{difG}
&&\partial_kG^{+}_{k}\nonumber\\&&=\partial_k\left(\frac{1}{4\pi}e^{ikx}\left(-x^{-1}(E_{k}+\sum_{j=1}^{3}\alpha_{j}k\frac{x_{j}}{x}+\beta
m) -ix^{-2}\sum_{j=1}^{3}\alpha_{j}\frac{x_{j}}{x}\right)\right)
\nonumber\\&&=\frac{1}{4\pi}e^{ikx}\left(-i(E_{k}+\sum_{j=1}^{3}\alpha_{j}k\frac{x_{j}}{x}+\beta
m)
+x^{-1}\sum_{j=1}^{3}\alpha_{j}\frac{x_{j}}{x}-x^{-1}(\frac{k}{E_k}+\sum_{j=1}^{3}\alpha_{j}\frac{x_{j}}{x})\right)
\nonumber\\&&=\frac{1}{4\pi}e^{ikx}\left(-i(E_{k}+\sum_{j=1}^{3}\alpha_{j}k\frac{x_{j}}{x}+\beta
m) -x^{-1}\frac{k}{E_k}\right)\;,
\end{eqnarray}

\begin{eqnarray}\label{difG2}
&&\partial_k^2G^{+}_{k}\nonumber\\&&=\partial_k\frac{1}{4\pi}e^{ikx}\left(-i(E_{k}+\sum_{j=1}^{3}\alpha_{j}k\frac{x_{j}}{x}+\beta
m) -x^{-1}\frac{k}{E_k}\right)
\nonumber\\&&=\frac{1}{4\pi}e^{ikx}\left(x(E_{k}+i\sum_{j=1}^{3}\alpha_{j}k\frac{x_{j}}{x}+\beta
m)
-i\frac{k}{E_k}-i\frac{k}{E_k}-i\sum_{j=1}^{3}\alpha_{j}\frac{x_{j}}{x}-x^{-1}\frac{m^2}{E_k^3}\right)
\nonumber\\&&=\frac{1}{4\pi}e^{ikx}\left(x(E_{k}+\sum_{j=1}^{3}\alpha_{j}k\frac{x_{j}}{x}+\beta
m)-2i\frac{k}{E_k}
-i\sum_{j=1}^{3}\alpha_{j}\frac{x_{j}}{x}-x^{-1}\frac{m^2}{E_k^3}\right)
\end{eqnarray}

and

\begin{eqnarray}\label{difG3}
\partial_k^3G^{+}_{k}&=&\partial_k\frac{1}{4\pi}e^{ikx}\left(x(E_{k}+\sum_{j=1}^{3}\alpha_{j}k\frac{x_{j}}{x}+\beta
m)-2i\frac{k}{E_k} -i\sum_{j=1}^{3}\alpha_{j}\frac{x_{j}}{x}-x^{-1}\frac{m^2}{E_k^3}\right)
\nonumber\\&=&\frac{1}{4\pi}e^{ikx}\big(ix^2(E_{k}+\sum_{j=1}^{3}\alpha_{j}k\frac{x_{j}}{x}+\beta
m)+2x\frac{k}{E_k}-i\frac{m^2}{E_k^3}
+\sum_{j=1}^{3}\alpha_{j}x_{j}\nonumber\\&&+x\frac{k}{E_k}+\sum_{j=1}^{3}\alpha_{j}x_{j}-2i\frac{m^2}{E_k^3}+3x^{-1}\frac{km^2}{E_k^5}\big)
\nonumber\\&=&\frac{1}{4\pi}e^{ikx}\big(ix^2(E_{k}+\sum_{j=1}^{3}\alpha_{j}k\frac{x_{j}}{x}+\beta
m)+3x\frac{k}{E_k}
\nonumber\\&&+2\sum_{j=1}^{3}\alpha_{j}x_{j}-3i\frac{m^2}{E_k^3}+3x^{-1}\frac{km^2}{E_k^5}\big)
\end{eqnarray}


By Taylors formula there exists a $k_0< k$ such that

\begin{eqnarray}\label{taylorsp}
\langle A\left(1-\mu T_{k}\right)\phi_{\lambda_c}\mid\phi_{\lambda_c}\rangle&=&
\langle A\left(1-\mu T_0\right)\phi_{\lambda_c}\mid\phi_{\lambda_c}\rangle
+k\left(\partial_k\langle A\mu T_{k}\phi_{\lambda_c}\mid\phi_{\lambda_c}\rangle\mid_{k=0}\right)
\nonumber\\&&+\frac{1}{2}k^2\left(\partial_k^2\langle A\mu
T_{k}\phi_{\lambda_c}\mid\phi_{\lambda_c}\rangle\mid_{k=0}\right)
\nonumber\\&&+\frac{1}{6}k^3\left(\partial_k^3\langle A\mu
T_{k}\phi_{\lambda_c}\mid\phi_{\lambda_c}\rangle\mid_{k=0}\right)
\nonumber\\&&+\frac{1}{24}k^4\left(\partial_k^4\langle A\mu
T_{k}\phi_{\lambda_c}\mid\phi_{\lambda_c}\rangle\mid_{k=k_0}\right)
\nonumber\\&=:&S_0+S_1+S_2+S_3+S_4\;.
\end{eqnarray}

We estimate these terms separately. For the first term we have using (\ref{taufn}) that

\begin{eqnarray}\label{firstsumm}
S_0=\langle A\left(1-\mu T_0\right)\phi_{\lambda_c}\mid\phi_{\lambda_c}\rangle&=&\frac{\lambda_c-\mu}{\lambda_c}
\langle A\phi_{\lambda_c}\mid\phi_{\lambda_c}\rangle>0
\end{eqnarray}

for $\lambda_c>\mu$ since $A$ is positive.

For $S_1$ we obtain with (\ref{difG}) that

\begin{eqnarray*}
\mu\partial_kT_{k}\mid_{k=0}\phi_{\lambda_c}&=&-i\int
 \frac{1}{4\pi}\left(m+\beta m
\right)A(\mathbf{x}-\mathbf{x'})
 \phi_{\lambda_c}(\mathbf{x}-\mathbf{x}')d^3x'\;.
\end{eqnarray*}

Hence by (\ref{wirdnull})

\begin{equation}\label{secondsumm}
S_1=0\;.
\end{equation}



For $S_3$ we obtain by (\ref{difG3})

\begin{eqnarray}\label{splits3}
S_3&=&\frac{1}{6}k^3\langle A\mu
\int
 \partial_k^3
 G^{+}_{k}(\mathbf{x}-\mathbf{x'})A(\mathbf{x'})\phi_{\lambda_c}(\mathbf{x}')d^3x'
\mid\phi_{\lambda_c}\rangle
\nonumber\\&=&\frac{1}{6}k^3\langle A\mu\frac{1}{4\pi}
\int
i(\mathbf{x}-\mathbf{x}')^2(m+\beta m)
 A(\mathbf{x'})\phi_{\lambda_c}(\mathbf{x}')d^3x'
\mid\phi_{\lambda_c}\rangle
\nonumber\\&&+\frac{1}{6}k^3\langle A\mu\frac{1}{4\pi}
\int
2\sum_{j=1}^{3}\alpha_{j}(x_{j}-x'_j)
 A(\mathbf{x'})\phi_{\lambda_c}(\mathbf{x}')d^3x'
\mid\phi_{\lambda_c}\rangle
\nonumber\\&&+\frac{1}{6}k^3\langle A\mu\frac{1}{4\pi}
\int
-3i\frac{m^2}{m^3}
 A(\mathbf{x'})\phi_{\lambda_c}(\mathbf{x}')d^3x'
\mid\phi_{\lambda_c}\rangle
%
%
\nonumber\\&=&\frac{\mu m i}{24\pi}k^3\int\int A(\mathbf{x})(\mathbf{x}-\mathbf{x}')^2
 A(\mathbf{x'})\phi_{\lambda_c}(\mathbf{x}')(1+\beta)\phi_{\lambda_c}^\dagger(\mathbf{x})d^3x'
d^3x
\nonumber\\&&+\frac{\mu}{12\pi}k^3\int\int A(\mathbf{x})\sum_{j=1}^{3}
 A(\mathbf{x'})\phi_{\lambda_c}(\mathbf{x}')\alpha_{j}(x_{j}-x'_j)\phi_{\lambda_c}^\dagger(\mathbf{x})d^3x'
d^3x
\nonumber\\&&-\frac{\mu i}{8\pi m}k^3\int\int A(\mathbf{x})
 A(\mathbf{x'})\phi_{\lambda_c}(\mathbf{x}')\phi_{\lambda_c}^\dagger(\mathbf{x})d^3x'
d^3x
\nonumber\\&=:&S_{3,1}+S_{3,2}+S_{3,3}\;.
\end{eqnarray}

For $S_{3,1}$ we can write

\begin{eqnarray}\label{s5split}
S_{3,1}&=&\frac{\mu m i}{24\pi}k^3\int\int A(\mathbf{x})(\mathbf{x}^2+\mathbf{x}'^2)
 A(\mathbf{x'})\phi_{\lambda_c}(\mathbf{x}')(1+\beta)\phi_{\lambda_c}^\dagger(\mathbf{x})d^3x'
d^3x
\nonumber\\&&-\frac{\mu m i}{12\pi}k^3\int\int A(\mathbf{x})\mathbf{x}\cdot\mathbf{x}'
 A(\mathbf{x'})\phi_{\lambda_c}(\mathbf{x}')(1+\beta)\phi_{\lambda_c}^\dagger(\mathbf{x})d^3x'
d^3x\;.
%
%
\end{eqnarray}

Using symmetry in exchanging $\mathbf{x}$ with $\mathbf{x}'$ on the first term it becomes

\begin{eqnarray*}
&&\frac{\mu m i}{12\pi}k^3\int\int A(\mathbf{x})\mathbf{x}^2
 A(\mathbf{x'})\phi_{\lambda_c}(\mathbf{x}')(1+\beta)\phi_{\lambda_c}^\dagger(\mathbf{x})d^3x'
d^3x
\\&=&\frac{\mu m i}{12\pi}k^3\int A(\mathbf{x})\mathbf{x}^2\phi_{\lambda_c}(\mathbf{x})\int(1+\beta)
 A(\mathbf{x'})\phi_{\lambda_c}^\dagger(\mathbf{x}')d^3x'
d^3x=0
\end{eqnarray*}

by (\ref{wirdnull}). Thus

\begin{eqnarray*}
S_{3,1}&=&-\frac{\mu m i}{12\pi}k^3\int A(\mathbf{x})\phi^\dagger_{\lambda_c}(\mathbf{x})\mathbf{x}d^3x(1+\beta)
\cdot\int\mathbf{x}'
 A(\mathbf{x'})\phi_{\lambda_c}(\mathbf{x}')d^3x'\;.
\end{eqnarray*}

Setting

\begin{equation}\label{neustern}
\xi:=\sqrt{\frac{\mu m}{12\pi}}\int A(\mathbf{x})\phi_{\lambda_c}(\mathbf{x})\mathbf{x}d^3x\;.
\end{equation}

we obtain

\begin{eqnarray}\label{neustern2}
S_{3,1}&=&-ik^3 \langle\xi (1-\beta) \xi\rangle\;,
\end{eqnarray}

Since $(\beta)$ is self adjoint it follows that
$\langle\xi(1-\beta) \xi\rangle\in\mathbb{R}$, since $\| \beta
\|=1$ it follows that $\langle\xi (1-\beta) \xi\rangle\geq0$ hence
there exists a $C_4\in\mathbb{R}_0^+$ such that

\begin{equation}\label{s5b}
S_{3,1}=-ik^3C_4\;.
\end{equation}

Due to symmetry in exchanging $\mathbf{x}$ with $\mathbf{x}'$ we have that

\begin{equation}\label{s6}
S_{3,2}=-S_{3,2}=0\;.
\end{equation}

For $S_{3,3}$ we can write

\begin{eqnarray}\label{s4gr01}
S_{3,3}&=&-\frac{\mu i}{8\pi m}k^3\left|\int A(\mathbf{x})\phi_{\lambda_c}(\mathbf{x})d^3x\right|^2
\end{eqnarray}

it follows that there exists a $C_5\geq0$ with

\begin{equation}\label{s4gr0}
S_{3,3}=-ik^3C_5\;.
\end{equation}

This (\ref{s5b}) and (\ref{s6}) in (\ref{splits3}) yield that there exists a $C_3\geq0$ such that

\begin{equation}\label{fourthsumm}
S_{3}=-ik^3C_3\;.
\end{equation}

Since $A$ was defined to satisfy either (\ref{restricta1}) or (\ref{restricta2}) it follows taking note of
(\ref{neustern}) and (\ref{neustern2}) as well as (\ref{s4gr01}) that $C_4$ or $C_5>0$.

For $S_4$ (see (\ref{taylorsp})) we have that there exists a $C\in\mathbb{R}$ such that

\begin{equation}\label{s4}\mid S_4 \mid \leq k^4 C\;.\end{equation}

(\ref{firstsumm}), (\ref{secondsumm}) and (\ref{fourthsumm}) in (\ref{taylorsp}) yield that

\begin{eqnarray*}
&&\langle A\left(1-\mu
T_{k}\right)\phi_{\lambda_c}\mid\phi_{\lambda_c}\rangle\\&&=\frac{\mu-\lambda_c}{\lambda_c}
\langle A\phi_{\lambda_c}\mid\phi_{\lambda_c}\rangle+
\frac{1}{2}\partial_k^2\langle A\mu
T_{k}\phi_{\lambda_c}\mid\phi_{\lambda_c}\rangle\mid_{k=0}k^2
-ik^3C_3+S_4\;.
\end{eqnarray*}

We split the equation into real and imaginary part (observing that $A$ is positive)

\begin{eqnarray}\label{realteilneu}
\Re(\langle A\left(1-\mu
T_{k}\right)\phi_{\lambda_c}\mid\phi_{\lambda_c}\rangle)&=&\frac{\mu-\lambda_c}{\lambda_c} \langle
A\phi_{\lambda_c}\mid\phi_{\lambda_c}\rangle\nonumber\\&&+\Re( \frac{1}{2}\partial_k^2\langle A\mu
T_{k}\phi_{\lambda_c}\mid\phi_{\lambda_c}\rangle\mid_{k=0}) k^2+\Re(S_4)
\\\label{imteil}\Im(\langle A\left(1-\mu
T_{k}\right)\phi_{\lambda_c}\mid\phi_{\lambda_c}\rangle)&=&\Im( \frac{1}{2}\partial_k^2\langle A\mu
T_{k}\phi_{\lambda_c}\mid\phi_{\lambda_c}\rangle\mid_{k=0}k^2) \nonumber\\&&-k^3C_5+\Im(S_4)
\end{eqnarray}

Since the first summand in (\ref{realteilneu}) is positive and the second summand in
(\ref{realteilneu})is of order $k^2$ 
it follows with (\ref{s4}) that there exists a $C>0$ and a $\alpha\in\mathbb{R}$ such that

\begin{eqnarray}\label{oncethis2}
\mid\Re(\langle A\left(1-\mu T_{k}\right)\phi_{\lambda_c}\mid\phi_{\lambda_c}\rangle)\mid\geq
C\left|\frac{\mu-\lambda_c}{\lambda_c} -\alpha k^2\right|\;.
\end{eqnarray}

For (\ref{imteil}) observe that since $k^3<k^2$ for $k\rightarrow0$ it follows that there exists a $C>0$ such
that for $k<k_0$ with appropriate $k_0$

\begin{eqnarray}\label{imk2}
\mid\Im(\langle A\left(1-\mu T_{k}\right)\phi_{\lambda_c}\mid\phi_{\lambda_c}\rangle)\mid\geq C k^3\;.
\end{eqnarray}

\begin{rem}
(\ref{imk2}) seems to be not optimal since the right hand side of (\ref{imteil}) seems to be of order of order
$k^2$. But

$$\Im( \frac{1}{2}\partial_k^2\langle A\mu
T_{k}\phi_{\lambda_c}\mid\phi_{\lambda_c}\rangle\mid_{k=0}k^2)=0\;,$$

the proof of which is not given since (\ref{imk2}) suffices.
\end{rem}\vspace{1cm}

Using now the fact that the absolute value of a complex number $z=u+iv$

\begin{eqnarray*}
\mid z \mid = \frac{1}{2}\mid z \mid+\frac{1}{2}\mid z \mid\geq \frac{1}{2}\mid u \mid+\frac{1}{2}\mid v \mid
\end{eqnarray*}

 with (\ref{oncethis2}) and (\ref{imk2}) equation (\ref{opnorma}) follows.

\newpage

\begin{rem}

Note that $\alpha$ is in fact greater than zero: For $\mu<\lambda_c$ there exist imaginary $k$ (in this case we
in fact have no imaginary part c.f. (\ref{imteil}): The imaginary $k$ makes all imaginary parts  which are of
odd order in $k$ real) - thus negative $k^2$ - such that
$B(\mathbf{k},\mu)\widetilde{f}(\mathbf{k},\mu,\cdot)=0$ (namely the respective eigenvalues). Hence we can find
$\mu<\lambda_c$ and $k^2<0$ such that $\frac{\mu-\lambda_c}{\lambda_c} -\alpha k^2\approx 0$, hence
$\alpha\geq0$. Since $\phi_\mu$ dives properly into the continuous spectrum, hence
$\frac{\mu-\lambda_c}{\lambda_c} -\alpha k^2\approx 0$ for $\frac{\mu-\lambda_c}{\lambda_c}$ proportional to
$E_k\approx \frac{k^2}{2m}$ it follows that $\alpha\neq0$.

Also in the case $\mu> \lambda_c$, $k\in\mathbb{R}$ it may happen that $\frac{\mu-\lambda_c}{\lambda_c} -\alpha
k^2=0$. This case is called ''resonance'' in the physical literature. Around the resonance the norm of $B$ is
governed by the imaginary part above which is of order $k^3$. Varying $k$ the real part above changes its sign
when crossing the resonance, so does the divergent part of the generalized eigenfunctions.

\end{rem}\vspace{1cm}

\begin{figure}\label{Bild4}
\begin{center}
\includegraphics*[width=200pt,height=100pt]{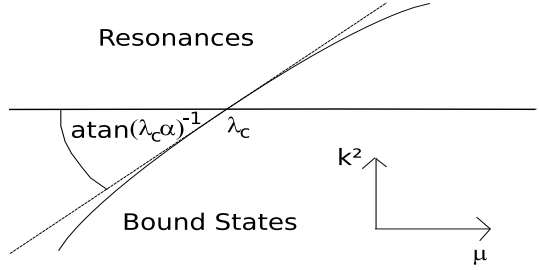}
\caption[alpha]{Resonances and eigenvalues. For $\frac{\mu-\lambda_c}{\lambda_c} -\alpha k^2\approx 0$
 we have eigenvalues for $\mu\leq\lambda_c$ and resonances in the continuous spectrum for $\mu>\lambda_c$. }
\end{center}
\end{figure}

We thus have proven (\ref{punkta}) and (\ref{punktb}) (recalling (\ref{opnorma}) yields (\ref{infmain});
(\ref{infmain}) yields (\ref{punkta}) and (\ref{punktb})). It is left to prove (\ref{punktc}).

Let $(\mathbf{k},\mu)\in\mathcal{P}_{k_0}$, $f\in\mathcal{N}$. Similar as above we have using Taylors formula
that

\begin{eqnarray*}
(\mu T_{k}-1)f&=& (\mu T_0-1)f+k\partial_k(\mu
T_{k})f\mid_{k=0}+\mathcal{O}(k^2)\|
f(\mathbf{k},\mu,\cdot)\|_\infty\;.
\end{eqnarray*}

Since $f\in\widetilde{\mathcal{N}}$

$$(\lambda_c T_0-1)f=0$$

and thus

\begin{equation*}
(\mu T_{0}-1)f= (\frac{\mu}{\lambda_c} -1)f=\frac{\mu-\lambda_c}{\lambda_c}f\;.
\end{equation*}

It follows that

\begin{eqnarray*}
(\mu T_{k}-1)f&=&\frac{\mu-\lambda_c}{\lambda_c}f+k\partial_k(\mu
T_{k}-1)f\mid_{k=0}+\mathcal{O}(k^2)\| f\|_\infty\nonumber\;.
\end{eqnarray*}

and

\begin{eqnarray}\label{taylor2}
\|(\mu T_{k}-1)f\|_\infty&\leq&\frac{\mu-\lambda_c}{\lambda_c}\|
f\|_\infty\nonumber+k\|[\partial_k(\mu
T_{k}-1)]_{k=0}f\mid_{k=0}\|_\infty\\&&+\mathcal{O}(k^2)\|
f\|_\infty\;.
\end{eqnarray}

With (\ref{difG}) and (\ref{deft}) we have that by virtue of (\ref{wirdnull})

\begin{eqnarray*}
[\partial_k(\mu T_{k}-1)]_{k=0}f&=&\frac{-im}{4\pi}\int
 (1+\beta)A(\mathbf{x'})f(\mathbf{x}')d^3x'\;.
\end{eqnarray*}

Using (\ref{wirdnull}) it follows that

\begin{eqnarray*}
[\partial_k(\mu T_{k}-1)]_{k=0}f&=&0\;.
\end{eqnarray*}

Thus we can estimate  (\ref{taylor2}) by

\begin{equation}\label{Bohneproj}
\|(\mu T_{k}-1)f\|_\infty <C\frac{\mu-\lambda_c}{\lambda_c}\|
f\|_\infty+Ck^2\| f\|_\infty
\end{equation}

with appropriate $C<\infty$ uniform in $(\mathbf{k},\mu)\in\mathcal{P}_{k_0}$.

Hence in view of (\ref{linop2})

\begin{eqnarray}\label{supnormB}
\|
    B^\bot(\mathbf{k},\mu)\|_\infty^{op}&=&\sup_{f\in\mathcal{N},\| f\|_\infty=1}\|
    B^\bot(\mathbf{k},\mu) f\|_\infty
    \nonumber\\&=&\sup_{f\in\mathcal{N},\| f\|_\infty=1}
     \| (1-P^\|)(\mu
T_{k}-1)f\|_\infty
\nonumber\\    &\leq&
\sup_{f\in\mathcal{N},\| f\|_\infty=1}
     \| (\mu
T_{k}-1)f\|_\infty
 \nonumber\\&& + \sup_{f\in\mathcal{N},\| f\|_\infty=1}
     \| P^\|(\mu
T_{k}-1)f\|_\infty
\end{eqnarray}

Using the equivalence of all norms in the finite dimensional vector space $\mathcal{M}$ it follows that there
exists a $C<\infty$ uniform in $(\mathbf{k},\mu)\in\mathcal{P}_{k_0}$ such that

\begin{eqnarray*}
\sup_{f\in\mathcal{N},\| f\|_\infty=1}
     \| P^\|(\mu
T_{k}-1)f\|_\infty&\leq&C\sup_{f\in\mathcal{N},\| f\|_\infty=1}
     \| P^\|(\mu
T_{k}-1)f\|
\\&\leq&C\sup_{f\in\mathcal{N},\| f\|_\infty=1}
     \| (\mu
T_{k}-1)f\|
\\&\leq& C\frac{\mu-\lambda_c}{\lambda_c}\| f\|_\infty+C_3k^2\|
f\|_\infty
\end{eqnarray*}

by (\ref{Bohneproj}).

This and (\ref{Bohneproj}) in (\ref{supnormB}) yield (\ref{punktc}).

\begin{flushright}$\Box$\end{flushright}

We turn now to the proof of Lemma \ref{senkrechtphic}, and we recall the Lemma for convenience.

\vspace{0.5cm}
\noindent\textbf{Lemma \ref{senkrechtphic}}\\

\it

\begin{itemize}

\item[(a)] For any $\mu\in[\lambda_c,\overline{\mu}]$ we have that

$$
h^\bot\in\mathcal{M}^\bot\Leftrightarrow (\mu T_0-1) h^\bot\in\mathcal{M}^\bot\;.
$$

\item[(b)] For any $\mu\in[\lambda_c,\overline{\mu}]$ the map $\mu T_0-1:
\mathcal{M}^\bot\rightarrow\mathcal{M}^\bot$ is invertible.

\item[(c)] There exists a $C<\infty$ such that for all $h^\bot\in\mathcal{M}^\bot$ and all
$\mu\in[\lambda_c,\overline{\mu}]$

\begin{equation*}
\| \left(\mu T_0-1\right)^{-1}h^\bot\|_\infty\leq C \|
h^\bot\|_\infty\;.
\end{equation*}

\end{itemize}

\rm

\vspace{0.5cm}
\noindent\textbf{Proof of part a) of Lemma \ref{senkrechtphic}}\\

Let $\mu\in[\lambda_c,\overline{\mu}]$.

We show first that for $h\in\mathcal{B}$ and $g\in\mathcal{B}\cap L_2$ with $T_0 g\in L_2$

\begin{equation}\label{tselbstad}
\langle A h,T_0 g\rangle = \langle AT_0 h, g\rangle
\end{equation}

by computing

\begin{eqnarray}\label{symmetric}
\langle Ah\mid T_0g \rangle&=&\int A(\mathbf{x})h^*(\mathbf{x}) T_0g(\mathbf{x})d^3x \nonumber\\&=&\int
A(\mathbf{x})h^*(\mathbf{x})\int G^{+}_{0}(\mathbf{x}-\mathbf{x'}) A(\mathbf{x'})g(\mathbf{x}')d^3x'd^3x
\nonumber\\&=&\int\int A(\mathbf{x})h^*(\mathbf{x}) G^{+}_{0}(\mathbf{x}-\mathbf{x'})d^3x
A(\mathbf{x'})g(\mathbf{x}')d^3x' \nonumber\\&=&\langle AT_0 h\mid g\rangle\;.
\end{eqnarray}

We may apply this to $h\in\mathcal{B}$ and $g=\phi\in\mathcal{N}$ to obtain

\begin{eqnarray*}
\langle Ah,\phi\rangle&=&\langle Ah,\lambda_cT_0\phi\rangle\\&=&\lambda_c\langle AT_0h,\phi\rangle
\end{eqnarray*}

and thus

\begin{eqnarray*}
\frac{\mu}{\lambda_c}\langle Ah,\phi\rangle&=&\mu\langle AT_0h,\phi\rangle
\end{eqnarray*}

and hence

\begin{eqnarray}\label{sternein}
\langle A \left(1-\mu T_0\right) h\mid \phi\rangle &=&\langle A\frac{\lambda_c-\mu}{\lambda_c} h\mid
\phi\rangle=\frac{\lambda_c-\mu}{\lambda_c}\langle Ah,\phi\rangle\;.
\end{eqnarray}

This equation directly implies part a) of Lemma \ref{senkrechtphic}: If $h\in\mathcal{M}^\bot$ (which means that
$\langle Ah,\phi_{\lambda_c}\rangle=0$) it follows that $\left(1-\mu T_0\right) h\in \mathcal{M}^\bot$ (which
means $\langle A \left(1-\mu T_0\right) h\mid \phi\rangle=0$) and vice versus.

\vspace{0.5cm}
\noindent\textbf{Proof of part b) of Lemma \ref{senkrechtphic} for $\mu\neq\lambda_c$}\\

First observe that there exists no bound state or resonance for $\mu\in]\lambda_c,\overline{\mu}]$ hence
following the proof of Lemma 3.4. in \cite{pickl} $(\mu T_0-1)^{-1}f$ exists for all $f\in\mathcal{B}$, so $\mu
T_0-1$ is invertible on $\mathcal{M}^\bot\subset\mathcal{B}$.

\vspace{0.5cm}
\noindent\textbf{Proof of part c) of Lemma \ref{senkrechtphic} for $\mu\neq\lambda_c$}\\

Let $\mu\in]\lambda_c,\overline{\mu}]$, $f\in\mathcal{M}^\bot$. Set

\begin{equation}\label{defhlem}
h:=\left(1-\mu T_0\right)^{-1}f\;.
\end{equation}

Then by Lemma \ref{senkrechtphic} (a)
\begin{equation}\label{hierzu}h\in\mathcal{M}^\bot\;.\end{equation}

We now show that there exists a $C<\infty$ such that for all $h\in\mathcal{M}^\bot$ and
$\mu\in]\lambda_c,\overline{\mu}]$

\begin{equation}\label{equivinvert}
\| h\|_\infty\leq C \| \left(1-\mu T_0\right)h\|_\infty
\end{equation}

from which (\ref{invert}) follows.

We will prove (\ref{equivinvert}) by contradiction. Hence assume that for every $C>0$ there exist a $\mu_C$ with
$\lambda_c<\mu_C\leq\overline{\mu}$ and a function $h_C\in\mathcal{M}^\bot$ such that

\begin{equation}\label{assumeinvert}
\| h_C\|_\infty> C \| \left(1-\mu_C T_0\right)h_C\|_\infty\;.
\end{equation}

Since $T_0$ is a linear operator we can restrict ourselves to
functions $h_C$ with $\| h_C\|_\infty=1$. Hence
(\ref{assumeinvert}) becomes

\begin{equation}\label{asumeinv2}
\| \left(1-\mu_C T_0\right)h_C\|_\infty<\frac{1}{C}\;.
\end{equation}
Consider a sequence $C_n\rightarrow\infty$. Hence there exists a series of elements
$(\mu_n,h_n)_{n\in\mathbb{N}}$ in $]\lambda_c,\overline{\mu}]\times\mathcal{M}^\bot$ such that

$$
\lim_{n\rightarrow\infty}\|\left(1-\mu_n
T_0\right)h_n\|_\infty=0\;,
$$

i.e. in sup norm

\begin{equation}\label{limhn}
\lim_{n\rightarrow\infty}\left(1-\mu_n T_0\right)h_n=0\;,
\end{equation}


But the sequence $\mu_nT_0 h_n$ is Arzela-Ascoli compact, since

\begin{equation}\label{defseta}
\mathcal{A}:=\{T_0 g\in\mathcal{B},\space\| g\|_\infty=1\}
\end{equation}

 is compact in the Arzela-Ascoli
sense, i.e.  for any $\delta>0$ there exists a $\zeta>0$ such that

\begin{equation}\label{arzasc}
\mid f(\mathbf{x})-f(\mathbf{y})\mid <\delta
\end{equation}

for all $\mathbf{x}, \mathbf{y}\in\mathbb{R}^3$ with $\|
\mathbf{x}-\mathbf{y}\|<\zeta$ and all $f\in\mathcal{A}$.

To prove this let $\delta>0$, $f\in\mathcal{A}$ and let $g$ be
such that $f=T_0 g$ ,$\| g\|_\infty=1$.

Then

\begin{eqnarray}
\mid f(\mathbf{x})-f(\mathbf{y})\mid&=& \mid T_0g(\mathbf{x})-T_0g(\mathbf{y})\mid\nonumber\\&=&\big|\int
G^{+}_{0}(\mathbf{x}-\mathbf{x'})A(\mathbf{x'})g(\mathbf{x}')d^3x'
\nonumber\\&&-\int G^{+}_{0}(\mathbf{y}-\mathbf{y}')A(\mathbf{y'})g(\mathbf{y'})d^3x'\big|
\nonumber\\&=&\big|\int
\left(G^{+}_{0}(\mathbf{x}-\mathbf{x'})-G^{+}_{0}(\mathbf{y}-\mathbf{x}')\right)A(\mathbf{x'})g(\mathbf{x}')d^3x'
\big|
\nonumber\\&\leq&\| A\|_\infty \| g \|_\infty \int_{\mathcal{S}_A}
\mid
G^{+}_{0}(\mathbf{x}-\mathbf{x'})-G^{+}_{0}(\mathbf{y}-\mathbf{x}')\mid
d^3x'
\nonumber\\\label{anddies}&=&\| A\|_\infty \int_{\mathcal{S}_A}
\mid
G^{+}_{0}(\mathbf{x}-\mathbf{x'})-G^{+}_{0}(\mathbf{y}-\mathbf{x}')\mid
d^3x'
\end{eqnarray}

Let $r>0$ and $K_r(\mathbf{x})$ the open ball around $\mathbf{x}$ with radius $r$. Since $\mid
G^{+}_{0}(\mathbf{x}-\mathbf{x'})\mid$ (see (\ref{kernel})) is integrable over any compact set we can choose $r$
so small that for all $\mathbf{x}$ and all $\mathbf{y}$

\begin{eqnarray}\label{chooser}
\int_{K_r(\mathbf{x})} \mid G^{+}_{0}(\mathbf{x}-\mathbf{x'})\mid
d^3x'&<&\frac{\delta}{3\| A\|_\infty}
\\\label{chooser2}\int_{K_r(\mathbf{x})} \mid G^{+}_{0}(\mathbf{y}-\mathbf{x'})\mid
d^3x'&<&\frac{\delta}{3\| A\|_\infty}\;.
\end{eqnarray}

Let

$$h(\mathbf{x}):=\mid
G^{+}_{0}(\mathbf{x}-\mathbf{x'})-G^{+}_{0}(\mathbf{y}-\mathbf{x}')\mid$$

For $\mathbf{y}\in K_{\frac{r}{2}}(\mathbf{x})$ the function is continuous on the compact set
$\mathcal{S}_A\backslash K_r(\mathbf{x})$ and thus uniformly continuous on $\mathcal{S}_A\backslash
K_r(\mathbf{x})$(recall that $G^{+}_{0}(\mathbf{x}')$ is continuous on $\mathbb{R}^3\backslash\{0\}$), i.e.
there exists a $\zeta_1>0$ such that

$$\mid
G^{+}_{0}(\mathbf{x}-\mathbf{x'})-G^{+}_{0}(\mathbf{y}-\mathbf{x}')\mid<\frac{\delta}{3\|
A\|_\infty \int_{\mathcal{S}_A\backslash K_r(\mathbf{x})}d^3x}$$

 for all $
\mathbf{x}'\in \mathcal{S}_A\backslash K_r(\mathbf{x})\mid$ and
for all $\| \mathbf{x}-\mathbf{y}\|<\zeta_1$, hence

\begin{eqnarray*}
\| A\|_\infty \int_{\mathcal{S}_A\backslash K_r(\mathbf{x})} \mid
G^{+}_{0}(\mathbf{x}-\mathbf{x'})-G^{+}_{0}(\mathbf{y}-\mathbf{x}')\mid
d^3x'&\leq&\frac{\delta}{3}
\end{eqnarray*}

for $\| \mathbf{x}-\mathbf{y}\|<\zeta_1$.

With (\ref{chooser}) and (\ref{chooser2}) we obtain for (\ref{anddies})

\begin{eqnarray*}
\mid f(\mathbf{x})-f(\mathbf{y})\mid\leq\| A\|_\infty
\int_{\mathcal{S}_A} \mid
G^{+}_{0}(\mathbf{x}-\mathbf{x'})-G^{+}_{0}(\mathbf{y}-\mathbf{x}')\mid
d^3x'&\leq&\delta
\end{eqnarray*}

for $\|
\mathbf{x}-\mathbf{y}\|<\zeta:=\min\{\zeta_1,\frac{r}{2}\}$.

It follows directly that
$\widetilde{\mathcal{A}}:=[\lambda_c,\overline{\mu}]\times\cup_{\mu\in[\lambda_c,\overline{\mu}]}\mu\mathcal{A}$
is compact.

Thus there exists a convergent subsequence $(\mu_{n(k)},\mu_{n(k)} T_0 h_{n(k)})_{k\in\mathbb{N}}$ of
$(\mu_n,\mu_n T_0 h_n)_{n\in\mathbb{N}}$ with
$\lim_{k\rightarrow\infty}T_0h_{n(k)}=\widetilde{h}^\bot\in\mathcal{A}$ and
$\lim_{k\rightarrow\infty}\mu_{n(k)}=\mu\in[\lambda_c,\overline{\mu}]$.

By virtue of (\ref{limhn}) $h_{n(k)}$ converges, hence there exists a $h\in\mathcal{B}$ with
$\lim_{k\rightarrow\infty}h_{n(k)}=h$. Since $T_0$ is continuous, $T_0h=\widetilde{h}^\bot$.

Hence by (\ref{limhn}) $(1-\mu T_0)h=0$. Since $(1-\mu T_0)h=0$ has nontrivial solutions only for
$\mu=\lambda_c$ it follows that $\mu=\lambda_c$ and $h\in\mathcal{N}$. Since $A$ is positive we have that

\begin{equation}\label{contra1}\langle
h,A h\rangle>0\;.\end{equation}

On the other hand since $h_n\in\mathcal{M}^\bot$

\begin{equation}\label{contra2}
\langle h,A h_n \rangle=0
\end{equation}

for all $n\in\mathbb{N}$. (\ref{contra1}) with (\ref{contra2}) is contradiction to the continuity of the scalar
product. So (\ref{equivinvert}) holds and part c) of the Lemma follows for $\mu\neq\lambda_c$.

\vspace{0.5cm}
\noindent\textbf{Proof of part (b) of Lemma \ref{senkrechtphic} for $\mu=\lambda_c$}\\

Let $f\in\mathcal{M}^\bot$. We develop $(1-\mu T_0)^{-1}$ into a Neumann series around $1+\frac{1}{2C}$, which
we show to be convergent on $M^\bot$ if $C$ is chosen such that (\ref{invert}) is satisfied for
$\mu\neq\lambda_c$ and $\frac{\lambda_c}{1+\frac{1}{2C}}\leq\overline{\mu}$ (note, that
$\lim_{C\rightarrow\infty}\frac{\lambda_c}{1+\frac{1}{2C}}=\lambda_c<\overline{\mu}$).

\begin{lem}\label{overlinef}
Let $f\in\mathcal{M}^\bot$ and let

\begin{equation}\label{deffinv}
f_n:= (1+\frac{1}{2C}-\lambda_c T_0)^{-1}\sum_{j=0}^n \left(\frac{1}{2C}(1+\frac{1}{2C}-\lambda_c
T_0)^{-1}\right)^jf\;.
\end{equation}

Then there exists a $\overline{f}\in\mathcal{B}$ such that

$$\lim_{n\rightarrow\infty}\| f_n-\overline{f}\|_\infty=0\;.$$

\end{lem}

\vspace{0.5cm}
\noindent\textbf{Proof}\\

Let $f\in\mathcal{M}^\bot$. We have that

\begin{eqnarray*}
1+\frac{1}{2C}-\lambda_c
T_0&=&(1+\frac{1}{2C})\left(1-\frac{\lambda_c}{1+\frac{1}{2C}}T_0\right)=(1+\frac{1}{2C})\left(1-\mu_0T_0\right)\;.
\end{eqnarray*}

where $\mu_0:=\frac{\lambda_c}{1+\frac{1}{2C}}$. Note that $\mu_0>\lambda_c$ and by our choice of $C$
$\mu_0\leq\overline{\mu}$.

Hence part (c) of the Lemma for $\mu=\mu_0\in]\lambda_c,\overline{\mu}]$ yields that on $\mathcal{M}^\bot$

\begin{eqnarray*}
\sup_{f\in\mathcal{M}^\bot}\frac{\|(1+\frac{1}{2C}-\lambda_c
T_0)^{-1}f\|_\infty}{\| f\|_\infty}
\leq C (1+\frac{1}{2C})^{-1}\;.
\end{eqnarray*}

It follows that

\begin{eqnarray*}
\sup_{f\in\mathcal{M}^\bot}\frac{\|(1+\frac{1}{2C}-\lambda_c
T_0)^{-1}f\|_\infty}{\| f\|_\infty}
\leq\frac{1}{2}(1+\frac{1}{2C})^{-1}<\frac{1}{2}\;.
\end{eqnarray*}

Using part (a) of the Lemma it follows, that for $f\in\mathcal{M}^\bot$ all the summands in (\ref{deffinv}) are
in $\mathcal{M}^\bot$. Hence

\begin{eqnarray*}
\sup_{f\in\mathcal{M}^\bot}\frac{\|\left((1+\frac{1}{2C}-\lambda_c
T_0)^{-1}\right)^jf\|_\infty}{\| f\|_\infty}
<\frac{1}{2^j}\;.
\end{eqnarray*}

Hence the series on the right hand side of (\ref{deffinv}) is majorized by a geometrical series and thus it
converges and since $\mathcal{B}$ is a banach space, $\overline{f}\in\mathcal{B}$ with

$$\overline{f}(\mathbf{x}):=\left((1+\frac{1}{2C}-\lambda_c
T_0)^{-1}\sum_{j=0}^\infty \left(\frac{1}{2C}(1+\frac{1}{2C}-\lambda_c T_0)^{-1}\right)^jf\right)(\mathbf{x})$$

exists.

\begin{flushright}$\Box$\end{flushright}

Furthermore we have for $\overline{f}$ that

\begin{eqnarray*}
(1-\lambda_c T_0)\overline{f}
&=&f+\frac{1}{2C}\overline{f}-\frac{1}{2C}\overline{f}=f\;.
\end{eqnarray*}

We have proven that for any $f\in\mathcal{M^\bot}$ there exists a $\overline{f}\in\mathcal{B}$ such that
$(1-\lambda_c T_0)\overline{f}=f$, hence $(1-\lambda_c T_0)$ is invertible on $\mathcal{M}^\bot$ which is part
(b) of Lemma \ref{senkrechtphic} for $\mu=\lambda_c$.

\vspace{0.5cm}
\noindent\textbf{Proof of part (c) of Lemma \ref{senkrechtphic} for $\mu=\lambda_c$}\\

Part c) of Lemma \ref{senkrechtphic} for $\mu=\lambda_c$ follows direct from part (b) of the Lemma and part (c)
of the Lemma for $\mu\neq\lambda_c$ using the continuity of the operator $T_0$ (see (\ref{zetatildelse2})).

\begin{flushright}$\Box$\end{flushright}

\vspace{0.5cm}
\noindent\textbf{Proof of Lemma \ref{ordnungtk}}\\

We  recall the Lemma for convenience

\vspace{0.5cm}
\noindent\textbf{Lemma \ref{ordnungtk}}\\

\it

 There exists a $C\in\mathbb{R}$ such that
for all $h\in\mathcal{B}$

\begin{equation}\label{abschtkh}
\|(T_k-T_0)h\|_\infty<Ck\| h\|_\infty\;,
\end{equation}

and

\begin{equation}\label{abschtkhsp}
\mid \langle A(T_k-T_0)h\mid\phi_{\lambda_c}\rangle\mid<Ck^2\|
h\|_\infty
\end{equation}

for all $\phi_{\lambda_c}\in\mathcal{N}$ with
$\|\phi_{\lambda_c}\|_\infty=1$.

\rm

\vspace{0.5cm}
\noindent\textbf{Proof}\\

Let $h\in\mathcal{B}$. We have using (\ref{deft})

\begin{eqnarray*}
(T_{k}-T_0)h&=&\int
 \left(G^{+}_{k}(\mathbf{x'})-G^{+}_{0}(\mathbf{x'})\right)A(\mathbf{x}-\mathbf{x'})h(\mathbf{x}-\mathbf{x}')d^3x'\;.
\end{eqnarray*}

It follows that

\begin{eqnarray*}
\|(T_{k}-T_0)h\|_\infty&\leq&\| h\|_\infty\int
 \mid G^{+}_{k}(\mathbf{x'})-G^{+}_{0}(\mathbf{x'})\mid
 A(\mathbf{x}-\mathbf{x'})d^3x'\\
 &\leq&\| h\|_\infty\| A\|_\infty
\int_{\mathbf{x}-\mathbf{x'}\in\mathcal{S}_A}
 \mid G^{+}_{k}(\mathbf{x'})-G^{+}_{0}(\mathbf{x'})\mid
d^3x' \;.
\end{eqnarray*}

Using the definition of $G^{+}_{k}$ (see (\ref{kernel})) we have that

\begin{eqnarray*}&&\|
G^{+}_{k}(\mathbf{x})-G^{+}_{0}(\mathbf{x})\|_\infty
\\&&\hspace{0.5cm}=\|\frac{1}{4\pi}e^{ikx}\left(-x^{-1}(E_{k}+\sum_{j=1}^{3}\alpha_{j}k\frac{x_{j}}{x}+\beta m)
-ix^{-2}\sum_{j=1}^{3}\alpha_{j}\frac{x_{j}}{x}\right)\\&&\hspace{0.7cm}-\frac{1}{4\pi}\left(-x^{-1}(m+\beta
m) -ix^{-2}\sum_{j=1}^{3}\alpha_{j}\frac{x_{j}}{x}\right)\|_\infty
\\&&\hspace{0.5cm}\leq\|\frac{1}{4\pi}e^{ikx}\left(-x^{-1}(E_{k}-m+\sum_{j=1}^{3}\alpha_{j}k\frac{x_{j}}{x})\right)\|_\infty
\\&&\hspace{0.7cm}+\|
\frac{1}{4\pi}(e^{ikx}-1))\left(-x^{-1}(m+\beta m)
-ix^{-2}\sum_{j=1}^{3}\alpha_{j}\frac{x_{j}}{x}\right)\|_\infty\;.
\end{eqnarray*}

The first summand is of order $k$. Since $e^{ikx}-1$ is of order $kx$, the second summand is of order $k$ and
(\ref{abschtkh}) follows.

For the left hand side of (\ref{abschtkhsp}) we use Taylors formula, i.e. that there exists a $k_1$ so that

\begin{eqnarray}\label{splitins1unds2}
&&\langle A(T_{k}-T_0)h,\phi_{\lambda_c}\rangle \nonumber\\&&\hspace{2cm}=
k \partial_k \langle A(\mathbf{x})( T_{k}
h)(\mathbf{x}),\phi_{\lambda_c}(\mathbf{x})\rangle\mid_{k=0}+\frac{1}{2}k^2\partial_k^2 \langle A(
T_{k})h,\phi_{\lambda_c}\rangle\mid_{k=k_1}
\nonumber\\&&\hspace{2cm}=
k  \langle A(\mathbf{x})(
\partial_kT_{k}\mid_{k=0}
h)(\mathbf{x}),\phi_{\lambda_c}(\mathbf{x})\rangle+\frac{1}{2}k^2 \langle A(\partial_k^2
T_{k})\mid_{k=k_1}h\mid\phi_{\lambda_c}\rangle
\nonumber\\&&\hspace{2cm}=:S_1+S_2
\end{eqnarray}

For $S_1$ we have using (\ref{deft}) and (\ref{difG})

\begin{eqnarray}\label{s1absch}
S_1&=&k\int \left(A(\mathbf{x})(
\partial_kT_{k}\mid_{k=0}
h)(\mathbf{x})\right)^*\phi_{\lambda_c}(\mathbf{x})d^3x
\nonumber\\&=&k \int\int A(\mathbf{x})\left(\partial_kG_{k}(\mathbf{x}-\mathbf{x}')\mid_{k=0}
A(\mathbf{x'})h(\mathbf{x}')\right)^*\phi_{\lambda_c}(\mathbf{x})d^3x'd^3x
\nonumber\\&=&k \int\int A(\mathbf{x})\frac{1}{4\pi}\left(i(m+\beta m)\right)
A(\mathbf{x'})h^*(\mathbf{x}')\phi_{\lambda_c}(\mathbf{x})d^3x'd^3x
\nonumber\\&=&k \int A(\mathbf{x})\left(i(m+\beta m)\right)\phi_{\lambda_c}(\mathbf{x})d^3x\frac{1}{4\pi}
A(\mathbf{x'})h^*(\mathbf{x}')d^3x'
=0
\end{eqnarray}

by virtue of (\ref{wirdnull}) (Recall that A is scalar).

For $S_2$ we have using (\ref{deft}) and (\ref{difG}) that

\begin{eqnarray*}
(\partial_k^2 T_{k})h\mid_{k=k_1}&=&\int
\frac{1}{4\pi}e^{ik_1\mid
\mathbf{x}-\mathbf{x'}\mid}\big(\mid\mathbf{x}-\mathbf{x'}\mid(E_{k_1}+\sum_{j=1}^{3}\alpha_{j}k_1\frac{x_{j}-x'_j}{x}
+\beta m)
\\&&-2i\frac{k_1}{E_{k_1}} -i\sum_{j=1}^{3}\alpha_{j}\frac{x_{j}-x'_j}{\mid\mathbf{x}-\mathbf{x'}\mid}
-\mid\mathbf{x}-\mathbf{x'}\mid^{-1}\frac{m^2}{E_{k_1}^3}\big)
A(\mathbf{x'})h(\mathbf{x}')d^3x'
\\&\leq&\frac{1}{4\pi}\int
\mid\mathbf{x}-\mathbf{x'}\mid\;\;\big|(E_{k_1}+\sum_{j=1}^{3}\alpha_{j}k_1\frac{x_{j}-x'_j}{x} +\beta
m)-2i\frac{k_1}{E_{k_1}}
\\&&-i\sum_{j=1}^{3}\alpha_{j}\frac{x_{j}-x'_j}{\mid\mathbf{x}-\mathbf{x'}\mid}
-\mid\mathbf{x}-\mathbf{x'}\mid^{-1}\frac{m^2}{E_{k_1}^3}\big)\big|\;\;
A(\mathbf{x'})\| h\|_\infty d^3x' \;.
\end{eqnarray*}

Since $A(\mathbf{x'})$ has compact support $\frac{(\partial_k^2
T_{k})h\mid_{k=k_1}}{\| h\|_\infty}$ is bounded in
$\mathbf{x}\in\mathbb{R}^3$. Going back to (\ref{splitins1unds2})
we see - again using that $A(\mathbf{x'})$ has compact support -
that there exists a $C<\infty$ such that

$$S_2\leq C k^2\| h\|_\infty$$

for all in $h\in\mathcal{B}$ and $\mathbf{k}\in\mathbb{R}^3$.

With (\ref{s1absch}) (\ref{abschtkhsp}) follows and Lemma \ref{ordnungtk} is proven.

\begin{flushright}$\Box$\end{flushright}

\vspace{0.5cm}
\noindent\textbf{Lemma \ref{equive} for $n>0$}\\

We exemplarily prove (e) for $n=1$.

Heuristically deriving (\ref{zetatildelse}) with respect to $k$ will yield
$\partial_{k}\zeta^{j}(\mathbf{k},\mu,\cdot)$. We denote the function we get by this formal method by
$\dot{\zeta}^{j}(\mathbf{k},\mu,\cdot)$.

\begin{eqnarray*}
\dot{\zeta}^{j}(\mathbf{k},\mu,\mathbf{x})&=&\mu\partial_k g(\mathbf{k},\mu,\mathbf{x})-\mu(\partial_k
T_k)\zeta^{j}(\mathbf{k},\mu,\mathbf{x})
-\mu T_{k}\dot{\zeta}^{j}(\mathbf{k},\mu,\mathbf{x})\;.
\end{eqnarray*}

 Using the definition
of $T_{k}$ (\ref{deft} we get

\begin{eqnarray}\label{ableitung}
\dot{\zeta}^{j}(\mathbf{k},\mu,\mathbf{x})&=&\mu\partial_k g(\mathbf{k},\mu,\mathbf{x})-\mu\int
A(\mathbf{x}')\partial_{k}G(\mathbf{k},\mu,\cdot)^{+}(\mathbf{x}-\mathbf{x}')\zeta^{j}(\mathbf{k},\mu,\mathbf{x}')d^{3}x'
\nonumber\\&&-\mu T_{k}\dot{\zeta}^{j}(\mathbf{k},\mu,\mathbf{x})\;.
\end{eqnarray}

In \cite{pickl} it is shown that (\ref{ableitung}) has a unique solution and that in fact
$\dot{\zeta}^{j}(\mathbf{k},\mu,\cdot)=\partial_{k}\zeta^{j}(\mathbf{k},\mu,\cdot)$.

In the present paper we want to go further and also establish the estimates  needed in (\ref{lemgeprobe1}) and
(\ref{lemgeprobe}). Set

\begin{eqnarray}\label{dkzetatilde}
\nonumber\widetilde{g}(\mathbf{k},\mu,\mathbf{x})&:=&\mu\partial_k g(\mathbf{k},\mu,\mathbf{x})\\&&-\mu\int
A(\mathbf{x}')\zeta^{j}(\mathbf{k},\mu,\mathbf{x}')
\partial_{k}G(\mathbf{k},\mu,\cdot)^{+}(\mathbf{x}-\mathbf{x}')d^{3}x'\;,
\end{eqnarray}

hence

$$\dot{\zeta}^{j}(\mathbf{k},\mu,\mathbf{x})=\widetilde{g}(\mathbf{k},\mu,\mathbf{x})-\mu
T_{k}\dot{\zeta}^{j}(\mathbf{k},\mu,\mathbf{x})\;.$$

Note that $\widetilde{g}(\mathbf{k},\mu,\mathbf{x})$ is in general not in $\mathcal{B}$. Hence we proceed as
above (see below (\ref{LSE})) and define

\begin{eqnarray*}
\overline{\zeta}^{j}(\mathbf{k},\mu,\mathbf{x})&:=&\dot{\zeta}^{j}(\mathbf{k},\mu,\mathbf{x})-\widetilde{g}(\mathbf{k},\mu,\mathbf{x})(\mathbf{x})\;.
\end{eqnarray*}
$\overline{\zeta}^{j}(\mathbf{k},\mu,\cdot)$ satisfies:

\begin{equation}\label{ueberzetatilde}
\overline{\zeta}^{j}(\mathbf{k},\mu,\mathbf{x})=\mu\overline{g}(\mathbf{k},\mu,\cdot)-\mu
T_{k}\overline{\zeta}^{j}(\mathbf{k},\mu,\cdot)
\end{equation}

with

\begin{equation}\label{gtilde}
\overline{g}(\mathbf{k},\mu,\cdot):=-T_{k}\widetilde{g}(\mathbf{k},\mu,\mathbf{x})\;.
\end{equation}

Since $T_k$ maps $C^\infty$ into $\mathcal{B}$ it follows that $\overline{g}\in\mathcal{B}$.

Multiplying both sides with of (\ref{ueberzetatilde})
$\|\overline{g}(\mathbf{k},\mu,\cdot)\|_\infty^{-1}$ it is
formally equivalent to (\ref{zetatildelse}). Hence showing that
$\overline{g}\in\widetilde{\mathcal{B}}$ and that

\begin{eqnarray}\label{gbarest}
\|\overline{g}(\mathbf{k},\mu,\cdot)\|_\infty< C
\left(1+k\left(\mid\mu-\lambda_c-\alpha
k^{2}\mid+k^{3}\right)^{-1}\right)\;.
\end{eqnarray}

for some appropriate constant $C<\infty$ we get (\ref{lemgeprobe1}) and (\ref{lemgeprobe}) for $n=1$.

We need some properties of $\overline{g}$, for that we control $\widetilde{g}$. We define

\begin{eqnarray}\label{g1und2}
\widetilde{g}_1(\mathbf{k},\mu,\mathbf{x})&:=&\mu\partial_k g(\mathbf{k},\mu,\mathbf{x})
\\\label{g1und2b}\widetilde{g}_2(\mathbf{k},\mu,\mathbf{x})&:=&\mu\int
A(\mathbf{x}-\mathbf{x}')\zeta^{j}(\mathbf{k},\mu,\mathbf{x}-\mathbf{x}')
\partial_{k}G(\mathbf{k},\mu,\cdot)^{+}(\mathbf{x}')d^{3}x'
\end{eqnarray}

hence

$$\widetilde{g}=\widetilde{g}_1+\widetilde{g}_2\;.$$

Using (\ref{defg}) we have that

\begin{eqnarray*}
\widetilde{g}_1(\mathbf{k},\mu,\mathbf{x})&=&-\mu\partial_k
T_\mathbf{k}\phi^{j}(\mathbf{k},0,\mathbf{x})\nonumber\\
\\&=&-\mu\int
A(\mathbf{x}-\mathbf{x'})\phi^{j,\mathbf{k}}_{0}(\mathbf{x}-\mathbf{x}')
 \partial_k
 G^{+}_{k}(\mathbf{x'})d^3x'
\nonumber\\&&-\mu\int G^{+}_{k}(\mathbf{x'})A(\mathbf{x}-\mathbf{x'})
 \partial_k\phi^{j,\mathbf{k}}_{0}(\mathbf{x}-\mathbf{x}')d^3x'\;.
\end{eqnarray*}

Using (\ref{difG}) it follows that

\begin{eqnarray*}
\widetilde{g}_1(\mathbf{k},\mu,\mathbf{x})&=& -\mu\int
A(\mathbf{x}-\mathbf{x'})\phi^{j,\mathbf{k}}_{0}(\mathbf{x}-\mathbf{x}')
 \frac{1}{4\pi}e^{ikx'}\\&&\left(-i(E_{k}+\sum_{j=1}^{3}\alpha_{j}k\frac{x'_{j}}{x'}+\beta m)
-x'^{-1}\frac{k}{E_k}\right)d^3x'
\nonumber\\&&-\mu\int G^{+}_{k}(\mathbf{x'})A(\mathbf{x}-\mathbf{x'})
 \partial_k\phi^{j,\mathbf{k}}_{0}(\mathbf{x}-\mathbf{x}')d^3x'\;.
\end{eqnarray*}

Since $A$ has compact support it follows that

\begin{eqnarray*}
\sup_{\mathbf{x}\in\mathbb{R}^3}\int A(\mathbf{x}-\mathbf{x'})d^3x'&<&\infty
\\\sup_{\mathbf{x}\in\mathbb{R}^3}\int
A(\mathbf{x}-\mathbf{x'})x'd^3x'&<&\infty
\\\sup_{\mathbf{x}\in\mathbb{R}^3}\int
A(\mathbf{x}-\mathbf{x'})x'^2d^3x'&<&\infty\;.
\end{eqnarray*}

Since $\phi^{j}(\mathbf{k},0,\mathbf{x})$ is normalized it follows in view of (\ref{kernel}) that

\begin{equation}\label{estp1}
\sup_{(\mathbf{k},\mu)\in\mathcal{P}}\|
\widetilde{g}_1(\mathbf{k},\mu,\cdot)\|_\infty<\infty\;.
\end{equation}

For $\widetilde{g}_2$ we have in view of (\ref{kernel})

\begin{eqnarray*}
\widetilde{g}_2(\mathbf{k},\mu,\mathbf{x})&=&\mu\int A(\mathbf{x}-\mathbf{x}')
\zeta^{j}(\mathbf{k},\mu,\mathbf{x}-\mathbf{x}')
\frac{1}{4\pi}e^{ikx'}
\\&&\hspace{2cm}\left(-i(E_{k}+\sum_{j=1}^{3}\alpha_{j}k\frac{x'_{j}}{x'}+\beta
m) -x'^{-1}\frac{k}{E_k}\right)
d^{3}x'
\end{eqnarray*}

With (\ref{ergebnis}) it follows that

\begin{eqnarray*}
\widetilde{g}_2&=&\mu\int A(\mathbf{x}-\mathbf{x}') \omega^{j}(\mathbf{k},\mu,\mathbf{x}-\mathbf{x}')
\frac{1}{4\pi}e^{ikx'}
\\&&\hspace{2cm}\left(-i(E_{k}+\sum_{j=1}^{3}\alpha_{j}k\frac{x'_{j}}{x'}+\beta
m) -x'^{-1}\frac{k}{E_k}\right)
d^{3}x'
\\&&+\mu\int A(\mathbf{x}-\mathbf{x}')
f(\mathbf{k},\mu,\mathbf{x})(\mathbf{x}-\mathbf{x}')
\frac{1}{4\pi}e^{ikx'}
\\&&\hspace{2cm}\left(-i(E_{k}+\sum_{j=1}^{3}\alpha_{j}k\frac{x'_{j}}{x'}+\beta
m) -x'^{-1}\frac{k}{E_k}\right)
d^{3}x'
\\&=:&\widetilde{g}_3+\widetilde{g}_4\;.
\end{eqnarray*}

Using Lemma \ref{properties} (e) for $n=0$ it follows, that

\begin{equation}\label{estp3}
\sup_{k\in\mathbb{R}^3;\mu\geq\lambda_c}\{\|
\widetilde{g}_3(\mathbf{k},\mu,\cdot)\|_\infty\}<\infty\;.
\end{equation}

For $\widetilde{g}_4$ we rewrite in view of (\ref{wirdnull})

\begin{eqnarray*}
\widetilde{g}_4&=&-i\mu\int A(\mathbf{x}-\mathbf{x}') f^{0}(\mathbf{k},\mu,\mathbf{x})(\mathbf{x}-\mathbf{x}')
\frac{1}{4\pi}e^{ikx'}\left(m+\beta m\right)
d^{3}x'
\\&&+\mu\int A(\mathbf{x}-\mathbf{x}')
f^{0}(\mathbf{k},\mu,\mathbf{x})(\mathbf{x}-\mathbf{x}')
\frac{1}{4\pi}e^{ikx'}
\\&&\hspace{2cm}\left(-i(E_{k}-m+\sum_{j=1}^{3}\alpha_{j}k\frac{x'_{j}}{x'})
-x'^{-1}\frac{k}{E_k}\right)
d^{3}x'
\\&=&\mu\int A(\mathbf{x}-\mathbf{x}')
f^{0}(\mathbf{k},\mu,\mathbf{x})(\mathbf{x}-\mathbf{x}')
\frac{1}{4\pi}e^{ikx'}
\\&&\hspace{2cm}\left(-i(E_{k}-m+\sum_{j=1}^{3}\alpha_{j}k\frac{x'_{j}}{x'})
-x'^{-1}\frac{k}{E_k}\right)
d^{3}x'
\end{eqnarray*}

Note that by relativistic dispersion relation $E_k-m$ is for small $k$ of order $k^2$. Furthermore $A$ is
compactly supported, so there exists a constant $C<\infty$ such that

\begin{eqnarray*}
\| \widetilde{g}_4(\mathbf{k},\mu,\cdot)\|_\infty &\leq&C k \|
f(\mathbf{k},\mu,\mathbf{x})\|_\infty
\end{eqnarray*}

using Lemma \ref{properties} (e) for $n=0$ it follows that

\begin{equation}\label{estp6}
\| \widetilde{g}_4\|_\infty \leq C
k\left(1+\left(\mid\mu-\lambda_c-\alpha
k^{2}\mid+k^{3}\right)^{-1}\right)\;.
\end{equation}

With (\ref{estp1}) and (\ref{estp3})  it follows that there exists a constant $C<\infty$ such that

\begin{equation}\label{estp}
\| \widetilde{g}(\mathbf{k},\mu,\mathbf{x})\|_\infty\leq C
\left(1+k\left(\mid\mu-\lambda_c-\alpha
k^{2}\mid+k^{3}\right)^{-1}\right)\;.
\end{equation}

With (\ref{gtilde}) we have that

\begin{eqnarray}\label{estoverlineg}
\| \overline{g}_\mathbf{k}\|_\infty\leq C
\left(1+k\left(\mid\mu-\lambda_c-\alpha
k^{2}\mid+k^{3}\right)^{-1}\right)
\nonumber\\&&\overline{g}_\mathbf{k}\in\mathcal{B}\;.
\end{eqnarray}

Similar as above we can show, that (\ref{ueberzetatilde}) has a unique solution in $\mathcal{B}$ and that

\begin{equation}\label{ueberzetatildegrossk2}
\lim_{k\rightarrow\infty}\|\partial_k\overline{\zeta}^{j}(\mathbf{k},\mu,\cdot)\|_\infty=0\;.
\end{equation}

Similar as above we are left with controlling

$$\lim_{(\mathbf{k},\mu)\rightarrow(\lambda_c,0)}\|\partial_k\overline{\zeta}^{j,0}_{\mu}\|_\infty\;.$$

We proceed as above, using (\ref{estoverlineg}). It follows that $\partial_k\zeta^{j}(\mathbf{k},\mu,\cdot)$ is
of order $k\left(\min\{\mid\mu-\lambda_c\mid^{-1},k^{-2}\}\right)^2$

Similarly we get (e) for $n>1$. The factors  $(x+1)^{-n}$ are needed to keep the functions
$\widetilde{g}(\mathbf{k},\mu,\mathbf{x})$ (see (\ref{dkzetatilde}) for the function
$\widetilde{g}(\mathbf{k},\mu,\mathbf{x})$ for $n=1$) and the $\partial_k^n\phi^{j}(\mathbf{k},0,\mathbf{x})$
 bounded.

\begin{flushright}$\Box$\end{flushright}

\vspace{0.5cm}
\noindent\textbf{Proof of Lemma \ref{equive1}}\\

We show that Lemma \ref{equive} implies Lemma \ref{equive1}.

Due to \cite{pickl} the $\zeta^{j}$ exist and are infinitely often continuously differentiable in $k$ for all
$(\mathbf{k},\mu)\in\mathcal{P}$. Hence the estimates on the generalized eigenfunctions (right hand side of
(\ref{lemgeprobe1}) and (\ref{lemgeprobe})) follow for any compact subset of
$\mathbb{R}^3\times[\lambda_c,\overline{\mu}]\backslash(0,\lambda_c)$.

We also verified (\ref{lemgeprobe1}) and (\ref{lemgeprobe}) for some subset of $\mathcal{P}$ ''around''
$(0,\lambda_c)$ (Lemma \ref{equive}), so it is left to verify the estimates for $k\rightarrow\infty$. Hence
Lemma \ref{equive1} follows from

\begin{eqnarray}\label{grossek}
\lim_{k\rightarrow\infty}\|(x+1)^{-n}\partial_{k}^{n}\zeta^{j}\|_\infty<C_n
\end{eqnarray}

for all $n\in\mathbb{N}_0$ with appropriate constants $C_n<\infty$. In the following we refer to \cite{stefan}
and show only how the proof of Theorem 2.4. in \cite{stefan} generalizes to our case.

We will show that $\phi^{j}$ can for sufficiently large $k$ be written as a Born series, i.e. that there exists
a $K<\infty$ such that

\begin{equation}\label{borngrosk}
\phi^{j}(\mathbf{k},\mu,\mathbf{x}):=\sum_{j=0}^\infty T_k^j \phi^{j}(\mathbf{k},0,\mathbf{x})
\end{equation}

exists for all $k>K$. To prove that the right hand side of (\ref{borngrosk}) exists we will first derive a
formula for $T_k^j$.

Using (\ref{deft}) we can write for any $\chi\in L^\infty$

\begin{eqnarray}\label{onceshow}
T_k^2\chi&=&
T_k\int
 G^{+}_{k}(\mathbf{x}-\mathbf{x}')A(\mathbf{x'})\chi(\mathbf{x'})d^3x'
\nonumber\\&=&\int
 G^{+}_{k}(\mathbf{x}-\mathbf{x}')A(\mathbf{x'})
\int
 G^{+}_{k}(\mathbf{x}'-\mathbf{x}'')A(\mathbf{x}'')\chi(\mathbf{x}'')d^3x''
 d^3x'
 \nonumber\\&=&\int\int
 G^{+}_{k}(\mathbf{x}-\mathbf{x}')A(\mathbf{x'})
 G^{+}_{k}(\mathbf{x}'-\mathbf{x}'') d^3x'A(\mathbf{x}'')\chi(\mathbf{x}'')d^3x''
\;.
\end{eqnarray}

Using (\ref{Green1}) we have that

\begin{eqnarray}\label{greenformel}
G^{+}_{k}(\mathbf{x})= (E_{k}+D^0)G^{KG}_{k}(\mathbf{x})
\end{eqnarray}

(remember that $D^0$ is the free dirac operator (\ref{dirac0}))
where $G^{KG}_{k}(\mathbf{x})$ is the Klein Gordon kernel which
solves

\begin{equation}\label{GreenKG}
(E_{k}^2-(D^0)^2)G_k^{KG,+}(\mathbf{x}-\mathbf{x'})=\delta(\mathbf{x}-\mathbf{x'})\;.
\end{equation}

(\ref{greenformel}) in (\ref{onceshow}) yields in view of (\ref{dirac0}) that

\begin{eqnarray*}
T_k^2\chi&=&
\int\int
 \left((E_{k}+D^0)G^{KG}_{k}(\mathbf{x}-\mathbf{x}')\right)A(\mathbf{x'})
\\&&\hspace{1cm}\left((E_{k}+D^0)G^{KG}_{k}(\mathbf{x}'-\mathbf{x}'')\right) d^3x'
A(\mathbf{x}'')\chi(\mathbf{x}'')d^3x''
\\&=&
\int\int
 \left((E_{k}-i\nabla\hspace{-0.3cm}/\hspace{0.03cm}+\beta m)G^{KG}_{k}(\mathbf{x}-\mathbf{x}')\right)A(\mathbf{x'})
\\&&\hspace{1cm} \left((E_{k}+D^0)G^{KG}_{k}(\mathbf{x}'-\mathbf{x}'')\right) d^3x'
A(\mathbf{x}'')\chi(\mathbf{x}'')d^3x''
\;.
\end{eqnarray*}

One partial integration yields

\begin{eqnarray*}
&&T_k^2\chi\\&&=
\int\int
 G^{KG}_{k}(\mathbf{x}-\mathbf{x}')
 \left((E_{k}+i\nabla\hspace{-0.3cm}/\hspace{0.03cm}+2\beta m)\left(A(\mathbf{x'})(E_{k}+D^0)G^{KG}_{k}(\mathbf{x}'-\mathbf{x}'')\right)\right) d^3x'
\\&&\hspace{2cm}A(\mathbf{x}'')\chi(\mathbf{x}'')d^3x''
\\&&=
\int\int
 G^{KG}_{k}(\mathbf{x}-\mathbf{x}')
 \left((E_{k}-D_0+2\beta m)\left(A(\mathbf{x'})(E_{k}+D^0)G^{KG}_{k}(\mathbf{x}'-\mathbf{x}'')\right)\right) d^3x'
\\&&\hspace{2cm}A(\mathbf{x}'')\chi(\mathbf{x}'')d^3x''
\\&&=
\int\int
 G^{KG}_{k}(\mathbf{x}-\mathbf{x}')
2\beta
mA(\mathbf{x'})\left((E_{k}+D^0)G^{KG}_{k}(\mathbf{x}'-\mathbf{x}'')\right)
d^3x'A(\mathbf{x}'')\chi(\mathbf{x}'')d^3x''
\\&&+
\int\int
 G^{KG}_{k}(\mathbf{x}-\mathbf{x}')
 \left((E_{k}-D_0)\left(A(\mathbf{x'})(E_{k}+D^0)G^{KG}_{k}(\mathbf{x}'-\mathbf{x}'')\right)\right) d^3x'
\\&&\hspace{2cm}A(\mathbf{x}'')\chi(\mathbf{x}'')d^3x''\;.
\end{eqnarray*}

With (\ref{GreenKG}) it follows that

\begin{eqnarray*}
&&T_k^2\chi\\&&=%
\int\int
 G^{KG}_{k}(\mathbf{x}-\mathbf{x}')
2\beta
mA(\mathbf{x'})(E_{k}+D^0)G^{KG}_{k}(\mathbf{x}'-\mathbf{x}'')
d^3x'A(\mathbf{x}'')\chi(\mathbf{x}'')d^3x''
\\&&+
\int\int
 G^{KG}_{k}(\mathbf{x}-\mathbf{x}')
 \left(\nabla\hspace{-0.3cm}/\hspace{0.03cm}A(\mathbf{x'})\right)
 (E_{k}+D^0)G^{KG}_{k}(\mathbf{x}'-\mathbf{x}'')d^3x'A(\mathbf{x}'')\chi(\mathbf{x}'')d^3x''
\\&&+
\int\int
 G^{KG}_{k}(\mathbf{x}-\mathbf{x}')
 \delta(\mathbf{x}'-\mathbf{x}'') d^3x'A(\mathbf{x}'')\chi(\mathbf{x}'')d^3x''
\\&&=%
\int\int
 G^{KG}_{k}(\mathbf{x}-\mathbf{x}')
2\beta mA(\mathbf{x'})G^{+}_{k}(\mathbf{x}'-\mathbf{x}'') A(\mathbf{x}'')\chi(\mathbf{x}'')d^3x'd^3x''
\\&&+
\int\int
 G^{KG}_{k}(\mathbf{x}-\mathbf{x}')
 \left(\nabla\hspace{-0.3cm}/\hspace{0.03cm}A(\mathbf{x'})\right)
 G^{+}_{k}(\mathbf{x}'-\mathbf{x}'')A(\mathbf{x}'')\chi(\mathbf{x}'')d^3x'd^3x''
\\&&+
\int\int
 G^{KG}_{k}(\mathbf{x}-\mathbf{x}')
A(\mathbf{x}')\chi(\mathbf{x}') d^3x'
\end{eqnarray*}

Defining the operator $T_k^{KG}(A):L^\infty\rightarrow L^\infty$ by

\begin{eqnarray}\label{deftkg}
T_k^{KG}(A)\chi=\int
 G^{KG}_{k}(\mathbf{x}-\mathbf{x}')A(\mathbf{x}')\chi(\mathbf{x}')d^3x'
\end{eqnarray}

we can write

\begin{equation}\label{tkformel}
T_k^2=T_k^{KG}(2\beta mA)T_k+T_k^{KG}(\nabla\hspace{-0.3cm}/\hspace{0.03cm}A)T_k
+T_k^{KG}(A)
\end{equation}

In a similar manner we write

\begin{eqnarray}\label{tkphi}
T_k\phi^{j}(\mathbf{k},0,\mathbf{x})&=&
\int
 (E_{k}+D^0)G^{KG}_{k}(\mathbf{x}-\mathbf{x}')A(\mathbf{x'})\phi^{j}(\mathbf{k},0,\mathbf{x}')d^3x'
 \nonumber\\&=&2\beta
m\int
G^{KG}_{k}(\mathbf{x}-\mathbf{x}')A(\mathbf{x'})\phi^{j}(\mathbf{k},0,\mathbf{x}')d^3x'
\nonumber\\&&+\int
G^{KG}_{k}(\mathbf{x}-\mathbf{x}')A(\mathbf{x'})(E_{k}-D^0)\phi^{j}(\mathbf{k},0,\mathbf{x}')d^3x'
\nonumber\\&=&2\beta m\int
G^{KG}_{k}(\mathbf{x}-\mathbf{x}')A(\mathbf{x'})\phi^{j}(\mathbf{k},0,\mathbf{x}')d^3x'
\\&=&T_k^{KG}(2\beta m A)\phi^{j}(\mathbf{k},0,\mathbf{x})
\end{eqnarray}

(\ref{tkformel}) and (\ref{tkphi}) yield, that all the $T_k^j\phi^{j}(\mathbf{k},0,\mathbf{x})$  can be written
as sums of powers of $T_k^{KG}(A)$, $T_k^{KG}(2m\beta A)$ and
$T_k^{KG}(\nabla\hspace{-0.3cm}/\hspace{0.03cm}A(\mathbf{x'}))$ acting on $\phi^{j}(\mathbf{k},0,\mathbf{x}')$.

Following the proof in \cite{teufel} (note, that the Klein Gordon
eigenvalue equation and the Schr\"odinger eigenvalue equation are
formally equivalent, with $E_k=\sqrt{m^2+k^2}$ in the Klein
Gordon, $E_k=\frac{k^2}{2m}$ in the Schr\"odinger case) we can
conclude that $\sup_{\mathbf{k}\in\mathbb{R}^3\backslash B_1}\|
T_k^{KG}\|_\infty^{op}<\infty$ and

\begin{equation*}
\lim_{k\rightarrow\infty}\| (T_k^{KG})^2\|_\infty^{op}=0\;.
\end{equation*}

Hence there exists a $K>0$ such that $\|
T_k^{KG}(A)\|_\infty^{op}<\frac{1}{2}$, $\| T_k^{KG}(2m\beta
A)\|_\infty^{op}<\frac{1}{2}$ and $\|
T_k^{KG}(\nabla\hspace{-0.3cm}/\hspace{0.03cm}A(\mathbf{x'}))\|_\infty^{op}<\frac{1}{2}$.
So the right hand side of (\ref{borngrosk}) is bounded by a
geometrical series. Furthermore it follows that

$$\|\phi^{j}(\mathbf{k},\mu,\mathbf{x})-\phi^{j}(\mathbf{k},0,\mathbf{x})\|_\infty
:=\|\sum_{l=1}^\infty T_k^l
\phi^{j}(\mathbf{k},0,\mathbf{x})\|_\infty$$

is bounded uniformly in $\mathbf{k}\in\mathbb{R}^3;k>K$ and (\ref{grossek}) follows for $n=0$.



\begin{flushright}$\Box$\end{flushright}

\newpage

\section{Derivative of $\phi_\mu$}

\vspace{0.5cm}

We will need some properties of the critical bound state $\phi_{\mu}$. We will only observe bound states which
dive into the continuous spectrum properly. Note that the switching factor satisfies (\ref{switch}), so
(\ref{diveproperly}) is satisfied if and only if there exists a $\mu_0<\lambda_c$ such that

\begin{equation}\label{diveproperly}
0<\partial_\mu E_\mu<C
\end{equation}

for all $\mu_0\leq \mu\leq \lambda_c$.

\begin{lem}\label{lemmacbder}

For every $\phi_{\lambda_c}\in \mathcal{N}$ (see (\ref{defmengen})) there exists a $\mu_0<\lambda_c$ and a
$C<\infty$ such that one can find a function $\phi_\mu$ for any $\mu\in[\mu_0,\lambda_c]$ with
$D_\mu\phi_\mu=E_\mu\phi_\mu$ such that

\begin{equation}\label{lemcbderb}\|\partial_\mu \phi_\mu\|\leq C
(\lambda_c-\mu)^{-\frac{25}{28}}
\end{equation}

for all $\mu\in[\mu_0,\lambda_c[$.

\end{lem}

\begin{rem}

This estimate is not optimal, but sufficient for what is needed later. It seems reasonable to conjecture that
the correct exponent is $-\frac{1}{2}$.

\end{rem}\vspace{1cm}

\vspace{0.5cm}
\noindent\textbf{Proof of Lemma \ref{lemmacbder}}\\

By assumption we have that only one eigenvalue dives into the upper continuous spectrum, hence there is a gap
between $m$ and the next smaller eigenvalue $E_{-1}$. For transparency of the proof we assume that
$D_{\lambda_c}$ has no further eigenvalues between $-m$ and $m$ (This assumption is merely convenient and can be
easily relaxed at the cost of more terms).

Let $\phi_{\lambda_c}\in\mathcal{N}$ with
$\|\phi_{\lambda_c}\|=1$. Using Gram Schmidt one can find a
orthonormal Basis
$B_\mathcal{N}:=\{\phi_{\lambda_c},\Phi^2,\Phi^3\ldots \Phi^n\}$
of $\mathcal{N}$. Let $\mathcal{N}_\mu$ be the set of
eigenfunctions of $D_\mu$ with energy eigenvalue $E_\mu$.

For any $\mu\leq\lambda_c$ we choose a normalized $\phi_\mu\in\mathcal{N}_\mu$ such that $\phi_\mu\bot\Phi^l$
for all $2\leq l\leq n$.

We first prove, that such a $\phi_\mu$ exists for any $\mathcal{N}_\mu$.

Let $P_\mathcal{N}$ be the projector onto $\mathcal{N}$, $\{\Phi^l_\mu;l=1,\ldots,n\}$ be a basis of
$\mathcal{N}_\mu$.

\vspace{0.5cm}
\noindent\textbf{1. Case:} The vectors $P_\mathcal{N}\Phi^l_\mu;l=1,\ldots,n$ are linearly independent\\

Choose a $\widetilde{\phi}_\mu\in\mathcal{N}_\mu$ such that $P_\mathcal{N}\widetilde{\phi}_\mu=\phi_{\lambda_c}$
Normalizing $\widetilde{\phi}_\mu$ yields $\phi_\mu$.

\vspace{0.5cm}
\noindent\textbf{2. Case:} The vectors $P_\mathcal{N}\Phi^l_\mu;l=1,\ldots,n$ are linearly dependent\\

Choose a nontrivial $\widetilde{\phi}_\mu\in\mathcal{N}_\mu$ such that $P_\mathcal{N}\widetilde{\phi}_\mu=0$.
Normalizing $\widetilde{\phi}_\mu$ yields $\phi_\mu$.

\vspace{0.5cm}

Now we show that $\phi_\mu$ satisfies the conditions of the Lemma. Therefore we first define

$$\zeta_\mu:=\frac{\phi_\mu}{\langle\phi_\mu,\phi_{\lambda_c}\rangle}\;,$$

where by definition of $\phi_\mu$

\begin{equation}\label{cf1}
\lim_{\mu\rightarrow\lambda_c}\langle\phi_\mu,\phi_{\lambda_c}\rangle=1\;,
\end{equation}
so that $\langle\phi_\mu,\phi_{\lambda_c}\rangle\neq0$ for $\mu_0$ close enough to $\lambda_c$, hence
$\zeta_\mu$ is well defined for $\mu\in[\mu_0,\lambda_c]$.

Thus

\begin{equation}\label{skpmitphi}(\varepsilon
D_{\mu}-E_\mu)\zeta_\mu=0\;,\;\;\;\;\langle\zeta_\mu,\phi_{\lambda_c}\rangle=1
\end{equation}

and using that $\varepsilon(D_\mu-D_\nu)=(\mu-\nu)A$ we get

\begin{eqnarray*}
0&=&\langle (\varepsilon D_{\mu}-E_\mu)\zeta_\mu,\phi_{\lambda_c}\rangle
\\&=&\langle (\varepsilon
D_{\lambda_c}-E_\mu)\zeta_\mu,\phi_{\lambda_c}\rangle+\langle (\mu-\lambda_c)A\zeta_\mu,\phi_{\lambda_c}\rangle
\\&=&\langle \zeta_\mu,(\varepsilon
D_{\lambda_c}-E_\mu)\phi_{\lambda_c}\rangle+\langle (\mu-\lambda_c)A\zeta_\mu,\phi_{\lambda_c}\rangle
\\&=&(
m-E_\mu)\langle \zeta_\mu,\phi_{\lambda_c}\rangle+\langle (\mu-\lambda_c)A\zeta_\mu,\phi_{\lambda_c}\rangle
\\&=&(
m-E_\mu)+\langle (\mu-\lambda_c)A\zeta_\mu,\phi_{\lambda_c}\rangle
\end{eqnarray*}

Hence

\begin{eqnarray*}
(m-E_\mu)\phi_{\lambda_c}&=&\langle (\lambda_c-\mu)A\zeta_\mu,\phi_{\lambda_c}\rangle\phi_{\lambda_c}
\\&=&(\lambda_c-\mu)A\zeta_\mu+\left(\langle
(\lambda_c-\mu)A\zeta_\mu,\phi_{\lambda_c}\rangle\phi_{\lambda_c}-(\lambda_c-\mu)A\zeta_\mu\right)
\\&=&\varepsilon\left(D_{\lambda_c}-D_\mu\right)\zeta_\mu+\left(\langle
(\lambda_c-\mu)A\zeta_\mu,\phi_{\lambda_c}\rangle\phi_{\lambda_c}-(\lambda_c-\mu)A\zeta_\mu\right)
\end{eqnarray*}

Using $m\phi_{\lambda_c}=D_{\lambda_c}\phi_{\lambda_c}$ and $D_\mu\zeta_\mu=E_\mu\zeta_\mu$ we get that

\begin{eqnarray*}
\left( D_{\lambda_c}-E_\mu\right)\phi_{\lambda_c}&=&\left( D_{\lambda_c}-E_\mu\right)\zeta_\mu+\left(\langle
(\lambda_c-\mu)A\zeta_\mu,\phi_{\lambda_c}\rangle\phi_{\lambda_c}-(\lambda_c-\mu)A\zeta_\mu\right)\;.
\end{eqnarray*}

Note that $\left( D_{\lambda_c}-E_\mu\right)^{-1}$ exists for all $\mu<\lambda_c$, hence

\begin{eqnarray*}
\phi_{\lambda_c}&=&\zeta_\mu-(\lambda_c-\mu)\left( D_{\lambda_c}-E_\mu\right)^{-1}\left(A\zeta_\mu-\langle
A\zeta_\mu,\phi_{\lambda_c}\rangle\phi_{\lambda_c}\right)\;.
\end{eqnarray*}

This leads us to define for any $\mu\in[\mu_0,\lambda_c[$ the linear operator $R_\mu:L^2\rightarrow L^2$ by

\begin{equation}\label{defrop}
R_\mu\chi:=\left(D_{\lambda_c}-E_\mu\right)^{-1}\left(A\chi-\langle
A\chi,\phi_{\lambda_c}\rangle\phi_{\lambda_c}\right)
\end{equation}

for all $\chi\in L^2$, so that

\begin{eqnarray*}
\phi_{\lambda_c}&=&\left(1-(\lambda_c-\mu)R_\mu\right)\zeta_\mu\;\;\;\text{or}
\\\zeta_\mu&=&\left(1-
(\lambda_c-\mu)R_\mu\right)^{-1}\phi_{\lambda_c}\;.
\end{eqnarray*}

Below we will show that there exists a $C<\infty$ such that

\begin{eqnarray}\label{rnorm}
\| R_\mu\|^{op}_2<C\left(\lambda_c-\mu\right)^{-\frac{25}{28}}
\end{eqnarray}

Hence taking $\mu_0$ close enough to $\lambda_c$ we have that for all $\mu\in[\mu_0,\lambda_c[$ there exist
$q<1$ so that

\begin{equation}\label{sternchen}
\|(\lambda_c-\mu)R_\mu\|_2^{op}\leq q<1\;,
\end{equation}

and hence

\begin{eqnarray}\label{defzetas}
L^2\ni\zeta_{\mu}&=&\sum_{j=0}^\infty (\mu-\lambda_c)^jR_\mu^j\phi_{\lambda_c}\;.
\end{eqnarray}

It is this series which we shall eventually differentiate with respect to $\mu$. First we prove (\ref{rnorm}).

Let $\chi\in L^2$ with $\|\chi\|=1$ and $\chi\bot\Phi^l$ for all
$2\leq l\leq n$. We set

\begin{equation}\label{defxi}
\xi:=A\chi-\langle A\chi,\phi_{\lambda_c}\rangle\phi_{\lambda_c}\;.
\end{equation}

Note that by construction $\xi\bot\phi_{\lambda_c}$ hence - since $\chi\bot\Phi^l$ and
$\phi_{\lambda_c}\bot\Phi^l$ for all $2\leq l\leq n$ - that $\xi\bot\mathcal{N}$

\begin{equation}\label{stern3}
 R_\mu\chi=\left(D_{\lambda_c}-\mu\right)^{-1}\xi\;.
\end{equation}

Let  $\mathcal{B}(r_\mu)$ be the ball around zero with Radius

\begin{equation}\label{cf2}
r_\mu=(\lambda_c-\mu)^{-\frac{3}{14}}\;.
\end{equation}

$r_\mu$ is defined such, that part of $\chi$ which lies outside $\mathcal{B}(r_\mu)$ is in $L^2$ sufficiently
small and the part of $\chi$ which lies inside $\mathcal{B}(r_\mu)$ is in $L^1$ sufficiently small. Furthermore
$\mathcal{S}_A\subset\mathcal{B}(r_\mu)$ for sufficiently large $r_\mu$, so the part of $\chi$ which lies
outside $\mathcal{B}(r_\mu)$ is a multiple of $\phi_{\lambda_c}$.Below we will have two different methods in our
estimates, using smallness in $L^2$ and smallness in $L^1$.

For large enough $r_\mu$ we have that the $P_{\mathcal{N}}(1_{\mathcal{B}(r_\mu)}\Phi^l)$ for $1\leq l \leq n$
are linear independent. Hence we can find a $\widetilde{\phi}_{\lambda_c}\in\mathcal{N}$ such that

$$P_{\mathcal{N}}(1_{\mathcal{B}(r_\mu)}\widetilde{\phi}_{\lambda_c})=P_{\mathcal{N}}(1_{\mathcal{B}(r_\mu)}\xi)\;.$$

Hence

\begin{eqnarray}\label{defxi1}
\xi_{1,\mu}&:=&1_{\mathcal{B}(r_\mu)}\xi-1_{\mathcal{B}(r_\mu)}\widetilde{\phi}_{\lambda_c}
\\\label{defxi2}
\xi_{2,\mu}&:=&\xi-\xi_{1,\mu}\;.
\end{eqnarray}

are orthogonal to $\mathcal{N}$.

$\xi_{1,\mu}$ has compact support
 $\mathcal{B}(r_\mu)$, so

\begin{equation}\label{l1normxi1}
\|\xi_{1,\mu}\|_1\leq \frac{4}{3}\pi r_\mu^3\|\xi_{1,\mu}\|\leq
Cr_\mu^3\;,
\end{equation}

for some appropriate $C<\infty$. Introducing (\ref{defxi1}) and (\ref{defxi2}) into (\ref{stern3})

\begin{eqnarray*}
R_\mu\chi=\left(D_{\lambda_c}-E_\mu\right)^{-1}\xi
&=&\left(D_{\lambda_c}-E_\mu\right)^{-1}\xi_{1,\mu}+\left(D_{\lambda_c}-E_\mu\right)^{-1}\xi_{2,\mu}
\end{eqnarray*}

we see that (\ref{rnorm}) holds if for some appropriate $K<\infty$

\begin{eqnarray}\label{xi1est}
\|\left(D_{\lambda_c}-E_\mu\right)^{-1}\xi_{1,\mu}\|<
K\left(\lambda_c-\mu\right)^{-\frac{25}{28}}
\\\label{xi2est0}\|\left(D_{\lambda_c}-E_\mu\right)^{-1}\xi_{2,\mu}\|<
K\left(\lambda_c-\mu\right)^{-\frac{25}{28}}\;.
\end{eqnarray}

We show (\ref{xi1est}). By Lemma \ref{properties}

\begin{eqnarray*}
\widehat{\xi}_{1,\mu}(\mathbf{k},j)&:=&\mathcal{F}_{\lambda_c}(\xi_{1,\mu})(\mathbf{k},j)=\int(2\pi)^{-\frac{3}{2}}[\phi^{j}(\mathbf{k},\lambda_c,\mathbf{x}),\xi_{1,\mu}(\mathbf{x})]
d^{3}x
\\&=&\int(2\pi)^{-\frac{3}{2}}[\phi^{j}(\mathbf{k},\lambda_c,\mathbf{x})-f^{0}(\mathbf{k},\mu,\mathbf{x}),\xi_{1,\mu}(\mathbf{x})]
d^{3}x
\\&&+\int(2\pi)^{-\frac{3}{2}}[f^{0}(\mathbf{k},\mu,\mathbf{x}),\xi_{1,\mu}(\mathbf{x})]
d^{3}x
\end{eqnarray*}

Since $\xi_{1,\mu}$ is orthogonal to $\phi_{\lambda_c}$ we have that

\begin{eqnarray*}
\mid\widehat{\xi}_{1,\mu}(\mathbf{k},j)\mid&:=&\mid\mathcal{F}_{\lambda_c}(\xi_{1,\mu})(\mathbf{k},j)\mid
\\&=&\left|\int(2\pi)^{-\frac{3}{2}}[\phi^{j}(\mathbf{k},\lambda_c,\mathbf{x})-f^{0}(\mathbf{k},\mu,\mathbf{x}),\xi_{1,\mu}(\mathbf{x})]
d^{3}x\right|
\\&\leq&\|\phi^{j}(\mathbf{k},\lambda_c,\cdot)-f^{0}(\mathbf{k},\mu,\cdot)\|_\infty\;\|\xi_{1,\mu}\|_1\;.
\end{eqnarray*}

With Lemma \ref{properties} (e) and (\ref{l1normxi1}) it follows that there exists a $C<\infty$ such that

\begin{eqnarray}\label{xihut}
\mid\widehat{\xi}_{1,\mu}(\mathbf{k},j)\mid&\leq&Cr_\mu^3
\end{eqnarray}

for all $\mathbf{k}\in\mathbb{R}^3$, $\mu\in[\mu_0,\lambda_c]$. Thus

\begin{eqnarray*}
\| \left(D_{\lambda_c}-E_\mu\right)^{-1}\xi_{1,\mu}\|&=&\|
\frac{1}{(-1)^jE_k-E_\mu}\widehat{\xi}_{1,\mu}\|
\\&=&\left(\sum_{j=1}^{2}\int \mid \frac{1}{(-1)^jE_k-E_\mu}\widehat{\xi}_{1,\mu}(\mathbf{k},j)\mid^2
 d^{3}k\right)^{\frac{1}{2}}
\\&\leq&\left(\sum_{j=1}^{2}\int_{k<1} \mid \frac{1}{E_k-E_\mu}\widehat{\xi}_{1,\mu}(\mathbf{k},j)\mid^2
 d^{3}k\right)^{\frac{1}{2}}
 \\&&+\left(\sum_{j=1}^{2}\int_{k>1} \mid \frac{1}{ E_k - E_\mu}\widehat{\xi}_{1,\mu}(\mathbf{k},j)\mid^2
 d^{3}k\right)^{\frac{1}{2}}\;.
\end{eqnarray*}

Since $ E_k=\sqrt{k^2+m^2}\geq m>E_\mu$ it follows that

\begin{equation}\label{plus1}
E_k- E_\mu \geq m-E_\mu>0
\end{equation}

and

\begin{equation*}
E_k -E_\mu \geq E_k-m\geq 0\;.
\end{equation*}

 Hence we have with (\ref{xihut})

\begin{eqnarray*}
\|
\left(D_{\lambda_c}-E_\mu\right)^{-1}\xi_{1,\mu}\|&\leq&Cr_\mu^3\left(\sum_{j=1}^{2}\int_{k<(m-E_\mu)^{\frac{1}{2}}}
\mid \frac{1}{m-E_\mu}\mid^2
 d^{3}k\right)^{\frac{1}{2}}
 \\&&+Cr_\mu^3\left(\sum_{j=1}^{2}\int_{1>k>(m-E_\mu)^{\frac{1}{2}}} \mid \frac{1}{E_k-m}\mid^2
 d^{3}k\right)^{\frac{1}{2}}\\
&&+ \left(\sum_{j=1}^{2}\int_{k>1} \mid \frac{1}{E_k-m}\widehat{\xi}_{1,\mu}(\mathbf{k},j)\mid^2
 d^{3}k\right)^{\frac{1}{2}}
\end{eqnarray*}

For $k>1$ we have that $\frac{1}{E_k-m}\leq\frac{1}{E_1-m}$. Thus

\begin{eqnarray*}\left(\sum_{j=1}^{2}\int_{k>1} \mid
\frac{1}{E_k-m}\widehat{\xi}_{1,\mu}(\mathbf{k},j)\mid^2
 d^{3}k\right)^{\frac{1}{2}}&\leq&\frac{1}{E_1-m}\|\xi_{1,\mu}\|
\leq \frac{C}{E_1-m}<\infty\;.
\end{eqnarray*}

It follows that

\begin{eqnarray*}
\|
\left(D_{\lambda_c}-E_\mu\right)^{-1}\xi_{1,\mu}\|&\leq&Cr_\mu^3\left(\sum_{j=1}^{2}\int_{k<(m-E_\mu)^{\frac{1}{2}}}
\mid \frac{1}{(m-E_\mu)}\mid^2
 d^{3}k\right)^{\frac{1}{2}}
 \\&&+Cr_\mu^3\left(\sum_{j=1}^{2}\int_{1>k>(m-E_\mu)^{\frac{1}{2}}} \mid \frac{1}{E_k-m}\mid^2
 d^{3}k\right)^{\frac{1}{2}}
+ C
\end{eqnarray*}

Since $\mid E_k-m\mid$ is for $k<1$ of order $k^2$  (non relativistic limit of the kinetic energy), there exists
a $C<\infty$ such that

$$\mid E_k-m\mid\geq\frac{1}{C}k^2\;\;\;\text{ for }1>k>(m-E_\mu)^{\frac{1}{2}}\;.$$

Hence

\begin{eqnarray}\label{xiestoben}
\|
\left(D_{\lambda_c}-E_\mu\right)^{-1}\xi_{1,\mu}\|&\leq&Cr_\mu^3(m-E_\mu)^{-1}
 \left(\frac{4\pi}{3}(m-E_\mu)^{\frac{1}{2}}\right)^{\frac{3}{2}}
 \nonumber\\&&+Cr_\mu^3\left(\sum_{j=1}^{2}\int_{k>(m-E_\mu)^{\frac{1}{2}}} k^{-4}
 d^3k\right)^{\frac{1}{2}}+C
 \\ \nonumber&=&Cr_\mu^3\left(\frac{4\pi}{3}\right)^{\frac{3}{2}}(m-E_\mu)^{-\frac{1}{4}}
 +Cr_\mu^3\left(4\pi (m-E_\mu)\right)^{-\frac{1}{4}}+C\;.
\end{eqnarray}

Hence there exists for $\mu_0$ close enough to $\lambda_c$ a $C<\infty$ such that

$$\| \left(D_{\lambda_c}-E_\mu\right)^{-1}\xi_{1,\mu}\|\leq Cr_\mu^3
\left(m-E_\mu\right)^{-\frac{1}{4}}$$

Note that (due to (\ref{diveproperly})) $m-E_\mu\leq C(\lambda_c-\mu)$. Using this and (\ref{cf2}) we obtain
(\ref{xi1est}) with some appropriate constant $K<\infty$.


Next we prove (\ref{xi2est0}). By (\ref{defxi1}) and (\ref{defxi2})

\begin{eqnarray*}
\|\xi_{2,\mu}\|=\|\xi-\xi_{1,\mu}\|&\leq&\|\xi-1_{\mathcal{B}(r_\mu)}\xi\|
+\|1_{\mathcal{B}(r_\mu)}\widetilde{\phi}_{\lambda_c}\|
\end{eqnarray*}

Using $\xi_{1,\mu}\bot\mathcal{N}$ it we have with  (\ref{defxi1})

$$\langle 1_{\mathcal{B}(r_\mu)}\xi,\widetilde{\phi}_{\lambda_c}\rangle=\langle1_{\mathcal{B}(r_\mu)}\widetilde{\phi}_{\lambda_c},\widetilde{\phi}_{\lambda_c}\rangle
=\|1_{\mathcal{B}(r_\mu)}\widetilde{\phi}_{\lambda_c}\|^2\approx
\|\widetilde{\phi}_{\lambda_c}\|^2\;.$$

and by Schwartz inequality

$$\mid\langle1_{\mathcal{B}(r_\mu)}\xi,\widetilde{\phi}_{\lambda_c}\rangle\mid=\mid\langle\xi-1_{\mathcal{B}(r_\mu)}\xi,\widetilde{\phi}_{\lambda_c}\rangle\mid\leq\|\xi-1_{\mathcal{B}(r_\mu)}\xi\|\;\;\|\widetilde{\phi}_{\lambda_c}\|\;.$$

hence for $R_0$ large enough there exists a $C<\infty$ uniform in $r_\mu>R_0$ such that
$$\|1_{\mathcal{B}(r_\mu)}\widetilde{\phi}_{\lambda_c}\|\leq C\|\xi-1_{\mathcal{B}(r_\mu)}\xi\|\;.$$

It follows that

\begin{eqnarray}
\|\xi_{2,\mu}\|
&\leq&\|\xi-1_{\mathcal{B}(r_\mu)}\xi\|+\|1_{\mathcal{B}(r_\mu)}\widetilde{\phi}_{\lambda_c}\|%
\nonumber\\\label{einplus}&=&2\|\xi-1_{\mathcal{B}(r_\mu)}\xi\|\;.
\end{eqnarray}

By (\ref{defxi}) and the fact that $A$ is compactely supported we have that for large enough $r_\mu$ $\xi$ is
outside the ball $\mathcal{B}(r_\mu)$ a multiple of $\phi_{\lambda_c}$. Hence $\xi-1_{\mathcal{B}(r_\mu)}\xi$ is
outside the ball $\mathcal{B}(r_\mu)$ a multiple of $\phi_{\lambda_c}$.

From (\ref{phicdecay}) $\mid\phi_{\lambda_c}\mid\leq Cx^{-2}$.
Hence $\|\xi-1_{\mathcal{B}(r_\mu)}\xi\|\leq C\left(\int_{x>r_\mu}
x^{-4}d^3x\right)^{\frac{1}{2}}$. So there exists a $C<\infty$
such that

$$\|\xi_{2,\mu}\|\leq2\|\xi-1_{\mathcal{B}(r_\mu)}\xi\|\leq Cr_\mu^{-\frac{1}{2}}\;.$$

It follows that by (\ref{plus1})

\begin{eqnarray}\label{xi2absch}
\| \left(D_{\lambda_c}-E_\mu\right)^{-1}\xi_{2,\mu}\|
&=&\|
\frac{1}{E_k-E_\mu}\widehat{\zeta}_{2,\mu}\|\leq\frac{1}{m-E_\mu}\|\widehat{\xi}_{2,\mu}\|
\nonumber\\&\leq&\mid m-E_\mu\mid^{-1}C r_\mu^{-\frac{1}{2}}
\end{eqnarray}

Hence (\ref{xi2est0}) follows as above. We have thus established (\ref{rnorm}) and we turn now to the
differentiation of $\phi_\mu$ respectively $\zeta_{\mu}$ (\ref{defzetas}).


Next we estimate $\|\partial_\mu R_\mu\|_2^{op}$. Using the
spectral decomposition the differentiation yields

\begin{eqnarray*}
\partial_\mu R_\mu \chi&=&\partial_\mu\left(D_{\lambda_c}-E_\mu\right)^{-1}\left(A\chi-\langle A\chi,\phi_{\lambda_c}\rangle\phi_{\lambda_c}\right)
\\&=&\left(\partial_\mu E_\mu\right)\left(D_{\lambda_c}-E_\mu\right)^{-2} \left(A\chi-\langle A\chi,\phi_{\lambda_c}\rangle\phi_{\lambda_c}\right)
\\&=&\left(\partial_\mu E_\mu\right)\left(D_{\lambda_c}-E_\mu\right)^{-1} R_\mu\chi\;.
\end{eqnarray*}

It follows using (\ref{diveproperly}) that there exists a $C<\infty$ such that

\begin{equation}\label{dsrnorm}
\|\partial_\mu R_\mu\|_2^{op}\leq C(m-E_\mu)^{-1}\| R_\mu\|_2^{op}
\end{equation}


Next we estimate $\partial_\mu\zeta_{\mu}$. Using (\ref{defzetas}) (Note that $\sum_{j=1}^\infty
j\left((\mu-\lambda_c)R_{\mu}\right)^{j-1}\phi_{\lambda_c}$ is majorized by a convergent power series uniformly
in $\mu_0\leq\mu\leq\lambda_c$ (see (\ref{sternchen}), hence we can exchange limit and differentiation)

\begin{eqnarray*}
\partial_\mu\zeta_{\mu}&=&\partial_\mu\sum_{j=0}^\infty \left((\mu-\lambda_c)R_{\mu}\right)^j\phi_{\lambda_c}
\\&=&(\partial_\mu (\mu-\lambda_c)R_{\mu})\sum_{j=1}^\infty j\left((\mu-\lambda_c)R_{\mu}\right)^{j-1}\phi_{\lambda_c}
\\&=&R_{\mu}\sum_{j=1}^\infty
j\left((\mu-\lambda_c)R_{\mu}\right)^{j-1}\phi_{\lambda_c}
\\&&+(\mu-\lambda_c)(\partial_\mu R_{\mu})\sum_{j=1}^\infty
j\left((\mu-\lambda_c)R_{\mu}\right)^{j-1}\phi_{\lambda_c}
\end{eqnarray*}

By (\ref{sternchen}) $\|\sum_{j=1}^\infty
j(\mu-\lambda_c)^{j-1}R_{\mu}^j\phi_{\lambda_c}\|\leq C<\infty$
uniform in $\mu_0\leq \mu< \lambda_c$. Thus by (\ref{rnorm}) and
(\ref{dsrnorm})

\begin{eqnarray}\label{dszeta}
\|\partial_\mu\zeta_{\mu}\|&\leq&C\left(\|
R_{\mu}\|_2^{op}+\|(\mu-\lambda_c)\partial_\mu
R_{\mu}\|_2^{op}\right)\leq C \| R_{\mu} \|_2^{op}
\end{eqnarray}

 with appropriate
$C<\infty$.

Finally, to prove (\ref{lemcbderb}), we observe that

$$\phi_\mu=\frac{\zeta_{\mu}}{\| \zeta_{\mu} \|}\;,$$

and thus

$$\partial_\mu\phi_\mu=\frac{\partial_\mu\zeta_{\mu}}{\| \zeta_{\mu}
\|}-\frac{\zeta_{\mu}}{\| \zeta_{\mu} \|^2}\partial_\mu \|
\zeta_{\mu} \|\;.$$

Due to (\ref{skpmitphi}) we have that $\| \zeta_{\mu}\|\geq 1$ for
all $\mu\in[\mu_0,\lambda_c[$. Furthermore by triangle inequality
$\partial_\mu\|\zeta_{\mu}\|\leq\|\partial_\mu\zeta_{\mu}\|$,
therefore

$$\| \partial_\mu\phi_\mu\|\leq2\|
\partial_\mu\zeta_{\mu}\|\;.$$

This, (\ref{rnorm}) and (\ref{dszeta}) yield (\ref{lemcbderb}).

\begin{flushright}$\Box$\end{flushright}


\vspace{0.5cm}

\section{Proof of Theorem \ref{onecrit}}\label{proofmainlemma}

\vspace{0.5cm}

In the following we will set $s_{m1}=0$. Since we shall employ often eigenfunction expansions we need the
following properties of the generalized eigenfunctions. We provide the major results of Lemma \ref{properties}
and Lemma \ref{lemmacbder} in our notation, i.e. $\mu=\lambda\varphi(s)$ with the restriction
$0<\partial_s\varphi(s)<\infty$ (see (\ref{switch})).

We will slightly abuse notation, writing $\phi_s$ for $\phi_{\lambda\varphi(s)}$.

\begin{cor}\label{geprob}

Let $A$ be compactly supported, $A>0$, $\widetilde{s}>0$ be such, that $\lambda_c$ is the only critical coupling
constant in $[\lambda_c,\lambda\varphi(\widetilde{s})]$ and $\partial_s \varphi(s)>0$ in
$[\lambda_c,\lambda\varphi(\widetilde{s})]$. Let $\mathcal{B}$ be the Banach space of all continuous functions
tending uniformly to zero as $x\rightarrow\infty$ equipped with the supremum norm. Then

\begin{description}
\item[(a)] there exist unique solutions $\phi^{j}(\mathbf{k},s,\mathbf{x})$ of (\ref{dgmp2})
 in $\mathcal{B}$  for
all $k\in \mathbb{R}^{3}$, $s\in(0,\widetilde{s}]$ such that

\item[(b)]

for any $s\in(0,\widetilde{s}]$ the set of $\phi^{j}(\mathbf{k},s,\mathbf{x})$
    define a generalized Fourier transform in the space of scattering
    states by

\begin{equation}\label{her}
\mathcal{F}_{s}(\psi)(\mathbf{k},j):=\int(2\pi)^{-\frac{3}{2}}\langle\phi^{j}(\mathbf{k},s,\mathbf{x}),\psi(\mathbf{x})\rangle
d^{3}x
\end{equation}

and

\begin{equation}\label{hin}
    \psi(\mathbf{x})=\sum_{j=1}^{4}\int(2\pi)^{-\frac{3}{2}}
  \phi^{j}(\mathbf{k},s,\mathbf{x})\mathcal{F}_{s}(\psi)(\mathbf{k},j)d^{3}k\;.
 \end{equation}

The so defined $\mathcal{F}_{s}(\psi)$ is isometric to $\psi$, i.e.

\begin{eqnarray*}
\|\psi\|=
\left(\int
\mid\psi(\mathbf{x})\mid^2d^3x\right)^{\frac{1}{2}}=\left(\sum_{j=1}^4\int
\mid\mathcal{F}_{s}(\psi)(\mathbf{k},j)\mid^2d^3k\right)^{\frac{1}{2}}
=\|\mathcal{F}_{s}(\psi)\|
\end{eqnarray*}

\item[(c)] the functions $\phi^{j}(\mathbf{k},s,\mathbf{x})$ are infinitely often continuously differentiable
with respect to $k$, furthermore there exist $\alpha,C\in\mathbb{R}^+$ uniform in
$(\mathbf{k},s)\in\mathbb{R}^3\times[0,s_{m2}]$ and for all $n\in\mathbb{N}_0$ functions
$f^{n}(\mathbf{k},s,\mathbf{x})\in\mathcal{N}$ (see (\ref{defmengen})) with

\begin{eqnarray}\label{geprobe1}
\| f^{n}(\mathbf{k},s,\cdot)\|_\infty<C\left(1+k^n\left(\mid
s-\alpha k^{2}\mid+k^{3}\right)^{-n-1}\right)
\end{eqnarray}

such that

\begin{equation}\label{defzetaaa}
\zeta^{j,n}(\mathbf{k},s,\mathbf{x}):=\partial_{k}^{n}\phi^{j}(\mathbf{k},s,\mathbf{x})
-f^{n}(\mathbf{k},s,\mathbf{x})\phi_0(\mathbf{x})
\end{equation}

satisfies

\begin{eqnarray}\label{geprobe}
\|(x+1)^{-n}\zeta^{j,n}(\mathbf{k},s,\mathbf{x})\|_{\infty}<C\left(1+(s+
k^2)\| f^{n}(\mathbf{k},s,\cdot)\|_\infty\right)\;.
\end{eqnarray}

in particular for $n=0$

\begin{eqnarray}\label{sternla}
\|\phi^{j}(\mathbf{k},s,\mathbf{x})\|_{\infty}<C\left(1+\left(\mid
s-\alpha k^{2}\mid+k^{3}\right)^{-1}\right)\;.
\end{eqnarray}

\item[(d)] Let $\phi_s$ be defined as above. Then there exists $s_{in}<0$ and $C<\infty$ such that

$$\|\partial_s \phi_s\|\leq C s^{-\frac{25}{28}}$$

for all $s_{in}\leq s\leq 0$.

  \end{description}

\end{cor}\vspace{1cm}

To prove Theorem \ref{onecrit} we have to control the time propagation of $\phi_{s_{in}}$. This propagation is
qualitatively different for $s<0$ and $s>0$. Hence we control the propagation for $s<0$ and $s>0$ separately.

\vspace{0.5cm}

\subsection{Control of $\psi_{s}^\varepsilon$ for $s_{in}\leq s\leq 0$}

\vspace{0.5cm}

The adiabatic theorem yields that for any $s<0$ the wave function will stay a multiple of the respective bound
state as $\varepsilon$ goes to zero.

We shall extend the assertion to $s=0$.

\begin{lem}\label{Lemmakleins}
Let $\psi_{s,s_{in}}^{\varepsilon}$ with $s\in[s_{in},0[$ be solution of the Dirac equation with
$\psi_{s_{in},s_{in}}^{\varepsilon}=\phi_{s_{in}}$.

Then uniform in $\varepsilon>0$

$$\lim_{s_{in}\rightarrow0}\mid\langle\psi_{0,s_{in}}^{\varepsilon},\phi_{0}\rangle\mid=1\;.$$

\end{lem}

\vspace{0.5cm}
\noindent\textbf{Proof}\\

We introduce

\begin{equation}\label{defpsi1k}
\psi_{s,s_{in}}^{\varepsilon,1}:= \exp\left(-\frac{i}{\varepsilon}\int_{s_{in}}^sE_vdv\right)\phi_{s} \;.
\end{equation}

Note, that $\psi_{s_{in},s{in}}^{\varepsilon,1}=\phi_{s_{in}}$ and
$\mid\langle\psi_{0,s_{in}}^{\varepsilon,1},\phi_0\rangle\mid=1$. Thus to prove the Lemma we need only show that
$\psi_{0,s_{in}}^\varepsilon$ will be equal to $\psi_{0,s_{in}}^{\varepsilon,1}$ in the limit
$\lim_{s_{in}\rightarrow0}$ uniform in $\varepsilon$, i.e.

\begin{equation}\label{kleins}
\lim_{s_{in}\rightarrow0}\sup_{0<\varepsilon\leq1}\|
\psi_{0,s_{in}}^\varepsilon-\psi_{0,s_{in}}^{\varepsilon,1}
\|=0\;.
\end{equation}

Let now $U^\varepsilon(s,u)$ be the propagator of the Dirac equation, i.e. $i\partial_s
U^{\varepsilon}(s,0)=\frac{1}{\varepsilon}D_s U^\varepsilon(s,0)$. Using that $\psi^\varepsilon_{s,s_{in}}$ is
solution of the Dirac equation it follows with (\ref{defpsi1k}) that

\begin{eqnarray*}
\psi_{s,s_{in}}^{\varepsilon,1}-\psi^\varepsilon_{s,s_{in}}&=&
\int_{s_{in}}^s\partial_u\left(U^\varepsilon(s,u)\psi_{u,s_{in}}^{\varepsilon,1}\right)du
\\&=&-i
\int_{s_{in}}^sU^\varepsilon(s,u)\left(\frac{D_{u}}{\varepsilon}-i\partial_{u}\right)\exp\left(-\frac{i}{\varepsilon}\int_{s_{in}}^uE_vdv\right)\phi_{u}du
\\&=&-\frac{i}{\varepsilon}
\int_{s_{in}}^sU^\varepsilon(s,u)\left(D_{u}-E_u\right)\exp\left(-\frac{i}{\varepsilon}\int_{s_{in}}^uE_vdv\right)\phi_{u}du
\\&&+i
\int_{s_{in}}^sU^\varepsilon(s,u)\exp\left(-\frac{i}{\varepsilon}\int_{s_{in}}^uE_vdv\right)\partial_u\phi_{u}du
\end{eqnarray*}

Since $(D_u-E_u)\phi_u=0$

\begin{eqnarray*}
\psi_{s,s_{in}}^{\varepsilon,1}-\psi_{s,s_{in}}^\varepsilon
&=&i
\int_{s_{in}}^sU^\varepsilon(s,u)\exp\left(-\frac{i}{\varepsilon}\int_{s_{in}}^uE_vdv\right)\partial_u\phi_{u}du\;.
\end{eqnarray*}

Hence by unitarity of $U^\varepsilon$ and by Corollary \ref{geprob} (d)

\begin{eqnarray*}
\|\psi_{0,s_{in}}^\varepsilon-\psi_{0,s_{in}}^{\varepsilon,1}\|&\leq&
\int_{s_{in}}^0\|\partial_u\phi_{u}\| du
\\&\leq&
C\int_{s_{in}}^0u^{-\frac{25}{28}} du=- \frac{25}{28}C [u^{\frac{3}{28}}]_{s_{in}}^0=
\frac{25}{28}Cs_{in}^{\frac{3}{28}}
\end{eqnarray*}

and (\ref{kleins}) follows.

\begin{flushright}$\Box$\end{flushright}

\begin{cor}\label{corvor0} (Adiabatic Theorem without a gap)

Let $s_i<s_{m1}$ and $s_f>s_{m2}$ be such that $\phi_{s_i}$ and $\phi_{s_f}$ already / still exist. Let
$U^\varepsilon(s,u)$ be the time evolution operator of the Dirac equation (\ref{dirac}) on the adiabatic time
scale. Let $\phi_{s}$ be an overcritical bound state of the Dirac operator with potential $A_{s}(\mathbf{x})$ of
the form (\ref{potential}). Let $\phi_{s}$ dive properly into the positive continuous spectrum (see
(\ref{diveproperlys})).

\begin{eqnarray}\label{corvoreq1}
\lim_{\varepsilon\rightarrow0}\mid\langle U^\varepsilon(0,s_i)\phi_{s_i},\phi_{0}\rangle\mid&=&1
\\\label{corvoreq2}\lim_{\varepsilon\rightarrow0}\mid\langle
U^\varepsilon(s_f,s_{m2})\phi_{0},\phi_{s_f}\rangle\mid&=&1\;.
\end{eqnarray}

\end{cor}

\vspace{0.5cm}
\noindent\textbf{Proof}\\

We only prove (\ref{corvoreq1}). (\ref{corvoreq2}) follows equivalently following the propagation of
$\phi_{s_f}$ backwards in time.

We will show that for any $\delta>0$ there exists a $\varepsilon_0>0$ and phase factors $\pi^\varepsilon_0$ such
that

\begin{equation}\label{corvoreqeig}\| U^\varepsilon(0,s_i)\phi_{s_i}-\pi^\varepsilon_0\phi_{0}\|<\delta
\end{equation}
for all $\varepsilon<\varepsilon_0$.

Using Lemma \ref{Lemmakleins} we choose $s_{in}>0$ such that

$$\|\psi_{0,s_{in}}^{\varepsilon}-\pi^\varepsilon_1\phi_{0}\|<\frac{1}{2}\delta$$

for all $\varepsilon<\varepsilon_0$ with an appropriate phase factor $\pi^\varepsilon_1$.

Then using the adiabatic Theorem (see \cite{teufel}) we choose $\varepsilon>0$ such that

$$\| U^\varepsilon(s_{in},s_i)\phi_{s_i}-\pi^\varepsilon_2\phi_{s_{in}}\|<\frac{1}{2}\delta$$

for all $\varepsilon<\varepsilon_0$ with an appropriate phase factor $\pi^\varepsilon_2$.

Using the triangle inequality we get that

\begin{eqnarray*}
&&\|
U^\varepsilon(0,s_i)\phi_{s_i}-\pi^\varepsilon_1\pi^\varepsilon_2\phi_{0}\|
\\&&\hspace{2cm}\leq\;\;\| \pi^\varepsilon_2
U^\varepsilon(0,s_{in})\phi_{s_i}-\pi^\varepsilon_1\pi^\varepsilon_2\phi_{0}\|
+
\|
U^\varepsilon(s_{in},s_i)\phi_{s_i}-\pi^\varepsilon_2\phi_{s_{in}}\|
\\&&\hspace{2cm}=\;\;\|\psi_{0,s_{in}}^{\varepsilon}-\pi^\varepsilon_1\phi_{0}\|+
\| U^\varepsilon(0,s_i)\phi_{s_i}-\pi^\varepsilon_2\phi_{s_{in}}\|
\\&&\hspace{2cm}<\;\;\frac{\delta}{2}+\frac{\delta}{2}=\delta
\end{eqnarray*}

and (\ref{corvoreqeig}) follows.

\begin{flushright}$\Box$\end{flushright}

\vspace{0.5cm}

\subsection{Control of $\psi_{s}^\varepsilon$ for $ s> 0$}

\vspace{0.5cm}

Due to Corollary \ref{corvor0} $\psi_{0}^\varepsilon$ is - in the given limits and up to a phase factor - equal
to the bound state $\phi_{0}$. So Theorem \ref{onecrit} is a direct result of Corollary \ref{corvor0} and

\begin{lem}\label{lemmagrosss}
Let $U^\varepsilon(s,u)$ be the time evolution operator of the Dirac equation (\ref{dirac}) on the adiabatic
time scale. Let $\phi_{s}$ and $\widetilde{\phi}_{s}$ be overcritical bound states of the Dirac operator with
potential $A_{s}(\mathbf{x})$ of the form (\ref{potential}) that dive properly into the positive continuous
spectrum (see (\ref{diveproperlys})). Then (remember that $D_0=D_{s_{m2}}$, hence
$\widetilde{\phi}_{s_{m2}}=\widetilde{\phi}_0$)

\begin{equation}\label{onecriteqb}
\lim_{\varepsilon\rightarrow0}\langle U^\varepsilon(s_{m2},0)\phi_{0},\widetilde{\phi}_{0}\rangle=0\;.
\end{equation}

\end{lem}

\vspace{0.5cm}
\noindent\textbf{Proof}\\

Set

\begin{equation}\label{psiii}
\psi^{\varepsilon}_s:=U^\varepsilon(s,0)\phi_{0}
\end{equation}

To begin with we introduce the time $s_\varepsilon$ which shall be specified later and which should be thought
of as a time much larger than $\varepsilon$ and smaller than $1$ (for example $\varepsilon^{\frac{1}{2000}}$.
$s_\varepsilon$ will serve for defining an appropriate approximating dynamics: Let
$V_{s_\varepsilon}^\varepsilon$ be the time evolution operator of the time independent Dirac operator
$\frac{1}{\varepsilon}D_{s_\varepsilon}$. This evolution is controllable since we have by Corollary \ref{geprob}
good control of the generalized eigenfunctions and

\begin{equation}\label{propagsa}
V_{s_\varepsilon}^\varepsilon(s,u)\phi^{j}(\mathbf{k},s_\varepsilon,\mathbf{x})=\exp\left(-\frac{i}{\varepsilon}E_k(s-u)\right)\phi^{j}(\mathbf{k},s_\varepsilon,\mathbf{x})
\end{equation}

In the following we will always use the notation

\begin{equation}\label{notation}
\widehat{\psi}:=\mathcal{F}_{s_\varepsilon}(\psi)
\end{equation}

 for the
generalized Fourier transform of $\psi$ in the $\phi^{j}(\mathbf{k},s_\varepsilon,\mathbf{x})$ eigenbasis.

 \hspace{1cm}\\

We first give some formulas for different propagators, we will need below.

Let $\overline{U}(s,s_{in})$ and $\widetilde{U}(s,s_{in})$ be time propagators, $\overline{D}_s$ and
$\widetilde{D}_s$ the respective - in general time dependent - generators, i.e.

$$\partial_s \overline{U}(s,s_{in})=\overline{D}_sU(s,s_{in})\;\;\;\partial_s \widetilde{U}(s,s_{in})=\widetilde{D}_s\widetilde{U}(s,s_{in})\;.$$

Then

\begin{eqnarray}\label{cook0}
\overline{U}(s,s_{in})-\widetilde{U}(s,s_{in})&=&-\int_{s_{in}}^s
\partial_u\left(\overline{U}^\varepsilon(s,u)\widetilde{U}(u,s_{in})\right)du\nonumber
\\&=&-i\int_{s_{in}}^s\overline{U}^\varepsilon(s,u)\left(\overline{D}_{u}-i\partial_u\right)\widetilde{U}(u,s_{in})du
\\\label{cook}&=&-i\int_{s_{in}}^s\overline{U}^\varepsilon(s,u)\left(\overline{D}_{u}-\widetilde{D}_{u}\right)\widetilde{U}(u,s_{in})du\;.
\end{eqnarray}

We shall use the following identity for the time  $U^\varepsilon$, which follows directly from (\ref{cook})
setting $\widetilde{U}=U^\varepsilon$ and $\overline{U}=V^\varepsilon_{s_\varepsilon}$.

\begin{equation}\label{replprop1}
U^\varepsilon(s,0)=V_{s_\varepsilon}^\varepsilon(s,0)+i\int_{0}^s
V_{s_\varepsilon}^\varepsilon(s,w)\frac{1}{\varepsilon}(D_w-D_{s_\varepsilon})U^\varepsilon(w,0)dw\;.
\end{equation}



We shall now approximate $\psi^{\varepsilon}_s$ in three steps by
a wave function $\psi^{\varepsilon,3}_s$ which is easier to
control and such that the difference
$\|\psi^{\varepsilon,3}_s-\psi^{\varepsilon}_s\|\rightarrow0$ as
$\varepsilon\rightarrow0$.

\subsubsection*{1. Step:}

We replace $\phi_0$ by $\phi^{\varepsilon}$ given by

\begin{eqnarray}\label{replsce1}
\phi^{\varepsilon}(\mathbf{x}):=
    \phi_{0}(\mathbf{x})(1-\rho_{\varepsilon^{-\frac{1}{10000}}}(\mathbf{x}))
\end{eqnarray}

where $\rho_\kappa\in C^\infty$ is a mollifier given by

\begin{equation}\label{rhodef2}
\rho(\mathbf{x}):=\left\{%
\begin{array}{ll}
    0, & \hbox{for $x\leq1$;} \\
    1, & \hbox{for $x\geq2$.} \\
\end{array}%
\right.
\end{equation}

and

\begin{equation}\label{defrho2}
\rho_{\kappa}(\mathbf{x}):=\rho(\frac{\mathbf{x}}{\kappa})
\end{equation}

for $\kappa>0$. Hence $\phi^\varepsilon(\mathbf{x})=0$ for $\mathbf{x}\geq 2\varepsilon^{-\frac{1}{1000}}$ and
$\phi^\varepsilon(\mathbf{x})=\phi_0(\mathbf{x})$ for $\mathbf{x}\leq \varepsilon^{-\frac{1}{1000}}$. So
$\phi^{\varepsilon}$ has compact support $\mathcal{T}_\varepsilon$ with

\begin{equation}\label{tmid}
\mid\mathcal{T}_\varepsilon\mid\leq\frac{4}{3}\pi 8\varepsilon^{-\frac{3}{10000}}
\end{equation}

and that

\begin{equation}\label{parallelphieps}
\|\phi^{\varepsilon}\| \leq1\;.
\end{equation}

We set

\begin{eqnarray}\label{psi1def}
\psi^{\varepsilon,1}_s:=U^\varepsilon(s,0)\phi^{\varepsilon}
\end{eqnarray}

and for the error

\begin{eqnarray}\label{defeta1}
\eta^{\varepsilon,1}_s&:=&\psi^{\varepsilon}_s-\psi^{\varepsilon,1}_s\;.
\end{eqnarray}

\subsubsection*{2. Step:}

Observing (\ref{replprop1}) we obtain for (\ref{psi1def})

\begin{eqnarray}\label{psi1tats}
\psi^{\varepsilon,1}_s=V_{s_\varepsilon}^\varepsilon(s,0)\phi^{\varepsilon}+i\int_{0}^{s}
V_{s_\varepsilon}^\varepsilon(s,w)\frac{1}{\varepsilon}(D_w-D_{s_\varepsilon})U^\varepsilon(w,0)\phi^{\varepsilon}dw
\end{eqnarray}

and setting

\begin{eqnarray}\label{defzeta}
%
\zeta^\varepsilon_w(\mathbf{x})&:=&\frac{1}{\varepsilon}(D_w-D_{s_\varepsilon})U^\varepsilon(w,0)\phi^{\varepsilon}=\frac{1}{\varepsilon}(D_w-D_{s_\varepsilon})
\psi^{\varepsilon,1}_w\;,
\end{eqnarray}

(\ref{psi1tats}) becomes

\begin{eqnarray}\label{defpsi1}
\psi^{\varepsilon,1}_s&:=&V_{s_\varepsilon}^\varepsilon(s,0)\phi^{\varepsilon}+i\int_{0}^{s}
V_{s_\varepsilon}^\varepsilon(s,w)\zeta^\varepsilon_w(\mathbf{x})dw
\end{eqnarray}

We write

\begin{eqnarray*}
\zeta^\varepsilon_w(\mathbf{x})&=&\sum_{j=1}^{4}\int(2\pi)^{-\frac{3}{2}}
  \phi^{j}(\mathbf{k},s_\varepsilon,\mathbf{x})\widehat{\zeta}^\varepsilon_w(\mathbf{k},j)d^{3}k
\end{eqnarray*}

and replace for $0\leq w\leq s_\varepsilon$ $\widehat{\zeta}^\varepsilon_w(\mathbf{k},j)$ - using some
\begin{equation}\label{formelnummer}\varepsilon\ll k_\varepsilon\ll1\ll K_\varepsilon\end{equation} which will
be specified later on - by

\begin{equation}\label{defzeta2}
\widehat{\zeta}^{\varepsilon,1}_{w}(\mathbf{k},j):=
\widehat{\zeta}^\varepsilon_w(\mathbf{k},j)\rho_{k_\varepsilon}(1-\rho_{K_\varepsilon})
\end{equation}

Furthermore we write (\ref{replsce1}) as

\begin{eqnarray*}
\phi^{\varepsilon}(\mathbf{x})&=&\sum_{j=1}^{4}\int(2\pi)^{-\frac{3}{2}}
  \phi^{j}(\mathbf{k},s_\varepsilon,\mathbf{x})\widehat{\phi}^{\varepsilon}(\mathbf{k},j)d^{3}k
\end{eqnarray*}

and replace $\widehat{\phi}^{\varepsilon}$  by

\begin{equation}\label{defphiers}
\widehat{\phi}^{\varepsilon,1}(\mathbf{k},j):=
\widehat{\phi}^{\varepsilon}(\mathbf{k},j)\rho_{\kappa}(\mathbf{k})(1-\rho_{K_\varepsilon})\;.
\end{equation}

These replacements define a new wave function, namely

\begin{eqnarray}\label{psi2def}
\psi^{\varepsilon,2}_s&:=&V_{s_\varepsilon}^\varepsilon(s,0) \phi^{\varepsilon,1} +i\int_{0}^{s}
V_{s_\varepsilon}^\varepsilon(s,w)\zeta^{\varepsilon,1}_{w}(\mathbf{x})dw
\end{eqnarray}

for $s\leq s_\varepsilon$ and

\begin{eqnarray}\label{defpsi2b}
\psi^{\varepsilon,2}_s&=&U(s,s_\varepsilon)\psi^{\varepsilon,2}_{s_\varepsilon}
\end{eqnarray}

for $s>s_\varepsilon$.

We shall need the difference

$$\eta^{\varepsilon,2}_s:=\psi^{\varepsilon,1}_s-\psi^{\varepsilon,2}_s$$

only at time $s=s_{m2}$ (see Lemma \ref{lemmarest} below) and we note here already that (see (\ref{psi1def}) and
(\ref{defpsi2b}))

\begin{eqnarray*} \eta^{\varepsilon,2}_{s_{m2}}&=&\psi^{\varepsilon,1}_{s_{m2}}-\psi^{\varepsilon,2}_{s_{m2}}
\nonumber\\&=&U^\varepsilon(s_{m2},s_\varepsilon)\left(\psi^{\varepsilon,1}_{s_\varepsilon}-\psi^{\varepsilon,2}_{s_\varepsilon}\right)\;.
\end{eqnarray*}

With (\ref{psi1tats}) and (\ref{psi2def}) we get that

\begin{eqnarray}\label{defeta2}
\eta^{\varepsilon,2}_{s_{m2}}
&=&U^\varepsilon(s_{m2},s_\varepsilon)V_{s_\varepsilon}^\varepsilon(s_\varepsilon,0)\left(
\phi^{\varepsilon}-\phi^{\varepsilon,1}\right)
\\\nonumber&&+iU^\varepsilon(s_{m2},s_\varepsilon)\int_{0}^{s_\varepsilon}
V_{s_\varepsilon}^\varepsilon(s_\varepsilon,w)\left(\zeta^\varepsilon_w(\mathbf{x})-\zeta^{\varepsilon,1}_{w}(\mathbf{x})\right)dw\;.
\end{eqnarray}

\subsubsection*{3. Step:}

In this last step we more or less assert that the wave function evolution after time $s_\varepsilon$ is close to
the auxiliary time evolution $V_{s_\varepsilon}^\varepsilon$, namely we replace $\psi^{\varepsilon,2}_s$ by

\begin{equation}\label{psi3def}
\psi^{\varepsilon,3}_s:=
  \left\{%
\begin{array}{ll}
    \psi^{\varepsilon,2}_{s}, & \hbox{for $s\leq s_\varepsilon$;} \\
    V_{s_\varepsilon}^\varepsilon(s,s_\varepsilon)\psi^{\varepsilon,2}_{s_\varepsilon}, & \hbox{else.} \\
\end{array}%
\right.
\end{equation}

Again we shall need the difference

$$\eta^{\varepsilon,3}_s:=\psi^{\varepsilon,2}_{s}-\psi^{\varepsilon,3}_{s}$$

only at time $s=s_{m2}$ (see Lemma \ref{lemmarest} below) and we note here already that (see (\ref{defpsi2b})
and (\ref{psi3def}))

\begin{eqnarray}\label{defeta3}
\eta^{\varepsilon,3}_{s_{m2}}&=&\left(U^\varepsilon(s_{m2},s_\varepsilon)-V_{s_\varepsilon}^\varepsilon(s_{m2},s_\varepsilon)\right)\psi^{\varepsilon,2}_{s_\varepsilon}\;.
\end{eqnarray}

We use (\ref{cook}) setting $\widetilde{U}^\varepsilon=V_{s_\varepsilon}^\varepsilon$ and
$\overline{U}=U^\varepsilon$ and get

\begin{eqnarray}\label{eta3glg}
\eta^{\varepsilon,3}_{s_{m2}} &=&-i
\int_{s_\varepsilon}^{s_{m2}}U^\varepsilon(s_{m2},w)\frac{1}{\varepsilon}(D_w-D_{s_\varepsilon})V_{s_\varepsilon}^\varepsilon(w,s_\varepsilon)
\psi^{\varepsilon,2}_{s_\varepsilon}dw
%
\;.
\end{eqnarray}

Now Lemma \ref{lemmagrosss} follows from

\begin{lem}\label{lemmarest}

Let $\psi^{\varepsilon,3}_{s}$ and $\eta^{\varepsilon,l}_s$ $l=1,2,3$ given by (\ref{psi3def}), (\ref{defeta1}),
(\ref{defeta2}) and (\ref{defeta3}). Then for $s_\varepsilon=\varepsilon^{\frac{1}{2000}}$,
$k_\varepsilon=\varepsilon^{\frac{4}{9}+\frac{1}{1000}}$ and $K_\varepsilon=\varepsilon^{-4}$

\begin{itemize}

\item[(a)]

\begin{eqnarray*}
\lim_{\varepsilon\rightarrow0}\langle\psi^{\varepsilon,3}_{s_{m2}},\widetilde{\phi}_{0}\rangle=0
\end{eqnarray*}

\item[(b)]

\begin{eqnarray*}
\lim_{s_\varepsilon\rightarrow0}\|\eta^{\varepsilon,l}_{s_{m2}}\|&=&0
\end{eqnarray*}

for $l=1,2,3$.

\end{itemize}
\end{lem}

\vspace{0.5cm}
\noindent\textbf{Proof}\\
\vspace{0.5cm}

To prove the Lemma we need to control wave functions which have $V_{s_\varepsilon}^\varepsilon$ as time
evolution operator. We note that the wave functions above which have $V_{s_\varepsilon}^\varepsilon$ as time
evolution operator have nice features, which we shall summarize below. We will use that such wave functions
(which are smooth in generalized momentum space and not to heavily peaked around $\mathbf{k}=0$ as
$s_\varepsilon\rightarrow0$) show a typical scattering behavior, i.e. they decay fast enough in time.

\begin{lem}\label{propzeta}

Let $\mathcal{R}_{\mathcal{S}}$ be defined by

\begin{equation}\label{schoenefkt}\widehat{\xi}(\mathbf{k},j)\in\mathcal{R}_{\mathcal{S}}\Leftrightarrow
 \|\xi\|=1\text{ and }\xi(\mathbf{x})\text{ has support }\mathcal{S}
\end{equation}

For any $n\in\mathbb{N}_0$ there exist $C_n<\infty$  such that for any
$1>\varepsilon,k_\varepsilon,s_\varepsilon,u>0$, $K_\varepsilon<\infty$ and any compact set
$\mathcal{S}_\varepsilon\subset\mathbb{R}^3$

\begin{itemize}





\item[(a)]

for all $\widehat{\chi}\in(1-\rho_{k_\varepsilon})\mathcal{R}_{\mathcal{S}_\varepsilon}$

\begin{eqnarray}\label{propzetaeq1}
\|\chi\|\leq C_0 \sup_{k<2k_\varepsilon}\left(\left(\mid
s_\varepsilon-\alpha k^{2}\mid+k^3\right)^{-1}\right)
\sqrt{\mid\mathcal{S}_\varepsilon\mid}k_\varepsilon^{\frac{3}{2}}\;.
\end{eqnarray}

and

\begin{eqnarray}\label{propzetaeq2}
\| V_{s_\varepsilon}^\varepsilon(u,0)\chi\|_\infty\leq
C_0\sup_{k<2k_\varepsilon}\left(\left(\mid s_\varepsilon-\alpha
k^{2}\mid+k^3\right)^{-2}\right)\sqrt{\mid\mathcal{S}_\varepsilon\mid}
k_\varepsilon^3\;.
\end{eqnarray}

%

\item[(b)] for all $\chi(\mathbf{x})$ with
$\widehat{\chi}\in\rho_{k_\varepsilon}(1-\rho_{K_\varepsilon})\mathcal{R}_{\mathcal{S}_\varepsilon}$



\begin{eqnarray}\label{l1absch}
\| V_{s_\varepsilon}^\varepsilon(u,0)\chi\|_\infty
&\leq&C_nK_\varepsilon^3\sqrt{\mid\mathcal{S}_\varepsilon\mid}\frac{\varepsilon^n}{u^n}s_\varepsilon^{-3}\left(k_\varepsilon^{-2n}
+s_\varepsilon^{-\frac{3}{2}n}\right) \;.
\end{eqnarray}

\end{itemize}

\end{lem}

\vspace{0.5cm}
\noindent\textbf{Proof of Lemma \ref{propzeta} (a) formula (\ref{propzetaeq1})}\\

Let $\widehat{\chi}\in(1-\rho_{k_\varepsilon})\mathcal{R}_{\mathcal{S}_\varepsilon}$. Writing
$\widehat{\chi}=\widehat{\eta}(\mathbf{k},j)(1-\rho_{k_\varepsilon})$ with $\widehat{\eta}\in
R_{\mathcal{S}_\varepsilon}$ in view of (\ref{defrho2}) we have

\begin{eqnarray}
\|\widehat{\chi}\|&=&\left(
\int
  \mid\widehat{\eta}(\mathbf{k},j)(1-\rho_{k_\varepsilon})\mid^2d^{3}k\right)^{\frac{1}{2}}
\nonumber\\&\leq&\label{esteta}
  \|\widehat{\chi}\|\leq\sup_{k<2k_\varepsilon}\{\mid\widehat{\eta}(\mathbf{k},j)\mid\}\|1-\rho_{k_\varepsilon}\|
\end{eqnarray}

Substituting $p=\frac{k}{k_\varepsilon}$ yields in view of (\ref{defrho2})

\begin{eqnarray}\label{substk0raus}
\|\rho_{k_\varepsilon}-1\|=k_\varepsilon^{\frac{3}{2}}\left(\int\mid\rho(p)-1\mid^2d^3p\right)^{\frac{1}{2}}
\end{eqnarray}

where by (\ref{rhodef2})  $\int\mid\rho(p)-1\mid^2d^3p$ is bounded.

Furthermore

\begin{eqnarray}\label{obenrein}
\sup_{k<2k_\varepsilon}\{\mid\widehat{\eta}(\mathbf{k},j)\mid\}&\leq&
\sup_{k<2k_\varepsilon}\{
\int(2\pi)^{-\frac{3}{2}}\mid\langle\phi^{j}(\mathbf{k},s_\varepsilon,\mathbf{x}),\eta(\mathbf{x})\rangle\mid
d^{3}x \}
\nonumber\\&\leq&
\sup_{k<2k_\varepsilon}\{\|\phi^{j}(\mathbf{k},s_\varepsilon,\cdot)\|\}
\int(2\pi)^{-\frac{3}{2}}\mid\eta(\mathbf{x})\mid d^3x
\end{eqnarray}

Using Schwartz (observing $\widehat{\eta}\in\mathcal{R}_{\mathcal{S}_\varepsilon}$)

\begin{eqnarray}\label{l1l2}
\|\eta\|_1&=& \left|\int\mid\eta(\mathbf{x})\mid d^{3}x\right|
\nonumber\\&=&\left|\int1_{\mathcal{S}_\varepsilon}\mid\eta(\mathbf{x})\mid d^{3}x\right|
\nonumber\\&\leq&\|\eta(\mathbf{x})\|\sqrt{\mid\mathcal{S}_\varepsilon\mid}=\sqrt{\mid\mathcal{S}_\varepsilon\mid}
\end{eqnarray}

For
$\sup_{k<2k_\varepsilon}\{\|\phi^{j}(\mathbf{k},s_\varepsilon,\cdot)\|_\infty\}$
we have by Corollary \ref{geprob} (c) formula (\ref{sternla}) that

\begin{equation}\label{phiklein}
\sup_{k<2k_\varepsilon}\{\|\phi^{j}(\mathbf{k},s_\varepsilon,\cdot)\|_\infty\}\leq
\sup_{k<2k_\varepsilon}\{C\left(\mid s_\varepsilon-\alpha
k^{2}\mid+k^{3}\right)^{-1}\}\;.
\end{equation}

Hence for (\ref{obenrein})

\begin{equation}\label{supchi}
\sup_{k<2k_\varepsilon}\{\mid\widehat{\eta}(\mathbf{k},j)\mid\}\leq
\sup_{k<2k_\varepsilon}\{C\sqrt{\mid\mathcal{S}_\varepsilon\mid}\left(\mid s_\varepsilon-\alpha
k^{2}\mid+k^{3}\right)^{-1}\}\;.
\end{equation}

This and (\ref{substk0raus}) in (\ref{esteta}) yield (\ref{propzetaeq1}).

\vspace{0.5cm}
\noindent\textbf{Proof of Lemma \ref{propzeta} (a) formula (\ref{propzetaeq2})}\\

As above and in view of Corollary \ref{geprob} (b)

\begin{eqnarray*}
&&\| V_{s_\varepsilon}^\varepsilon(u,0)\chi\|_\infty
\\&&\hspace{0.5cm}\leq\nonumber\|\int(2\pi)^{-\frac{3}{2}}\mid
V_{s_\varepsilon}^\varepsilon(u,0)
\widehat{\eta}(\mathbf{k},j)(1-\rho_{k_\varepsilon}(\mathbf{k}))\phi^{j}(\mathbf{k},s_\varepsilon,\mathbf{x})\mid
d^3k\|_\infty
\\&&\hspace{0.5cm}=\nonumber\|\int(2\pi)^{-\frac{3}{2}}\mid e^{-\frac{i}{\varepsilon}E_ku}
\widehat{\eta}(\mathbf{k},j)(1-\rho_{k_\varepsilon}(\mathbf{k}))\phi^{j}(\mathbf{k},s_\varepsilon,\mathbf{x})\mid
d^3k\|_\infty
\\&&\hspace{0.5cm}\leq\sup_{k<2k_\varepsilon}\{\mid\widehat{\eta}(\mathbf{k},j)\mid\}\sup_{k<2k_\varepsilon}\{\|\phi^{j}(\mathbf{k},s_\varepsilon,\cdot)\|_\infty\}\int(2\pi)^{-\frac{3}{2}}\mid(1-\rho_{k_\varepsilon}(\mathbf{k}))\mid
d^3k\;.
\end{eqnarray*}

By (\ref{phiklein}) and (\ref{supchi}) and substituting $p=\frac{k}{k_\varepsilon}$ we can find a $C<\infty$
such that

\begin{eqnarray}\label{l1l2b}
\|
V_{s_\varepsilon}^\varepsilon(u,0)\chi\|_\infty&\leq&C\sup_{k<2k_\varepsilon}\{\left(\mid
s_\varepsilon-\alpha
k^{2}\mid+k^{3}\right)^{-2}\}\sqrt{\mid\mathcal{S}_\varepsilon\mid}k_\varepsilon^3\;.
\end{eqnarray}

\vspace{0.5cm}
\noindent\textbf{Proof of Lemma \ref{propzeta} (b)}\\

Using Corollary \ref{geprob} (b) and (\ref{propagsa}) we have that

\begin{eqnarray*}
V_{s_\varepsilon}^\varepsilon(u,0)\chi(\mathbf{x})&=&V_{s_\varepsilon}^\varepsilon(u,0)\sum_{j=1}^{4}\int(2\pi)^{-\frac{3}{2}}
  \phi^{j}(\mathbf{k},s_\varepsilon,\mathbf{x})\widehat{\chi}(\mathbf{k},j)d^{3}k
\\&=&\sum_{j=1}^{4}\int(2\pi)^{-\frac{3}{2}}
 \exp\left(-\frac{i}{\varepsilon} uE_k\right) \phi^{j}(\mathbf{k},s_\varepsilon,\mathbf{x})\widehat{\chi}(\mathbf{k},j)d^{3}k
\end{eqnarray*}

We estimate the right hand side via stationary phase method, i.e. we integrate by parts. Using
$\frac{i\varepsilon E_k}{ku}\partial_k\exp\left(-\frac{i}{\varepsilon}
uE_k\right)=\exp\left(-\frac{i}{\varepsilon} uE_k\right)$ n partial integrations yield
 - writing

$$\left(\partial_k\frac{E_k}{k}\right)^n:=\partial_k\frac{E_k}{k} \partial_k\frac{E_k}{k} \ldots$$

where $\partial_k$ acts on everything to the right -

\begin{eqnarray*}
V_{s_\varepsilon}^\varepsilon(u,0)\chi(\mathbf{x})&=&(-i\frac{\varepsilon}{u})^n\sum_{j=1}^{4}\int_{0}^\infty\int(2\pi)^{-\frac{3}{2}}
\exp\left(-\frac{i}{\varepsilon} uE_k\right)
\\&&\left(\left(\partial_k\frac{E_k}{k}\right)^n\phi^{j}(\mathbf{k},s_\varepsilon,\mathbf{x})
\widehat{\chi}(\mathbf{k},j)k^2\right)d\Omega dk
\\&=&(-i\frac{\varepsilon}{u})^n\sum_{j=1}^{4}\int k^{-2}(2\pi)^{-\frac{3}{2}}
\exp\left(-\frac{i}{\varepsilon} uE_k\right)
\\&&\left(\left(\partial_k\frac{E_k}{k}\right)^n\phi^{j}(\mathbf{k},s_\varepsilon,\mathbf{x})
\widehat{\chi}(\mathbf{k},j)k^2\right) d^3k
\end{eqnarray*}

Since $\rho_{k_\varepsilon}(\mathbf{k})=0$ for $k\leq k_\varepsilon$ and $k\geq K_\varepsilon$

\begin{eqnarray*}
&&\| V_{s_\varepsilon}^\varepsilon(u,0)\chi\|_\infty
\\&&\hspace{0.5cm}\leq(\frac{\varepsilon}{u})^n \frac{4}{3}\pi K_\varepsilon^3 \sup_{k>k_\varepsilon}\|
k^{-2}\left(\left(
\partial_k\frac{E_k}{k}\right)^n\sum_{j=1}^{4}\phi^{j}(\mathbf{k},s_\varepsilon,\cdot)
\widehat{\chi}(\mathbf{k},j)k^2\right)\|_\infty\;.
\end{eqnarray*}

Since
$\widehat{\chi}_\varepsilon\in\rho_{k_\varepsilon}(1-\rho_{K_\varepsilon})\mathcal{R}_{\mathcal{S}_\varepsilon}$
(\ref{propzetaeq2}) holds, if for any $n\in\mathbb{N}_0$ there exists a $C_n<\infty$ such that

\begin{eqnarray}\label{difR}
&&\sup_{k>k_\varepsilon,\widehat{\eta}\in\mathcal{R}_{\mathcal{S}}}\|
k^{-2}\left(\left(\partial_k\frac{E_k}{k}\right)^n\rho_{k_\varepsilon}
(1-\rho_{K_\varepsilon})\sum_{j=1}^{4}\widehat{\eta}(\mathbf{k},j)
\phi^{j}(\mathbf{k},s_\varepsilon,\cdot)k^2\right)\|_\infty
\nonumber\\&&\hspace{2cm}\leq\;\;
C_n\sqrt{\mid\mathcal{S}_\varepsilon\mid}s_\varepsilon^{-3}\left(k_\varepsilon^{-2n}
+s_\varepsilon^{-\frac{3}{2}n}\right)\;.
\end{eqnarray}

For this we first show that for any $j,l,r\in\mathbb{N}_0$ there exist $C_{j,l,r}$ with

\begin{eqnarray}\label{va1a}
\left(\partial_k\frac{E_k}{k}\right)^nk^2 f(k)&=&\sum_{j+l+r=n}C_{j,l,r}E_k^{n-2r} k^{-n-l+r+2}
\partial_k^j f(k)\;.
\end{eqnarray}

We prove this equation by induction over $n$. For $n=0$ (\ref{va1a}) follows trivially. Assume that (\ref{va1a})
holds for some $n\in\mathbb{N}$. It follows that

\begin{eqnarray*}
\left(\partial_k\frac{E_k}{k}\right)^{n+1}k^2 f(k)&=&
\partial_k\frac{E_k}{k}\left(\partial_k\frac{E_k}{k}\right)^nk^2 f(k)
\\&=&\partial_k\frac{E_k}{k}\sum_{j+l+r=n}C_{j,l,r}E_k^{n-2r} k^{-n+2-l+r}\partial_k^j f(k)
\\&=&\partial_k\sum_{j+l+r=n}C_{j,l,r}E_k^{n-2r+1} k^{(-n-1+2)-l+r}\partial_k^j f(k)
\\&=&\sum_{j+l+r=n}C_{j,l,r}\left(\partial_kE_k^{n-2r+1}\right) k^{(-n-1+2)-l+r}\partial_k^j f(k)
\\&&+\sum_{j+l+r=n}C_{j,l,r}E_k^{n-2r+1} \left(\partial_kk^{(-n-1+2)-l+r}\right)\partial_k^j f(k)
\\&&+\sum_{j+l+r=n}C_{j,l,r}E_k^{n-2r+1} k^{-n+3-l+r}\partial_k^{j+1} f(k)
\end{eqnarray*}

Using that $E_k=\sqrt{k^2+m^2}$ we have that

$$\partial_k E_k ^n = n E_k^{n-1} \partial_k\sqrt{k^2+m^2}=n E_k^{n-2}k\;.$$

Setting $\widetilde{n}=n+1$, $\widetilde{j}=j+1$, $\widetilde{l}=l+1$ and $\widetilde{r}=r+1$ yields

\begin{eqnarray*}
\left(\partial_k\frac{E_k}{k}\right)^{n+1}k^2 f(k)
&=&\sum_{j+l+\widetilde{r}=\widetilde{n}}C_{j,l,r}E_k^{\widetilde{n}-2\widetilde{r}}
k^{-\widetilde{n}+2-l+\widetilde{r}}\partial_k^j f(k)
\\&&+\sum_{j+l+r=\widetilde{n}}C_{j,l,r}E_k^{\widetilde{n}-2r} k^{-\widetilde{n}+2-\widetilde{l}+r}\partial_k^j
f(k)
\\&&+\sum_{\widetilde{j}+l+r=\widetilde{n}}C_{\widetilde{j},l,r}E_k^{\widetilde{n}-2r}
k^{-\widetilde{n}+2-l+r}\partial_k^{j+1} f(k)
\end{eqnarray*}

for appropriate $C_{\widetilde{j},l,r}<\infty$, $C_{j,\widetilde{l},r}<\infty$ and
$C_{j,l,\widetilde{r}}<\infty$, and (\ref{va1a}) follows  for $\widetilde{n}=n+1$. Induction yields that
(\ref{va1a}) holds for all $n\in\mathbb{N}_0$.

Note that for $k\rightarrow 0 $\\$k^{-2}E_k^{n-2r} k^{-n+2-l+r}$ is of order $k^{-n-l+r}$. For
$k\rightarrow\infty$ $E_k$ is of order $k$, hence $k^{-2} E_k^{n-2r} k^{-n+2-l+r}$ is of order $k^{-l-r}$ (hence
bounded for large $k$). Since we only observe $k_\varepsilon\rightarrow0$ it follows with (\ref{va1a}) that for
any $n,j\in\mathbb{N}_0$ there exist $C_{n,j}<\infty$ such that

\begin{eqnarray}\label{va1}
\mid k^2\left(\partial_k\frac{E_k}{k}\right)^nk^2 f(k)\mid&
\leq&\sum_{j=0}^n C_{n,j} k^{-2n+j} \mid \partial_k^j f(k)\mid\;.
\end{eqnarray}

Remember that due to Corollary \ref{geprob} (c),

\begin{eqnarray}
\label{va5}\|\partial_k^n\phi^{j}(\mathbf{k},s_\varepsilon,\cdot)\|_\infty&\leq&
C_n \left(1+k^n\left(\mid s_\varepsilon-\alpha
k^{2}\mid+k^{3}\right)^{-n-2}\right)
\;.
\end{eqnarray}

Next we show that

\begin{eqnarray}
\label{va2}\sup_{k>k_\varepsilon}\{\mid \partial_k^n
\rho_{k_\varepsilon}(\mathbf{k})\mid\}&\leq&C_nk_\varepsilon^n
%
%
%
\\\sup_{\widehat{\eta}\in\mathcal{R}_{\mathcal{S}}}\label{va4}\mid \partial_k^n\widehat{\eta}(\mathbf{k},j)\mid&\leq&C_n\sqrt{\mathcal{S}_\varepsilon}
\left(1+k^n\left(\mid s_\varepsilon-\alpha k^{2}\mid+k^{3}\right)^{-n-2}\right) \;.
\end{eqnarray}

We start with (\ref{va2}). Using the definition of $\rho_{k_\varepsilon}$ (\ref{defrho2}) and substituting
$\mathbf{k}=k_\varepsilon\mathbf{p}$ yields

$$\sup_{k>k_\varepsilon}\{\mid \partial_k^n \rho_{k_\varepsilon}(\mathbf{k})\mid\}\leq
k_\varepsilon^{-n}\sup_{\mathbf{p}\in\mathbb{R}^3}\{\mid \partial_p^n \rho(\mathbf{p})\mid\}$$

Since $\rho(p)\in C^\infty$  (\ref{va2}) follows.




it is left to prove (\ref{va4}). Let $\widehat{\eta}\in\mathcal{R}_{\mathcal{S}_\varepsilon}$. Using Corollary
\ref{geprob} (b) we have that

\begin{eqnarray*}
\mid\partial_k^n\widehat{\eta}(\mathbf{k},j)\mid&=&\left|\partial_k^n\int(2\pi)^{-\frac{3}{2}}\langle\phi^{j}(\mathbf{k},s_\varepsilon,\mathbf{x}),\eta(\mathbf{x})\rangle
d^{3}x\right|
\nonumber\\&=&\left|\int(2\pi)^{-\frac{3}{2}}\langle\partial_k^n\phi^{j}(\mathbf{k},s_\varepsilon,\mathbf{x}),\eta(\mathbf{x})\rangle
d^{3}x\right|
\nonumber\\&\leq&\|\partial_k^n\phi^{j}(\mathbf{k},s_\varepsilon,\cdot)\|_\infty
\left|\int(2\pi)^{-\frac{3}{2}}\mid\eta(\mathbf{x})\mid
d^{3}x\right| \;.
\end{eqnarray*}

Using (\ref{l1l2}) and (\ref{va5}) (\ref{va4}) follows.


It is left to show how (\ref{va1}) - (\ref{va4}) imply (\ref{difR}).

Using the product rule of differentiation, (\ref{va5}) - (\ref{va4}) yield that for any $n,j\in \mathbb{N}$
there exist $C_{n,j}<\infty$ such that

\begin{eqnarray*}
&&\|
\left(\partial_k^n\rho_{k_\varepsilon}\widehat{\eta}(\mathbf{k},j)\phi^{j}(\mathbf{k},s_\varepsilon,\cdot)\right)\|_\infty
\\&&\hspace{2cm}\leq\;\;\sum_{l=0}^n C_{n,l} \sqrt{\mid\mathcal{S}_\varepsilon\mid}\left(1+k^l\left(\mid s_\varepsilon-\alpha
k^{2}\mid+k^{3}\right)^{-l-2}\right)k_\varepsilon^{-n+l}\;.
%
\end{eqnarray*}

The sum will now be estimated by $n$-times the largest summand which however depends on $k_\varepsilon$. Hence
there exists for any $n\in\mathbb{N}$ a $C_n<\infty$ such that

\begin{eqnarray*}
&&\sup_{\widehat{\eta}\in\mathcal{R}_{\mathcal{S}}}\|
\partial_k^n\rho_{k_\varepsilon}\widehat{\eta}(\mathbf{k},j)\phi^{j}(\mathbf{k},s_\varepsilon,\cdot)\|_\infty
\\&&\leq C_n\sqrt{\mid\mathcal{S}_\varepsilon\mid}\left(1+\left(k_\varepsilon^{-n}\left(\mid s_\varepsilon-\alpha
k^{2}\mid+k^{3}\right)^{-2}\right)+\left(k^n\left(\mid s_\varepsilon-\alpha k^{2}\mid+k^{3}\right)^{-2-n}\right)
\right)
\end{eqnarray*}

With (\ref{va1}) it follows that for any $n,j\in\mathbb{N}_0$ there exist $C_{n,j}<\infty$ such that

\begin{eqnarray*}
&&\sup_{k>k_\varepsilon,\widehat{\eta}\in\mathcal{R}_{\mathcal{S}}}\|
k^{-2}\left(\partial_k\frac{E_k}{k}\right)^nk^2
\rho_{k_\varepsilon}\widehat{\eta}(\mathbf{k},j)\phi^{j}(\mathbf{k},s_\varepsilon,\cdot)\|_\infty
\\&&\hspace{2cm}\leq\sqrt{\mid\mathcal{S}_\varepsilon\mid}\sum_{j=0}^n C_{n,j}k_\varepsilon^{-2n}\sup_{k>k_\varepsilon}\left(\mid
s_\varepsilon-\alpha k^{2}\mid+k^{3}\right)^{-2}
\\&&\hspace{2.5cm}+\sqrt{\mid\mathcal{S}_\varepsilon\mid}\sum_{j=0}^n
C_{n,j}k_\varepsilon^{-2n+2j}\sup_{k>k_\varepsilon}\left(1+\left(\mid s_\varepsilon-\alpha
k^{2}\mid+k^{3}\right)^{-2-j}\right) \;.
\end{eqnarray*}

Note that the supremum $\sup_{k>k_\varepsilon}\left(\mid s_\varepsilon-\alpha k^{2}\mid+k^{3}\right)^{-1}$ is
realized at the resonance $\alpha k\approx s_\varepsilon$. Hence there exists a constant $C<\infty$ such that

$$\sup_{k<2k_\varepsilon}\left(\mid s_\varepsilon-\alpha k^{2}\mid+k^{3}\right)^{-1}\leq\sup_{\mathbf{k}\in\mathbb{R}^3}\left(\mid s_\varepsilon-\alpha k^{2}\mid+k^{3}\right)^{-1}<C
s_\varepsilon^{-\frac{3}{2}}$$

It follows that

\begin{eqnarray*}
&&\sup_{k>k_\varepsilon,\widehat{\eta}\in\mathcal{R}_{\mathcal{S}}}\|
k^{-2}\left(\partial_k\frac{E_k}{k}\right)^nk^2
\rho_{k_\varepsilon}\widehat{\eta}(\mathbf{k},j)\phi^{j}(\mathbf{k},s_\varepsilon,\cdot)\|_\infty
\\&&\hspace{2cm}\leq\sqrt{\mid\mathcal{S}_\varepsilon\mid}\sum_{j=0}^n C_{n,j}
\left(1+k_\varepsilon^{-2n}s_\varepsilon^{-3}+k_\varepsilon^{-2n+2j}s_\varepsilon^{-3-\frac{3}{2}j} \right)\;.
\end{eqnarray*}

As above the sum will be estimated by $n$-times the largest summand which again depends on $k_\varepsilon$.
Using that $1<s_\varepsilon^{-\frac{3}{2}n}$ (\ref{difR}) follows.

\begin{flushright}$\Box$\end{flushright}

\begin{lem}\label{lemprop}

Let $\psi^{\varepsilon,1}_s$ be defined as above (see (\ref{psi1def})), $K^\varepsilon=\varepsilon^{-4}$. Then
there exists a $C>0$ such that

$$\|\mathcal{F}_w( \psi^{\varepsilon,1}_s)\rho_{K_\varepsilon}\| <C\varepsilon^2$$

and
$$\|\mathcal{F}_w(A \psi^{\varepsilon,1}_s)\rho_{K_\varepsilon}\| <C\varepsilon^2$$

for all $w,s>0$.

\end{lem}

\vspace{0.5cm}
\noindent\textbf{Proof of Lemma \ref{lemprop}}\\

We first show, that the energy of $\psi^{\varepsilon,1}_s$ is bounded uniform in $\varepsilon>0$ and $s>0$.

Let $B$ be either $1$ or $A$. We have that

\begin{eqnarray*}
\partial_s\langle B\psi^{\varepsilon,1}_s ,D_sB\psi^{\varepsilon,1}_s\rangle&=&\langle B\left(\partial_s\psi^{\varepsilon,1}_s\right) ,D_sB\psi^{\varepsilon,1}_s\rangle
+\langle B\psi^{\varepsilon,1}_s ,\left(\partial_s D_s\right)B\psi^{\varepsilon,1}_s\rangle
\\&&+\langle B\psi^{\varepsilon,1}_s  ,D_sB\partial_s\psi^{\varepsilon,1}_s\rangle
\\&=&\langle B\frac{i}{\varepsilon}D_s\psi^{\varepsilon,1}_s ,D_sB\psi^{\varepsilon,1}_s\rangle
+\langle B \psi^{\varepsilon,1}_s ,\left(\partial_s D_s\right)B\psi^{\varepsilon,1}_s\rangle
\\&&+\langle B\psi^{\varepsilon,1}_s  ,D_sB\frac{i}{\varepsilon}D_s\psi^{\varepsilon,1}_s\rangle
\\&=&\langle B\psi^{\varepsilon,1}_s ,\left(\partial_s D_s\right)B\psi^{\varepsilon,1}_s\rangle
\\&=&\langle B\psi^{\varepsilon,1}_s \left(\partial_s\varphi(s)A\right),B\psi^{\varepsilon,1}_s\rangle
\\&\leq&C\| A\|_\infty^3\partial_s\varphi(s)
\end{eqnarray*}

Hence observing $\psi^{\varepsilon,1}_0=\phi^\varepsilon$ and $D_0\phi_0=m\phi_0$

\begin{eqnarray*}
\langle B\psi^{\varepsilon,1}_s, D_s
B\psi^{\varepsilon,1}_s\rangle&\leq&C\|
A\|_\infty^3\left(\varphi(s)-\varphi(0)\right)+\langle
B\phi^\varepsilon, D_0 B\phi^\varepsilon\rangle
\\&\leq&C\| A\|_\infty^3\left(\varphi(s)-\varphi(0)\right)+\langle
B\phi^\varepsilon, D_0 B(1-\rho_{\varepsilon^{\frac{1}{100}}})\phi_0\rangle
\\&=&C\| A\|_\infty^3\left(\varphi(s)-\varphi(0)\right)+\langle
B\phi^\varepsilon ,B(1-\rho_{\varepsilon^{\frac{1}{100}}})D_0\phi_0\rangle\\&&+\langle B\phi^\varepsilon
,\nabla\hspace{-0.3cm}/\hspace{0.03cm}\left(B(1-\rho_{\varepsilon^{\frac{1}{100}}})\right)\phi_0\rangle
\\&=&C\| A\|_\infty^3\left(\varphi(s)-\varphi(0)\right)+m\| A\|_\infty^2\langle
\phi^\varepsilon ,\phi^\varepsilon\rangle\\&&+\langle B\phi^\varepsilon
,\nabla\hspace{-0.3cm}/\hspace{0.03cm}\left(B(1-\rho_{\varepsilon^{\frac{1}{100}}})\right)\phi_0\rangle\;.
\end{eqnarray*}

Since $D_s-D_w=(\varphi(s)-\varphi(w))A$ it follows that

\begin{eqnarray*} \mid\langle B\psi^{\varepsilon,1}_s
,D_w B\psi^{\varepsilon,1}_s\rangle\mid&\leq&\mid\langle
B\psi^{\varepsilon,1}_s
(\varphi(s)-\varphi(w))A,B\psi^{\varepsilon,1}_s\rangle\mid+ \|
A\|_\infty^3\left(\varphi(s)-\varphi(0)\right)\\&&+m\|
A\|_\infty^2\langle \phi^\varepsilon
,\phi^\varepsilon\rangle+\langle B\phi^\varepsilon
,\nabla\hspace{-0.3cm}/\hspace{0.03cm}\left(B(1-\rho_{\varepsilon^{\frac{1}{100}}})\right)\phi_0\rangle\;.
\end{eqnarray*}

Noting that

$$\|\nabla\hspace{-0.3cm}/\hspace{0.03cm}(1-\rho_{\varepsilon^{\frac{1}{100}}})\|_\infty=
\varepsilon^{\frac{1}{100}}\|\nabla\hspace{-0.3cm}/\hspace{0.03cm}(1-\rho)\|_\infty\;,$$

and that $\|\nabla\hspace{-0.3cm}/\hspace{0.03cm}
A\|_\infty<\infty$ we have that

$$\mid\langle B\psi^{\varepsilon,1}_s D_w,B\psi^{\varepsilon,1}_s\rangle\mid<C\;\;\text{for}\;\varepsilon<1\;\;\text{and for all }s,w\;.$$

We calculate the scalar product on the left hand side in generalized Fourier space using that
$E_k=\sqrt{m^2+k^2}>k$

\begin{eqnarray*}
C>\langle B \psi^{\varepsilon,1}_s D_w,B \psi^{\varepsilon,1}_s\rangle&=&\int\mathcal{F}_w(B
\psi^{\varepsilon,1}_s)(\mathbf{k},j) E_k \mathcal{F}_w(B \psi^{\varepsilon,1}_s)(\mathbf{k},j)d^3k
\\&\geq&\int\rho_{K_\varepsilon}(\mathbf{k})\mathcal{F}_w(B \psi^{\varepsilon,1}_s)(\mathbf{k},j) E_k
\rho_{K_\varepsilon}(\mathbf{k})\mathcal{F}_w(B \psi^{\varepsilon,1}_s)(\mathbf{k},j)d^3k
\\&\geq&K_\varepsilon\int\rho_{K_\varepsilon}(\mathbf{k})\mathcal{F}_w(B \psi^{\varepsilon,1}_s)(\mathbf{k},j)
\rho_{K_\varepsilon}(\mathbf{k})\mathcal{F}_w(B \psi^{\varepsilon,1}_s)(\mathbf{k},j)d^3k
\\&=& K_\varepsilon\|\rho_{K_\varepsilon}\mathcal{F}_w(B \psi^{\varepsilon,1}_s)\|^2\;.
\end{eqnarray*}

Hence $\varepsilon^{-2}\|\rho_{K_\varepsilon}\mathcal{F}_w(B
\psi^{\varepsilon,1}_s)\|$ is bounded and Lemma \ref{lemprop}
follows.

\begin{flushright}$\Box$\end{flushright}

We shall now provide an estimate, how long the wave function $\psi^{\varepsilon,1}_s$ will stay in the range of
the potential. That time is roughly of the order of $s=\varepsilon^{\frac{1}{3}-\delta}$ (see below). For times
larger than $s=\varepsilon^{\frac{1}{3}-\delta}$ the part of the wave function which is affected by the
potential goes to zero with $\varepsilon\rightarrow0$. Note that we establish the estimate (\ref{decayofpsi2})
only for times $s<\widetilde{s}$. But this suffices already to establish the main result since for times larger
than $\widetilde{s}$ the potential has no more influence on the motion of the wave function, i.e. it evolves
freely and thus behaves like a scattering state going off to infinity.

\begin{lem}\label{corprop}

 Let $\psi^{\varepsilon,1}_s$ be given by (214), $K^\varepsilon=\varepsilon^{-4}$, $\widetilde{s}$ as in
Corollary 8.1. Then there exists a $C>0$ such that

\begin{eqnarray}\label{decayofpsi2}
\|
1_{\mathcal{S}_A}\psi^{\varepsilon,1}_s\|&\leq&\min\{1,Cs^{-\frac{5}{2}}\varepsilon^{\frac{5}{6}-\frac{1}{1000}}\}
\end{eqnarray}

for all $0\leq s\leq \widetilde{s}$.

\end{lem}

\vspace{0.5cm}
\noindent\textbf{Proof of Lemma \ref{corprop}}\\

We shall use that

\begin{eqnarray}\label{einsparallelinf}
\|
1_{\mathcal{S}_A}\chi\|&\leq&\|1_{\mathcal{S}_A}\|=\sqrt{\mid\mathcal{S}_A\mid}\;\;
\|\chi \|_\infty\;\;\;\;\;\;\;\;\;\;\;\;\,\text{for } \chi\in
L^\infty
\\\label{einsparallel}\| 1_{\mathcal{S}_A}\chi\|&\leq&\|1_{\mathcal{S}_A}\|_\infty\;\; \|\chi
\|=\|\chi \|\;\;\text{for } \chi\in L^2\;.
\end{eqnarray}

By (213), (214) and the unitarity of $U^\varepsilon$ - that

\begin{eqnarray}\label{parallelpsi1}
\|1_{\mathcal{S}_A}\psi^{\varepsilon,1}_s\|\leq\|\psi^{\varepsilon,1}_s\|\leq1\;.
\end{eqnarray}

Since for $s<2\varepsilon^{\frac{1}{3}}$ and $\varepsilon$ small enough
$s^{-\frac{5}{2}}\varepsilon^{\frac{5}{6}-\frac{1}{1000}}\gg1$ the difficult part is to show that
(\ref{decayofpsi2}) holds for $s\geq2\varepsilon^{\frac{1}{3}}$.


Let $\widetilde{s}\geq s\geq2\varepsilon^{\frac{1}{3}}$. Using (208) with $s_\varepsilon=s$ and

\begin{equation}\label{ddifferenz}
D_w-D_s=D_0+\varphi(w)A-(D_0+\varphi(s))=(\varphi(w)-\varphi(s))
\end{equation}

we have that

\begin{eqnarray}\label{splitpsipsi}
%
1_{\mathcal{S}_A}\psi^{\varepsilon,1}_s
&=&1_{\mathcal{S}_A}V_{s}^\varepsilon(s,0)\phi^{\varepsilon}\nonumber\\&&+i1_{\mathcal{S}_A}\varepsilon^{-1}\int_{0}^{s}
V_{s}^\varepsilon(s,w)(\varphi(w)-\varphi(s))A\psi^{\varepsilon,1}_w(\mathbf{x})dw
\;.
\end{eqnarray}

Now 

\begin{eqnarray*}
\|
1_{\mathcal{S}_A}V_{s}^\varepsilon(s,0)\phi^{\varepsilon}\|&\leq&
\|
1_{\mathcal{S}_A}V_{s}^\varepsilon(s,0)(1-\rho_{\kappa_{\varepsilon,s}}\mathcal{F}_s(\phi^{\varepsilon})\|
\nonumber\\&&+\|1_{\mathcal{S}_A}
V_{s}^\varepsilon(s,0)(1-\rho_{K_\varepsilon})\rho_{\kappa_{\varepsilon,s}}\mathcal{F}_s(\phi^{\varepsilon})\|
\nonumber\\&&+\|1_{\mathcal{S}_A}
V_{s}^\varepsilon(s,0)\rho_{K_\varepsilon}\mathcal{F}_s(\phi^{\varepsilon}\|)
\end{eqnarray*}

We use (\ref{einsparallelinf}) and (\ref{einsparallel}),the unitarity of $V_s^\varepsilon$ and the isometry in
ordinary and generalized momentum space (see Corollary 8.1 (b)) to obtain

\begin{eqnarray*}
\|
1_{\mathcal{S}_A}V_{s}^\varepsilon(s,0)\phi^{\varepsilon}\|&\leq&C
\|
V_{s}^\varepsilon(s,0)(1-\rho_{\kappa_{\varepsilon,s}}\mathcal{F}_s(\phi^{\varepsilon})\|_\infty
\\&&+C\|
V_{s}^\varepsilon(s,0)(1-\rho_{K_\varepsilon}\rho_{\kappa_{\varepsilon,s}}\mathcal{F}_s(\phi^{\varepsilon})\|_\infty
+\| \rho_{K_\varepsilon}\mathcal{F}_s(\phi^{\varepsilon})\|
\\&=&C
\|
V_{s}^\varepsilon(s,0)(1-\rho_{\kappa_{\varepsilon,s}}\mathcal{F}_s(\phi^{\varepsilon})\|_\infty
\\&&+C\|
V_{s}^\varepsilon(s,0)(1-\rho_{K_\varepsilon})\rho_{\kappa_{\varepsilon,s}}\mathcal{F}_s(\phi^{\varepsilon})\|_\infty
+\| \rho_{K_\varepsilon}\mathcal{F}_s(\phi^{\varepsilon})\|\;.
\end{eqnarray*}

Now we use Lemma \ref{propzeta} (a) (setting $s_\varepsilon=s$,
$k_\varepsilon=\kappa_{\varepsilon,s}$, $u=s$ and
$\widehat{\chi}=(1-\rho_{k_\varepsilon})\frac{\mathcal{F}_s(\phi^{\varepsilon})}{\|
\phi^{\varepsilon}\|}$) on the first, Lemma \ref{propzeta} (b)
(setting $s_\varepsilon=s$,
$k_\varepsilon=\kappa_{\varepsilon,s}$, $u=s$ and
$\widehat{\chi}=\rho_{k_\varepsilon}(\mathbf{k}))(1-\rho_{K_\varepsilon}(\mathbf{k}))\frac{\phi^{\varepsilon}}{\|
\phi^{\varepsilon}\|}$) on the second and Lemma \ref{lemprop} on
the third summand. Hence there exists for any $n\in\mathbb{N}$ a
$C_n<\infty$ and a $C<\infty$ such that

\begin{eqnarray*}
\|
1_{\mathcal{S}_A}V_{s}^\varepsilon(s,0)\phi^{\varepsilon}\|&\leq&
C\sup_{k<2\kappa_{\varepsilon,s}}\left(\left(\mid s-\alpha
k^2\mid+k^3\right)^{-2}\right)\sqrt{\mid\mathcal{T}_\varepsilon\mid}\kappa_{\varepsilon,s}^3
\\&&+C_nK_\varepsilon^3\sqrt{\mid\mathcal{T}_\varepsilon\mid}\frac{\varepsilon^n}{s^n}s^{-3}\left(\kappa_{\varepsilon,s}^{-2n}
+s^{-\frac{3}{2}n}\right)
+C\varepsilon^2
\end{eqnarray*}

where we recall (c.f.(\ref{replsce1})) that $\mathcal{T}_\varepsilon$ is the support of $\phi\varepsilon$.

We now choose

\begin{equation}\label{kappase}
\kappa_{\varepsilon,s}=\varepsilon^{\frac{4999}{10000}}s^{-\frac{1}{2}}\;.
\end{equation}

so that

\begin{eqnarray}\label{zerfalleps1}
\frac{\varepsilon}{s\kappa_{\varepsilon,s}^2}&=&\varepsilon^{\frac{2}{10000}}
\end{eqnarray}

and for $s>2 \varepsilon^{\frac{1}{3}}$

\begin{eqnarray}\label{zerfalleps2}\frac{\varepsilon}{s}s^{-\frac{3}{2}}&=&\varepsilon
s^{-\frac{5}{2}}<2^{-\frac{5}{2}}\varepsilon^{\frac{1}{6}}<\varepsilon^{\frac{2}{10000}}
\\\kappa_{\varepsilon,s}^2&=&\varepsilon^{\frac{9998}{10000}}s^{-1}\ll s\;.
\end{eqnarray}

Hence there exists a $C<\infty$ such that

\begin{eqnarray}\label{untenverb}
\inf_{k<2\kappa_{\varepsilon,s}^2}\left(\mid s-\alpha \kappa_{\varepsilon,s}^{2}\mid+s^{\frac{3}{2}}\right)&>&Cs
%
\end{eqnarray}

It follows that there exists for any $n\in\mathbb{N}$ a $C_n<\infty$ and a $C<\infty$ such that

\begin{eqnarray*}\label{dreisummanden}
\|
1_{\mathcal{S}_A}V_{s}^\varepsilon(s,0)\phi^{\varepsilon}\|&\leq&
Cs^{-2}\sqrt{\mid\mathcal{T}_\varepsilon\mid}\kappa_{\varepsilon,s}^3
+C_nK_\varepsilon^3\sqrt{\mid\mathcal{T}_\varepsilon\mid}s^{-3}\varepsilon^{\frac{2}{10000}n}
+C\varepsilon^2\;.
\end{eqnarray*}

Choosing $n$ large enough the second term decays faster than any polynomial in $\varepsilon$. Using (\ref{tmid})
noting that $\sqrt{\mid T_\varepsilon\mid}<\mid T_\varepsilon\mid$ for small enough $\varepsilon$ we can find a
$C<\infty$ such that

\begin{equation}\label{estphi4}
\| 1_{\mathcal{S}_A}V_{s}^\varepsilon(s,0)\phi^{\varepsilon}\|\leq
Cs^{-\frac{7}{2}}\varepsilon^{-\frac{3}{10000}}\varepsilon^{\frac{14997}{10000}}=Cs^{-\frac{7}{2}}\varepsilon^{\frac{14994}{10000}}
\end{equation}


Next we estimate the second summand in (\ref{splitpsipsi}). Below we will introduce the
$\kappa_{\varepsilon,s-w}$-cutoff. For $w\rightarrow s$ $\kappa_{\varepsilon,s-w}$ goes to infinity. So to keep
the $\kappa_{\varepsilon,s-w}$ cutoff small we use the above estimate only for sufficiently large $s-w$ and
handle $s-w<\sigma_\varepsilon$ for some $\sigma_\varepsilon$ which will be specified later separately. Hence we
split

\begin{eqnarray}\label{splitpsi12}&&1_{\mathcal{S}_A}\int_{0}^{s}
V_{s}^\varepsilon(s,w)(\varphi(w)-\varphi(s))\varepsilon^{-1}A\psi^{\varepsilon,1}_w(\mathbf{x})dw
\nonumber\\&&\hspace{0.5cm}\leq\varepsilon^{-1}\int_{\sigma_\varepsilon}^{s}
\|
1_{\mathcal{S}_A}V_{s}^\varepsilon(s,w)(\varphi(w)-\varphi(s))A\psi^{\varepsilon,1}_w\|
dw
\nonumber\\&&\hspace{0.7cm}+
\varepsilon^{-1}\int_{0}^{\sigma_\varepsilon} \|
1_{\mathcal{S}_A}V_{s}^\varepsilon(s,w)(\varphi(w)-\varphi(s))A\psi^{\varepsilon,1}_w\|
dw
\nonumber\\&&\hspace{0.5cm}=:S_1+S_2\;.
\end{eqnarray}

For $S_1$ we have using (\ref{einsparallel}), (6) and (\ref{parallelpsi1})

\begin{eqnarray}\label{formalnummernau}
S_1&\leq&\int_{\sigma_\varepsilon}^{s} \|
V_{s}^\varepsilon(s,w)\varepsilon^{-1}(\varphi(w)-\varphi(s))A\psi^{\varepsilon,1}_w\|
dw
\nonumber\\&=&\varepsilon^{-1}\int_{\sigma_\varepsilon}^{s}
\|(\varphi(w)-\varphi(s))A\psi^{\varepsilon,1}_w\| dw
\nonumber\\&\leq&\varepsilon^{-1}\int_{\sigma_\varepsilon}^{s}\|
A\|_\infty
\|(\varphi(w)-\varphi(s))1_{\mathcal{S}_A}\psi^{\varepsilon,1}_w\|
dw
\nonumber\\&\leq&
C\varepsilon^{-1}\int_{\sigma_\varepsilon}^{s}(s-w)
\|1_{\mathcal{S}_A}\psi^{\varepsilon,1}_w\| dw\;.
\nonumber\\&\leq& C\varepsilon^{-1}\int_{\sigma_\varepsilon}^{s}(s-w)
 dw=\frac{C}{2}\varepsilon^{-1}(s-\sigma_\varepsilon)^2
\end{eqnarray}

\begin{eqnarray}\label{splitpsi123}
S_2&\leq&\varepsilon^{-1}\int_{0}^{\sigma_\varepsilon}\|
1_{\mathcal{S}_A}V_{s}^\varepsilon(s,w)(1-\rho_{K_\varepsilon})\rho_{\kappa_{\varepsilon,s-w}}(\varphi(w)-\varphi(s))\mathcal{F}_s(A\psi^{\varepsilon,1}_w)\|
dw
\nonumber\\&&+\varepsilon^{-1\int_{0}^{\sigma_\varepsilon}\|
1_{\mathcal{S}_A}V_{s}^\varepsilon(s,w)\rho_{K_\varepsilon}(\varphi(w)-\varphi(s))}\mathcal{F}_s(A\psi^{\varepsilon,1}_w)\|
dw
\nonumber\\&&+\varepsilon^{-1}\int_{0}^{\sigma_\varepsilon}\|
 1_{\mathcal{S}_A}V_{s}^\varepsilon(s,w)(1-\rho_{\kappa_{\varepsilon,s-w}}(\varphi(w)-\varphi(s))\mathcal{F}_s(A\psi^{\varepsilon,1}_w)\|
dw
\nonumber\\&=:&S_3+S_4+S_5\;.
\end{eqnarray}

For $S_3$ we have by (\ref{einsparallelinf})

\begin{eqnarray*}
S_3&\leq&\sqrt{\mid\mathcal{S}_A\mid}\varepsilon^{-1}\int_{0}^{\sigma_\varepsilon}\|
V_{s}^\varepsilon(s,w)(1-\rho_{K_\varepsilon})\rho_{\kappa_{\varepsilon,s-w}}
(\varphi(w)-\varphi(s))\mathcal{F}_s(A\psi^{\varepsilon,1}_w)\|_\infty
dw\;.
\end{eqnarray*}

Applying Lemma \ref{propzeta} (b) (choosing
$k_\varepsilon=\kappa_{\varepsilon,s-w}$, $s_\varepsilon=s$,
$u=s-w$ and
$\widehat{\chi}=(1-\rho_{K_\varepsilon})\rho_{\kappa_{\varepsilon,s-w}}\frac{\mathcal{F}_s(A\psi^{\varepsilon,1}_w)}{\|
A\psi^{\varepsilon,1}_w\|}$) there exists a $C<\infty$ so that

\begin{eqnarray*}
S_3&\leq&C\varepsilon^{-1}\int_{0}^{\sigma_\varepsilon}(\varphi(w)-\varphi(s))
C_nK_\varepsilon^3\sqrt{\mid\mathcal{S}_A\mid}\frac{\varepsilon^n}{(s-w)^n}s^{-3}\\&&\hspace{2cm}\left(\kappa_{\varepsilon,s-w}^{-2n}
+s^{-\frac{3}{2}n}\right)\| A\psi^{\varepsilon,1}_w\| dw\;.
\end{eqnarray*}

Now we choose $\sigma_\varepsilon$ such that this integral decays faster than any polynomial in $\varepsilon$,
i.e. setting

\begin{equation}\label{setsigma}
\sigma_\varepsilon:=s-\varepsilon^{\frac{9998}{10000}} s^{-\frac{3}{2}}\;,
\end{equation}

we get for $w\leq \sigma_\varepsilon$

\begin{equation}\label{sgrk}
s-w\geq \varepsilon^{\frac{9998}{10000}} s^{-\frac{3}{2}}\;\;\Rightarrow\;\;s^{-\frac{3}{2}}\leq
\frac{s-w}{\varepsilon^{\frac{9998}{10000}}}=\kappa_{\varepsilon,s-w}^{-2}\;.
\end{equation}

So there exists a $C<\infty$ such that

\begin{eqnarray*}
S_3&\leq&C\varepsilon^{-1}\int_{0}^{\sigma_\varepsilon}(\varphi(w)-\varphi(s))
C_nK_\varepsilon^3\sqrt{\mid\mathcal{S}_A\mid}\left(\frac{\varepsilon}{(s-w)\kappa_{\varepsilon,s-w}^{2}}\right)^n
s^{-3}\;.
\end{eqnarray*}

Since $\frac{\varepsilon}{(s-w)\kappa_{\varepsilon,s-w}^{2}}=\varepsilon^{\frac{2}{10000}}$ (see
(\ref{zerfalleps1})) it follows that the integrand of $S_3$ decays faster than any polynomial in $\varepsilon$
and all $w\leq \sigma_\varepsilon$, hence there exists a $C<\infty$ such that

\begin{equation}\label{s3dec}
S_3\leq C \varepsilon\;.
\end{equation}

Furthermore (\ref{setsigma}) yields that

\begin{equation}\label{furhtermorefol}
S_1\leq\frac{C}{2}\varepsilon^{\frac{9996}{10000}}s^{-3}
\end{equation}

Next we estimate $S_4$ (see (\ref{splitpsi123})). Using (\ref{einsparallelinf}) and unitarity of
$V_{s}^\varepsilon$ we have

\begin{eqnarray*}
S_4&\leq&\varepsilon^{-1}\int_{0}^{\sigma_\varepsilon}\|
V_{s}^\varepsilon(s,w)\rho_{K_\varepsilon}(\varphi(w)-\varphi(s))\mathcal{F}_s(A\psi^{\varepsilon,1}_w
)\| dw
\\&\leq&\varepsilon^{-1}\int_{0}^{\sigma_\varepsilon}\|
\rho_{K_\varepsilon}(\varphi(w)-\varphi(s))\mathcal{F}_s(A\psi^{\varepsilon,1}_w
)\| dw
\end{eqnarray*}

Lemma \ref{lemprop} yields that there exists a $C<\infty$ such that

\begin{eqnarray}\label{s4dec}
S_4&\leq&C\varepsilon^{-1}\int_{0}^{\sigma_\varepsilon}(\varphi(s)-\varphi(w))\varepsilon^2 dw
\nonumber\\&\leq&C \varepsilon\;.
\end{eqnarray}

For $S_5$ by (\ref{einsparallelinf})

\begin{eqnarray*}
S_5&\leq&\sqrt{\mid\mathcal{S}_A\mid}\varepsilon^{-1}\int_{0}^{\sigma_\varepsilon}\|
V_{s}^\varepsilon(s,w)
(1-\rho_{\kappa_{\varepsilon,s-w}})(\varphi(w)-\varphi(s))\mathcal{F}_s(A\psi^{\varepsilon,1}_w)\|_\infty
dw\;,
\end{eqnarray*}

and using (6) ($\mid\partial_{s}\varphi(s)\mid<C$, hence $\mid\varphi(w)-\varphi(s)\mid<C(w-s)$)

\begin{eqnarray*}
S_5&\leq&C\varepsilon^{-1}\int_{0}^{\sigma_\varepsilon}(s-w)\|
V_{s}^\varepsilon(s,w)
(1-\rho_{\kappa_{\varepsilon,s-w}})\mathcal{F}_s(A\psi^{\varepsilon,1}_w)\|_\infty
dw\;.
\end{eqnarray*}

Applying Lemma \ref{propzeta} (a) (with $s_\varepsilon=s$,
$k_\varepsilon=\kappa_{\varepsilon,s}$, $u=s-w$ and
$\widehat{\chi}=(1-\rho_{\kappa_{\varepsilon,s-w}}\frac{\mathcal{F}_s(A\psi^{\varepsilon,1}_w
)}{\| A\psi^{\varepsilon,1}_w\|}$), there exists a $C<\infty$,
such that

\begin{eqnarray*}
S_5&\leq&C\varepsilon^{-1}\int_{0}^{\sigma_\varepsilon}(s-w)\|
A\psi^{\varepsilon,1}_w\|
\\&&\sup_{k<2\kappa_{\varepsilon,s-w}}\left(\left(\mid s-\alpha
k^{2}\mid+k^{3}\right)^{-2}\right)\sqrt{\mid\mathcal{S}_A\mid} \kappa_{\varepsilon,s-w}^3dw
\end{eqnarray*}

With (256)  it follows that

$$\kappa_{\varepsilon,s-w}^2\ll s-w<s\;,$$

hence there exists a $C<\infty$ such that

$$\inf_{k<2\kappa_{\varepsilon,s-w}^2}\left(\mid s-\alpha
\kappa_{\varepsilon,s}^{2}\mid+s^{\frac{3}{2}}\right)>Cs\;.$$

This and (253) yield that

\begin{eqnarray*}
S_5&\leq&C\varepsilon^{-1}\int_{0}^{\sigma_\varepsilon}(s-w)\|
A\psi^{\varepsilon,1}_w\| s^{-2} \kappa_{\varepsilon,s-w}^3dw
\nonumber\\&=&C\int_{0}^{\sigma_\varepsilon}\| A
\psi^{\varepsilon,1}_w\| s^{-2}
\varepsilon^{\frac{4997}{10000}}(s-w)^{-\frac{1}{2}}dw
\\&\leq&C\int_{0}^{\sigma_\varepsilon}\|
1_{\mathcal{S}_A}\psi^{\varepsilon,1}_w\| s^{-2}
\varepsilon^{\frac{4997}{10000}}(s-w)^{-\frac{1}{2}}dw
\end{eqnarray*}

Summarizing (\ref{splitpsipsi}), (\ref{estphi4}) - (\ref{splitpsi123}),  (\ref{furhtermorefol}) - (\ref{s4dec})
we get that there exists a $C<\infty$ such that for all $2\varepsilon^{\frac{1}{3}}\leq s \leq \widetilde{s}$

\begin{eqnarray*}
\|
1_{\mathcal{S}_A}\psi^{\varepsilon,1}_s\|&\leq&Cs^{-\frac{7}{2}}\varepsilon^{\frac{14994}{10000}}
+C\varepsilon^{\frac{9996}{10000}}s^{-3}
\nonumber\\&&+C \int_{0}^{\sigma_\varepsilon}\|
1_{\mathcal{S}_A}\psi^{\varepsilon,1}_w\| s^{-2}
\varepsilon^{\frac{4997}{10000}}(s-w)^{-\frac{1}{2}}dw\;.
\end{eqnarray*}

For $s\geq2\varepsilon^{\frac{1}{3}}$ we have that
$s^{-\frac{7}{2}}\varepsilon^{\frac{14994}{10000}}=(s^{-\frac{^1}{2}}\varepsilon^{\frac{1}{2}})s^{-3}
\varepsilon^{\frac{9994}{10000}}\ll s^{-3} \varepsilon^{\frac{9994}{10000}}$. Hence there exists a $C<\infty$
such that

\begin{eqnarray}\label{iterzeta}
\| 1_{\mathcal{S}_A}\psi^{\varepsilon,1}_s\|&\leq&C s^{-3}
\varepsilon^{\frac{9994}{10000}}
+C\int_{0}^{s}\| 1_{\mathcal{S}_A}\psi^{\varepsilon,1}_w\| s^{-2}
\varepsilon^{\frac{4997}{10000}}(s-w)^{-\frac{1}{2}}dw\;.
\end{eqnarray}

Furthermore $\| \psi^{\varepsilon,1}_w\|\leq 1$, so

\begin{eqnarray*}
\|
1_{\mathcal{S}_A}\psi^{\varepsilon,1}_s\|&\leq&C\varepsilon^{\frac{9994}{10000}}s^{-3}
+C\int_{0}^{s}s^{-2}
\varepsilon^{\frac{4997}{10000}}(s-w)^{-\frac{1}{2}}dw
\\&=&C\varepsilon^{\frac{9994}{10000}}s^{-3}
+2Cs^{-2} \varepsilon^{\frac{4997}{10000}}[(s-w)^{\frac{1}{2}}]_{0}^{s}
\\&=&C\varepsilon^{\frac{9994}{10000}}s^{-3}
+2Cs^{-\frac{3}{2}} \varepsilon^{\frac{4997}{10000}}
\end{eqnarray*}

Observing that $s>2\varepsilon^{\frac{1}{3}}$ we have

$$\varepsilon^{\frac{9994}{10000}}s^{-3}=s^{-\frac{3}{2}}\varepsilon^{\frac{1}{2}}s^{-\frac{3}{2}} \varepsilon^{\frac{4994}{10000}}<s^{-\frac{3}{2}} \varepsilon^{\frac{4994}{10000}}$$

and hence

\begin{eqnarray}\label{psierstab}
\| 1_{\mathcal{S}_A}\psi^{\varepsilon,1}_s\|&\leq&C
s^{-\frac{3}{2}} \varepsilon^{\frac{4994}{10000}}
\end{eqnarray}

Once more using (\ref{iterzeta}) yields that

\begin{eqnarray*}
\| 1_{\mathcal{S}_A}\psi^{\varepsilon,1}_s\|&\leq&C s^{-3}
\varepsilon^{\frac{9994}{10000}}
+C\int_{0}^{\varepsilon^{\frac{1}{3}} }\|
1_{\mathcal{S}_A}\psi^{\varepsilon,1}_w\| s^{-2}
\varepsilon^{\frac{4997}{10000}}(s-w)^{-\frac{1}{2}}dw
\\&&+C\int_{\varepsilon^{\frac{1}{3}} }^{s}\| 1_{\mathcal{S}_A}\psi^{\varepsilon,1}_w\| s^{-2}
\varepsilon^{\frac{4997}{10000}}(s-w)^{-\frac{1}{2}}dw \;.
\end{eqnarray*}

Inserting (\ref{parallelpsi1}) into the second, (\ref{psierstab}) into the third summand yields

\begin{eqnarray*}
\| 1_{\mathcal{S}_A}\psi^{\varepsilon,1}_s\|
&\leq&C\varepsilon^{\frac{9994}{10000}}s^{-3}
+C\int_{0}^{\varepsilon^{\frac{1}{3}} }s^{-2} \varepsilon^{\frac{4997}{10000}}(s-w)^{-\frac{1}{2}} dw
\nonumber\\&&+C\int_{\varepsilon^{\frac{1}{3}}}^{s}s^{-2} \varepsilon^{\frac{4994}{10000}}(s-w)^{-\frac{1}{2}}
w^{-\frac{3}{2}} \varepsilon^{\frac{4997}{10000}} dw
\end{eqnarray*}

Note that $w\leq\varepsilon^{\frac{1}{3}}$ and $s\geq 2\varepsilon^{\frac{1}{3}}$, hence $s-w\geq \frac{s}{2}$.
Hence

\begin{eqnarray*}
\int_{0}^{\varepsilon^{\frac{1}{3}} }s^{-2} \varepsilon^{\frac{4997}{10000}}(s-w)^{-\frac{1}{2}} dw
&\leq&2^{-\frac{1}{2}}s^{-\frac{5}{2}}\varepsilon^{\frac{4997}{10000}}\int_{0}^{\varepsilon^{\frac{1}{3}} } dw
\\&=&2^{-\frac{1}{2}}s^{-\frac{5}{2}}\varepsilon^{\frac{4997}{10000}+\frac{1}{3}}\;.
\end{eqnarray*}

It follows that

\begin{eqnarray*}
\|
1_{\mathcal{S}_A}\psi^{\varepsilon,1}_s\|&\leq&C\varepsilon^{\frac{9994}{10000}}s^{-3}+Cs^{-\frac{5}{2}}\varepsilon^{\frac{4997}{10000}+\frac{1}{3}}
+Cs^{-2}\varepsilon^{\frac{9991}{10000}}\int_{\varepsilon^{\frac{1}{3}}}^{s} (s-w)^{-\frac{1}{2}}
w^{-\frac{3}{2}} dw
\\&=&C\varepsilon^{\frac{9994}{10000}}s^{-3}+Cs^{-\frac{5}{2}}\varepsilon^{\frac{4997}{10000}+\frac{1}{3}}
+Cs^{-2}\varepsilon^{\frac{9991}{10000}}\int_{\varepsilon^{\frac{1}{3}}}^{\frac{s}{2}} (s-w)^{-\frac{1}{2}}
w^{-\frac{3}{2}} dw
\\&&+Cs^{-2}\varepsilon^{\frac{9991}{10000}}\int_{\frac{s}{2}}^{s}
(s-w)^{-\frac{1}{2}} w^{-\frac{3}{2}} dw
\end{eqnarray*}

Using that

\begin{eqnarray*}
\int_{\varepsilon^{\frac{1}{3}}}^{\frac{s}{2}} (s-w)^{-\frac{1}{2}} w^{-\frac{3}{2}} dw&\leq&
\sqrt{2}s^{-\frac{1}{2}}\int_{\varepsilon^{\frac{1}{3}}}^{\frac{s}{2}}  w^{-\frac{3}{2}}
dw\leq2^{-\frac{1}{2}}s^{-\frac{1}{2}}\varepsilon^{-\frac{1}{6}}
\end{eqnarray*}

and

\begin{eqnarray*}
\int_{\frac{s}{2}}^{s} (s-w)^{-\frac{1}{2}} w^{-\frac{3}{2}}
dw&\leq&(\frac{s}{2})^{-\frac{3}{2}}\int_{\frac{s}{2}}^{s} (s-w)^{-\frac{1}{2}}dw
\leq2^{\frac{1}{2}}s^{-1}
\end{eqnarray*}

it follows that there exists a $C<\infty$ such that

\begin{eqnarray*}
\|
1_{\mathcal{S}_A}\psi^{\varepsilon,1}_s\|&\leq&C\varepsilon^{\frac{9994}{10000}}s^{-3}+Cs^{-\frac{5}{2}}\varepsilon^{\frac{4997}{10000}+\frac{1}{3}}
+Cs^{-\frac{5}{2}}\varepsilon^{\frac{5}{6}-\frac{9}{10000}}
\\&&+Cs^{-3}\varepsilon^{1-\frac{9}{10000}}
\end{eqnarray*}

For $s>\varepsilon^{\frac{1}{3}}$ we have that

\begin{eqnarray*}
\varepsilon^{\frac{9994}{10000}}s^{-3}&=&s^{-\frac{1}{2}}\varepsilon^{\frac{1}{6}+\frac{4}{10000}}s^{-\frac{5}{2}}\varepsilon^{\frac{5}{6}-\frac{1}{1000}}<s^{-\frac{5}{2}}\varepsilon^{\frac{5}{6}-\frac{1}{1000}}
\\s^{-\frac{5}{2}}\varepsilon^{\frac{4997}{10000}+\frac{1}{3}}&=&\varepsilon^{\frac{7}{10000}}s^{-\frac{5}{2}}\varepsilon^{\frac{5}{6}-\frac{1}{1000}}<s^{-\frac{5}{2}}\varepsilon^{\frac{5}{6}-\frac{1}{1000}}
\\s^{-3}\varepsilon^{1-\frac{9}{10000}}&=&s^{-\frac{1}{2}}\varepsilon^{\frac{1}{6}}s^{-\frac{5}{2}}\varepsilon^{\frac{5}{6}-\frac{1}{1000}}<s^{-\frac{5}{2}}\varepsilon^{\frac{5}{6}-\frac{1}{1000}}
\end{eqnarray*}

and (\ref{decayofpsi2}) and thus Lemma \ref{corprop} follows.

\begin{flushright}$\Box$\end{flushright}


\vspace{0.5cm}
\noindent\textbf{Proof of Lemma \ref{lemmarest} (a)}\\

We need to show that

$$\lim_{\varepsilon\rightarrow0}\langle\psi^{\varepsilon,3}_{s_{m2}},\widetilde{\phi}_{0}\rangle=0\;.$$

Using (227) and (224) we have that

\begin{eqnarray*}
\psi^{\varepsilon,3}_{s_{m2}}= V_{s_\varepsilon}(s_{m2},s_\varepsilon)\psi^{\varepsilon,2}_{s_\varepsilon}
&=&V_{s_\varepsilon}^\varepsilon(s_{m2},0)\phi^{\varepsilon,1}+i\int_{0}^{s_\varepsilon}
V_{s_\varepsilon}^\varepsilon(s_{m2},w)\zeta^{\varepsilon,1}_{w}(\mathbf{x})dw\;.
\end{eqnarray*}

Note that by construction
$\widehat{\zeta}^{\varepsilon,1}_{w}(\mathbf{k},j)\in\rho_{k_\varepsilon}(1-\rho_{K_\varepsilon})\mathcal{R}_{\mathcal{S}_A}$
(see (220)) and
$\widehat{\phi}^{\varepsilon,1}=\rho_{k_\varepsilon}(1-\rho_{K_\varepsilon})\mathcal{F}_s(\phi^\varepsilon)$
(see (223)). Hence by Lemma \ref{propzeta} (c) (with
$\chi=\frac{\zeta^{\varepsilon,2}_{w}(\mathbf{x})}{\|\zeta^{\varepsilon,2}_{w}(\mathbf{x})\|}$
and $\chi=\|\phi^{\varepsilon}\|$) we get as above that

\begin{eqnarray}\label{psi3split}
\|\psi^{\varepsilon,3}_{s_{m2}}\|_\infty&\leq&
C_nK_\varepsilon^3\sqrt{\mid\mathcal{T}_\varepsilon\mid}\frac{\varepsilon^n}{s_{m2}^n}s_\varepsilon^3\left(k_\varepsilon^{-2n}
+s_\varepsilon^{-\frac{3}{2}n}\right)
\\&&+\int_{0}^{s_\varepsilon}
C_nK_\varepsilon^3\sqrt{\mid\mathcal{S}_A\mid}\frac{\varepsilon^n}{(s_{m2}-w)^n}s_\varepsilon^3\left(k_\varepsilon^{-2n}
+s_\varepsilon^{-\frac{3}{2}n}\right) dw
\end{eqnarray}

Furthermore since $k_\varepsilon=\varepsilon^{\frac{4}{9}+\frac{1}{1000}}$ and
$s_\varepsilon=\varepsilon^{\frac{1}{2000}}$ (c.f. Lemma \ref{lemmarest})

$$\frac{\varepsilon}{s_{m2}-s_\varepsilon}\left(k_\varepsilon^{-2}+s_{\varepsilon}^{-\frac{3}{2}}\right)<C\varepsilon^{\frac{1}{9}-\frac{2}{1000}}\;.$$

By choosing $n$ large enough

$$\lim_{\varepsilon\rightarrow0}\|\psi^{\varepsilon,3}_{s_{m2}}\|_\infty=0\;,$$
and hence

$$\lim_{\varepsilon\rightarrow0}\langle\psi^{\varepsilon,3}_{s_{m2}},\widetilde{\phi}_{0}\rangle=0\;.$$

\vspace{0.5cm}
\noindent\textbf{Proof of Lemma \ref{lemmarest} (b) for $l=1$}\\

Using unitarity of $U^\varepsilon$ we get in view of (214), (203) and (215)

\begin{eqnarray*}
\lim_{\varepsilon\rightarrow 0}\|\eta^{\varepsilon,1}_{s{m2}}\|&=&
\lim_{\varepsilon\rightarrow
0}\|\psi^{\varepsilon}_{s_{m2}}-\psi^{\varepsilon,1}_{s_{m2}}\|
\\&=&
\lim_{\varepsilon\rightarrow 0}\|\phi^{\varepsilon}-\phi_0\|\;.
\end{eqnarray*}

With (\ref{replsce1}), (221) and (211) it follows that since $\phi_0\in L^2$

\begin{eqnarray*}
\lim_{\varepsilon\rightarrow 0}\|\eta^{\varepsilon,1}_{s{m2}}\|&=&
\lim_{\varepsilon\rightarrow
0}\|\phi_{0}\rho_{\varepsilon^{-\frac{1}{10000}}}\|
\\&\leq&\lim_{\varepsilon\rightarrow 0}\|\phi_{0}(1-1_{\varepsilon^{-\frac{1}{10000}}})\|
=0\;.
\end{eqnarray*}

\vspace{0.5cm}
\noindent\textbf{Proof of Lemma \ref{lemmarest} (b) for $l=2$}\\

Using (226) and the unitarity of the propagators $U^\varepsilon$ and $V_{s_\varepsilon}$ we get that

\begin{eqnarray*}
\|\eta^{\varepsilon,2}_{s_{m2}}\| &\leq&\|\left(
\phi^{\varepsilon}-\phi^{\varepsilon,1}\right)\|
+\int_{0}^{s_\varepsilon}
\|\zeta^\varepsilon_w-\zeta^{\varepsilon,1}_{w}\| dw
\;.
\end{eqnarray*}

Using (223) and (220) we have that

\begin{eqnarray*}
\|\eta^{\varepsilon,2}_{s_{m2}}\|
&\leq&\|\widehat{\phi}^{\varepsilon}\left(1-\rho_{\kappa}(1-\rho_{K_\varepsilon})\right)\|
\\&&+\int_{0}^{s_\varepsilon}
\|\widehat{\zeta}^\varepsilon_w
\left(1-\rho_{k_\varepsilon}(1-\rho_{K_\varepsilon})\right)\| dw
\\&\leq&\|\widehat{\phi}^{\varepsilon}\left(1-\rho_{\kappa}\right)\|
+\|\widehat{\phi}^{\varepsilon}\rho_{K_\varepsilon}\|
\\&&+\int_{0}^{s_\varepsilon}
\|\widehat{\zeta}^\varepsilon_w\left(1-\rho_{k_\varepsilon}\right)\|
dw
+\int_{0}^{s_\varepsilon}
\|\widehat{\zeta}^\varepsilon_w\rho_{K_\varepsilon}\| dw\;.
\end{eqnarray*}

Using Lemma \ref{propzeta} (a)(with
$\chi=\frac{\phi^{\varepsilon}}{\|\phi^{\varepsilon}\|}$) on the
first summand and (with
$\chi=\frac{\zeta^\varepsilon_w}{\|\zeta^\varepsilon_w\|}$) on the
third summand and Lemma \ref{lemprop} on the second and the fourth
summand it follows that there exists a $C<\infty$ such that

\begin{eqnarray*}
\|\eta^{\varepsilon,2}_{s_{m2}}\|&\leq& C
\sup_{k<2k_\varepsilon}\left(\left(\mid s_\varepsilon-\alpha
k^{2}\mid+k^{3}\right)^{-1}\right)
\sqrt{\mid\mathcal{T}_\varepsilon\mid}k_\varepsilon^{\frac{3}{2}}\|\phi^{\varepsilon}\|+C\varepsilon^2
\\&&+C\int_{0}^{s_\varepsilon}
\| \zeta^\varepsilon_w\| \sup_{k<2k_\varepsilon}\left(\left(\mid
s_\varepsilon-\alpha k^{2}\mid+k^{3}\right)^{-1}\right)
\sqrt{\mid\mathcal{S}_A\mid}k_\varepsilon^{\frac{3}{2}} dw
\\&&+C\int_{0}^{s_\varepsilon}(\varphi(s_\varepsilon)-\varphi(w))\varepsilon dw\;.
\end{eqnarray*}

Using (217) and (6)

$$\|\zeta^\varepsilon_w\|\leq C(s_\varepsilon-w)\varepsilon^{-1}
\| 1_{\mathcal{S}_A}\psi^\varepsilon_w\|\;.$$

Furthermore we have that $k_\varepsilon^2\ll s_\varepsilon$, hence there exists a $C<\infty$ such that

$$\sup_{k<2k_\varepsilon}\left(\left(\mid s_\varepsilon-\alpha k^{2}\mid+k^{3}\right)^{-1}\right)<Cs_\varepsilon^{-1}$$

and it follows that

\begin{eqnarray*}
\|\eta^{\varepsilon,2}_{s_{m2}}\|&\leq& C s_\varepsilon^{-1}
\sqrt{\mid\mathcal{T}_\varepsilon\mid}k_\varepsilon^{\frac{3}{2}}\|\phi^{\varepsilon}\|+C\varepsilon^2
\\&&+C\int_{0}^{s_\varepsilon}(s_\varepsilon-w)\varepsilon^{-1}
 \| 1_{\mathcal{S}_A}\psi^\varepsilon_w\|
s_\varepsilon^{-1} \sqrt{\mid\mathcal{S}_A\mid}k_\varepsilon^{\frac{3}{2}} dw
\\&&+C\int_{0}^{s_\varepsilon}(\varphi(s_\varepsilon)-\varphi(w))\varepsilon dw
\;.
\end{eqnarray*}

Note that $\|\phi_0\|=1$, hence with (\ref{einsparallel})
$\|\phi^\varepsilon\|\leq1$, so

\begin{eqnarray*}
\lim_{\varepsilon\rightarrow0}\|\eta^{\varepsilon,2}_{s_{m2}}\|&\equiv&
\lim_{\varepsilon\rightarrow0}C\int_{0}^{s_\varepsilon}(s_\varepsilon-w)\varepsilon^{-1}
 \| 1_{\mathcal{S}_A}\psi^\varepsilon_w\|
s_\varepsilon^{-1} \sqrt{\mid\mathcal{S}_A\mid}k_\varepsilon^{\frac{3}{2}} dw
\\&&+\lim_{\varepsilon\rightarrow0}C\int_{0}^{s_\varepsilon}(\varphi(s_\varepsilon)-\varphi(w))\varepsilon dw
\\&=& \lim_{\varepsilon\rightarrow0}C\int_{0}^{s_\varepsilon}(s_\varepsilon-w)\varepsilon^{-1}
 \| 1_{\mathcal{S}_A}\psi^\varepsilon_w\|
s_\varepsilon^{-1} \sqrt{\mid\mathcal{S}_A\mid}k_\varepsilon^{\frac{3}{2}} dw \;.
\end{eqnarray*}

Next we split the integration and apply on the first integral $\|
1_{\mathcal{S}_A}\psi^\varepsilon_w\|\leq1$ and on the second
integral Lemma \ref{corprop}

\begin{eqnarray*}
\lim_{\varepsilon\rightarrow0}\|\eta^{\varepsilon,2}_{s_{m2}}\|
&=&
\lim_{\varepsilon\rightarrow0}C\int_{0}^{\varepsilon^{\frac{1}{3}}}(s_\varepsilon-w)\varepsilon^{-1}
s_\varepsilon^{-1}\sqrt{\mid\mathcal{S}_A\mid}k_\varepsilon^{\frac{3}{2}} dw
\\&&+\lim_{\varepsilon\rightarrow0}C\int_{\varepsilon^{\frac{1}{3}}}^{s_\varepsilon}(s_\varepsilon-w)
w^{-\frac{5}{2}}\varepsilon^{-\frac{1}{1000}-\frac{1}{6}} s_\varepsilon^{-1}
\sqrt{\mid\mathcal{S}_A\mid}k_\varepsilon^{\frac{3}{2}} dw
\\&\leq&
\lim_{\varepsilon\rightarrow0}C\int_{0}^{\varepsilon^{\frac{1}{3}}}\varepsilon^{-1} k_\varepsilon^{\frac{3}{2}}
dw
+\lim_{\varepsilon\rightarrow0}C[-w^{-\frac{3}{2}}\varepsilon^{-\frac{1}{1000}-\frac{1}{6}}
k_\varepsilon^{\frac{3}{2}} ]_{\varepsilon^{\frac{1}{3}}}^{s_\varepsilon}
\\&\leq&\lim_{\varepsilon\rightarrow0}C\varepsilon^{-\frac{2}{3}} k_\varepsilon^{\frac{3}{2}}
+\lim_{\varepsilon\rightarrow0}C\varepsilon^{-\frac{1}{2}}\varepsilon^{-\frac{1}{1000}-\frac{1}{6}}
k_\varepsilon^{\frac{3}{2}}\;.
\end{eqnarray*}

Using that $k_\varepsilon=\varepsilon^{\frac{4}{9}+\frac{1}{1000}}$ it follows that

$$\varepsilon^{-\frac{2}{3}} k_\varepsilon^{\frac{3}{2}}
=\varepsilon^{\frac{3}{2000}}
$$

and

$$\varepsilon^{-\frac{1}{2}}\varepsilon^{-\frac{1}{1000}-\frac{1}{6}}k_\varepsilon^{\frac{3}{2}}
=\varepsilon^{\frac{1}{2000}}\;,$$

hence

\begin{equation}\label{lims12b}
\lim_{\varepsilon\rightarrow0}\|\eta^{\varepsilon,2}_{s_{m2}}\|=0
\end{equation}

and Lemma \ref{lemmarest} (b) follows for $l=2$.

\vspace{0.5cm}
\noindent\textbf{Proof of Lemma \ref{lemmarest} (b) for $l=3$}\\

Using  (229) we have

\begin{eqnarray*}
\eta^{\varepsilon,3}_{s_{m2}}&=&i\varepsilon^{-1}
\int_{s_\varepsilon}^{s_{m2}}U^\varepsilon(s_{m2},w)(D_w-D_{s_\varepsilon})V_{s_\varepsilon}^\varepsilon(w,s_\varepsilon)
\psi^{\varepsilon,2}_{s_\varepsilon}dw
\\&=&i\varepsilon^{-1}
\int_{s_\varepsilon}^{s_\varepsilon+\varepsilon^{\frac{1}{10}}}U^\varepsilon(s_{m2},w)(D_w-D_{s_\varepsilon})V_{s_\varepsilon}^\varepsilon(w,s_\varepsilon)
\psi^{\varepsilon,2}_{s_\varepsilon}dw
\\&&+i\varepsilon^{-1}
\int_{s_\varepsilon+\varepsilon^{\frac{1}{10}}}^{s_{m2}}U^\varepsilon(s_{m2},w)(D_w-D_{s_\varepsilon})
V_{s_\varepsilon}^\varepsilon(w,s_\varepsilon) \psi^{\varepsilon,2}_{s_\varepsilon}dw
\\&=:&S_1+S_2
\end{eqnarray*}

For $S_1$ we can write (doing the last steps in section ''3. Step'' backwards)

\begin{eqnarray*}
S_1&=&\psi^{\varepsilon,2}_{s_\varepsilon+\varepsilon^{\frac{1}{10}}}-\psi^{\varepsilon,3}_{s_\varepsilon+\varepsilon^{\frac{1}{10}}}
\end{eqnarray*}

Using the definition of $\psi^{\varepsilon,3}_s$ (see (227)) yields in view of (224) that

\begin{eqnarray*}
\psi^{\varepsilon,3}_{s_\varepsilon+\varepsilon^{\frac{1}{10}}}&=&V_{s_\varepsilon}^\varepsilon(s_\varepsilon+\varepsilon^{\frac{1}{10}},s_\varepsilon)\psi^{\varepsilon,2}_{s_\varepsilon}
\\&=&V_{s_\varepsilon}^\varepsilon(s_\varepsilon+\varepsilon^{\frac{1}{10}},0) \phi^{\varepsilon,1} +i\int_{0}^{s_\varepsilon}
V_{s_\varepsilon}^\varepsilon(s_\varepsilon+\varepsilon^{\frac{1}{10}},w)\zeta^{\varepsilon,1}_{w}(\mathbf{x})dw
\end{eqnarray*}

Hence with (224)

$$S_1=i\int_{s_\varepsilon}^{s_\varepsilon+\varepsilon^{\frac{1}{10}}}
V_{s_\varepsilon}^\varepsilon(s_\varepsilon+\varepsilon^{\frac{1}{10}},w)\zeta^{\varepsilon,1}_{w}(\mathbf{x})dw\;.$$

and hence (by unitarity of $V^\varepsilon_{s_\varepsilon}$)

\begin{equation}\label{s11111}
\| S_1\|
\leq\int_{s_\varepsilon}^{s_\varepsilon+\varepsilon^{\frac{1}{10}}}
\|\zeta^{\varepsilon,1}_{w}\| dw\;.
\end{equation}

Using (220), (217) and Lemma \ref{corprop} we have that

\begin{eqnarray*}\|\zeta^{\varepsilon,1}_{w}\|&=&
\|\rho_{k_\varepsilon}(1-\rho_{K_\varepsilon})\mathcal{F}_s(\zeta^\varepsilon_w)\|
\leq\|\zeta^\varepsilon_w\|
\\&=&
\|(\varphi(s_\varepsilon)-\varphi(w))\varepsilon^{-1}
A\psi^{\varepsilon,1}_{w}\|
\leq C \mid w-s_\varepsilon\mid w^{-\frac{5}{2}}\varepsilon^{-\frac{1}{6}-\frac{1}{1000}}\end{eqnarray*}

This in (\ref{s11111}) yields that

\begin{eqnarray*}
\| S_1\|
&\leq&C\int_{s_\varepsilon}^{s_\varepsilon+\varepsilon^{\frac{1}{10}}}
(w-s_\varepsilon)w^{-\frac{5}{2}}\varepsilon^{-\frac{1}{6}-\frac{1}{1000}}dw
\\&\leq&C\varepsilon^{\frac{1}{10}}(\varepsilon^{\frac{1}{10}})s_\varepsilon^{-\frac{5}{2}}\varepsilon^{-\frac{1}{6}-\frac{1}{1000}}\;.
\end{eqnarray*}

Since $s_\varepsilon=\varepsilon^{\frac{1}{2000}}$

$$\| S_1\|\leq C\varepsilon^{\frac{1}{30}-\frac{9}{4000}}\;,$$

hence

\begin{equation}\label{limSfinal}
\lim_{\varepsilon\rightarrow0} \| S_1\| =0\;.
\end{equation}

It is left to estimate $S_2$. We write using (227), the triangle inequality and the unitarity of $U^\varepsilon$

\begin{eqnarray}\label{hierrein}
\| S_2\|&\leq&\varepsilon^{-1}
\int_{s_\varepsilon+\varepsilon^{\frac{1}{10}}}^{s}\|
U^\varepsilon(s,w)(D_w-D_{s_\varepsilon})\psi^{\varepsilon,3}_{w}\|
 dw
\nonumber\\&=&\varepsilon^{-1}
\int_{s_\varepsilon+\varepsilon^{\frac{1}{10}}}^{s}(\varphi(w)-\varphi(s_\varepsilon))\|
A\psi^{\varepsilon,3}_{w}\|
 dw
 \\&\leq&\varepsilon^{-1}\int_{s_\varepsilon+\varepsilon^{\frac{1}{10}}}^{s}(\varphi(w)-\varphi(s_\varepsilon))\| A\|\;\|
\psi^{\varepsilon,3}_{w}\|_\infty
 dw
\end{eqnarray}

To control this we estimate $\|
\psi^{\varepsilon,3}_{s}\|_\infty\;.$

We write using (227) and (224)

\begin{eqnarray*}
\psi^{\varepsilon,3}_{s}&=&V_{s_\varepsilon}^\varepsilon(s,0) \phi^{\varepsilon,1} +i\int_{0}^{s_\varepsilon}
V_{s_\varepsilon}^\varepsilon(s,w)\zeta^{\varepsilon,1}_{w}(\mathbf{x})dw
\end{eqnarray*}

Using Lemma \ref{propzeta} (b) (with
$\chi=\frac{\phi^{\varepsilon}}{\|\phi^{\varepsilon}\|}$) on the
first summand and (with
$\chi=\frac{\zeta^\varepsilon_w}{\|\zeta^\varepsilon_w\|}$) on the
second summand yields

\begin{eqnarray*}
\|\psi^{\varepsilon,3}_{s}\|_\infty&\leq&
C_n\|\phi^{\varepsilon}\|
K_\varepsilon^3\sqrt{\mid\mathcal{T}_\varepsilon\mid}\frac{\varepsilon^n}{s^n}s_\varepsilon^{-3}\left(k_\varepsilon^{-2n}
+s_\varepsilon^{-\frac{3}{2}n}\right)
\\&&+C_n\int_{0}^{s_\varepsilon}
\| \zeta^\varepsilon_w\|
K_\varepsilon^3\sqrt{\mid\mathcal{S}_A\mid}\frac{\varepsilon^n}{(s-w)^n}s_\varepsilon^{-3}\left(k_\varepsilon^{-2n}
+s_\varepsilon^{-\frac{3}{2}n}\right) dw
\end{eqnarray*}

Note that

$$\frac{\varepsilon}{(s-u) k_\varepsilon^2}=\frac{\varepsilon^{\frac{1}{9}-\frac{1}{500}}}{s-u}<\varepsilon^{\frac{1}{90}-\frac{1}{500}}$$

for any $u<s_\varepsilon$ and $s>s_\varepsilon+\varepsilon^{\frac{1}{10}}$  and that $k_\varepsilon^{-2}\gg
s_\varepsilon^{-\frac{3}{2}}$. Hence

$$\|\psi^{\varepsilon,3}_{s}\|_\infty$$

decays for all $s>s_\varepsilon+\varepsilon^{\frac{1}{10}}$ faster than any polynomial in $\varepsilon$.

In view of (\ref{hierrein}) it follows that

\begin{equation}\label{limSfinal2}
\lim_{\varepsilon\rightarrow0} \| S_2\| =0\;.
\end{equation}

This and (\ref{limSfinal}) yield the Lemma.

\begin{flushright}$\Box$\end{flushright}

\vspace{0.5cm}

\section{Acknowledgements}

\vspace{0.5cm}

I am very grateful for all the support and advice by Prof. Dr. Detlef D\"urr. His skills as a scientist and his
helpfulness and patience helped a lot to develop my abilities. I have learned a lot from him. The effort he
takes to support the students and junior scientist in his group are exemplary.

I want to thank Prof. Dr. Gheorghe Nenciu for his interesting talk he gave about spontaneous pair creation. His
talk gave me the motivation to work on this topic and gave me hints, how the problem can be solved. He also
revealed a serious mistake in an older version of this work which made the statement of the Theorem worthless.
Furthermore his own papers and the papers of other authors he suggested to read were very helpful.

I want also to thank Prof. Dr. Herbert Spohn, Prof. Dr. Stefan Teufel and Dr. Roderich Tumulka for helpful
discussions on the topic.

I wish to thank Tilo Moser for his company as a scientist and a friend and for all the discussion we had in the
recent years.


\newpage




\begin{thebibliography}{}

\bibitem{pickl}D\"urr, D. and Pickl, P.:
Flux-across-surfaces theorem for a Dirac-particle, J.\ Math.\
Phys.\ {\bf 44}, 423-465 (2003).

\bibitem{greiner}Greiner, W., M\"uller, B., Rafelski, J.:
Quantum Electrodynamics of Strong Fields, Springer Verlag, Berlin (1985).


\bibitem{ikebe}Ikebe, T.: Eigenfunction expansions assoziated with
the Schr\"odinger operators and their application to scattering
theory, Arch. Rat. Mech. Anal. {\bf 5}, 1-34 (1960)



\bibitem{klaus}Klaus, M.: On couplin constant thresholds and related eigenvalue properties of Dirac operators,
J. Reine Angew.Math. {\bf 362} 197-212 (1985)

\bibitem{nenciu1}Nenciu, G.:
On the adiabatic limit for Dirac particles in external fields,
Commun. Math. Phys. {\bf 76}, 117-128 (1980).

\bibitem{nenciu2}Nenciu, G.:
Existence of spontaneous pair creation in the external field approximation of Q.E.D., Commun. Math. Phys. {\bf
109}, 303-312 (1987).


\bibitem{riesz} Riesz, F., von Sz.-Nagy, B.: Functional Analysis.
New York: F. Ungar. Publ. Co. (1955).



\bibitem{stefan}Teufel, S.:
Adiabatic Perturbation Theory in Quantum Dynamics, Springer Verlag, Berlin (2000).

\bibitem{teufel}Teufel, S.:
The flux-across-surfaces theorem and its implications for scattering theory, Dissertiation an der
Ludwig-Maximilians-Universit\"at, M\"unchen (1999).

\bibitem{thaller} Thaller, B.:The Dirac equation, Springer Verlag,
Berlin (1992).



\end{thebibliography}
\end{document}